\documentclass[twocolumn, superscriptaddress, amsmath,amssymb, amsfonts, aps, prapplied]{revtex4-2}

\usepackage{enumitem} 
\usepackage{graphicx}
\usepackage{color}
\usepackage{dcolumn}
\usepackage{bm}
\usepackage{hyperref}

\newcommand{\up}[1]{\textsuperscript{#1}}
\newcommand{\etal}{\textit{et al.}}
\newcommand{\beq}{\begin{equation}}
\newcommand{\eeq}{\end{equation}}
\newcommand{\beqn}{\begin{eqnarray}}
\newcommand{\eeqn}{\end{eqnarray}}
\newcommand{\he}[1]{\up{#1}He}

\newcommand{\bs}{\textbackslash}

\begin{document}
\title{Superfluid Optomechanics with Phononic Nanostructures}

\author{S. Spence}
\affiliation{Department of Physics, Royal Holloway University of London, Egham, Surrey, TW20 0EX, UK}
\author{Z. X. Koong}
\affiliation{SUPA, Institute of Photonics and Quantum Sciences, Heriot-Watt University,Edinburgh EH14 4AS, Scotland, United Kingdom}
\author{S. A. R. Horsley}
\affiliation{Department of Physics and Astronomy, Stocker Road, University of Exeter, Exeter EX4 4QL, United Kingdom}
\author{X. Rojas}
\email{xavier.rojas@rhul.ac.uk}
\affiliation{Department of Physics, Royal Holloway University of London, Egham, Surrey, TW20 0EX, UK}
\date{\today}

\begin{abstract}
In quantum optomechanics, finding materials and strategies to limit losses has been crucial to the progress of the field. Recently, superfluid \he4 was proposed as a promising mechanical element for quantum optomechanics. This quantum fluid shows highly desirable properties (e.g. extremely low acoustic loss) for a quantum optomechanical system.
In current implementations, superfluid optomechanical systems suffer from external sources of loss, which spoils the quality factor of resonators. In this work, we propose a new implementation, exploiting nanofluidic confinement. Our approach, based on acoustic resonators formed within phononic nanostructures, aims at limiting radiation losses to preserve the intrinsic properties of superfluid \he4.
In this work, we estimate the optomechanical system parameters. Using recent theory, we derive the expected quality factors for acoustic resonators in different thermodynamic conditions. We calculate the sources of loss induced by the phononic nanostructures with numerical simulations.
Our results indicate the feasibility of the proposed approach in a broad range of parameters, which opens new prospects for more complex geometries.
\end{abstract}


\maketitle


\section{\label{sec:1}Introduction}
In recent years, progress in cavity optomechanics, which exploits the coupling of light or microwave fields to mechanical motion, has enabled the development of a wide range of designs and applications~\cite{Aspelmeyer2014}. By allowing the detection and control of non-classical states of light and mechanical motion in the quantum regime, quantum optomechanics~\cite{BowenBook} has pushed the limits of sensing capabilities, and offered interesting prospects for novel quantum technology applications. Recent achievements in the field, include the measurement of mechanical motion below the standard quantum limit~\cite{Teufel2009,Anetsberger2010}, cooling to the mechanical ground state~\cite{Teufel2011,Chan2011}, realising quantum coherent state transfer~\cite{Palomaki2013a,Reed2017}, quantum entanglement~\cite{Palomaki2013b,Riedinger2018}, quantum non-demolition measurements~\cite{Suh2014}, and quantum squeezing of mechanical motion~\cite{Wollman2015,Pirkkalainen2015,Lecoq2015}.

In general, mechanical resonators are fabricated from solid materials, however, there has been a recent interest in using superfluid helium 4 as a mechanical element in cavity optomechanical systems.  As a natural quantum fluid, superfluid \he4 holds several advantages for a mechanical system over classical materials, which include an absence of viscosity, a naturally high purity and thermal conductivity, and quantized vorticity. Of particular interest to quantum optomechanics, are its vanishing acoustic and dielectric loss at low temperature. For instance, it was shown that, in theory, losses in superfluid \he4 can lead to an acoustic quality factor of the order of $Q_a\sim10^{10}$ at 10 mK~\cite{DeLorenzo2014}, and a dielectric loss tangent $\tan{\delta}<10^{-10}$ at about 1.5 K~\cite{Hartung2006}. In a first attempt to exploit these remarkable properties in a superfluid optomechanical setup, superfluid \he4 was used in a gram-scale, ultra-high quality factor acoustic resonator~\cite{DeLorenzo2014,DeLorenzo2017}. It was shown that such system, if brought up to the kilogram-scale or more, can lead to highly sensitive gravitational wave detectors~\cite{Singh2017}. Furthermore, at much smaller scale (i.e. picogram and femtogram), superfluid \he4 resonators have shown great potential for quantum optomechanics experiments ~\cite{Kashkanova2016,Shkarin2019}, the study of quantized vorticity in thin films~\cite{Forstner2019,Sachkou2019} and levitating droplets~\cite{Childress2017}, the enhancement of Brillouin interaction~\cite{He2020,Harris2020}, and the realisation of qubits mechanical systems~\cite{Sfendla2020}.

To improve the capabilities of superfluid optomechanical systems as quantum resources, it is key to enhance the coherent coupling between the light field and the acoustic field by maximising the optomechanical coupling strength ($g_0$), while limiting the sources of acoustic loss ($\Gamma_a$) and optical loss ($\kappa$). While superfluid \he4 holds remarkable intrinsic properties (i.e. vanishing losses), these can be spoiled by external factors causing heating or radiation losses.

In the present work, we propose a novel architecture for superfluid optomechanics, based on engineered nanostructures allowing a better control over superfluid phonon propagation, preserving superfluid \he4's exceptional intrinsic properties, and leading to enhanced quality factors and coupling strengths. Exploiting recent progress in quantum nanofluidics, concerning the confinement at the nanoscale of quantum fluids (liquid helium-4~\cite{Gasparini2008,Duh2012,Levitin2013,Rojas2014,Rojas2015,Souris2017,Perron2019,Shook2020} and liquid helium-3~\cite{Levitin2013,Zhelev2017,Zhelev2018,Levitin2019,Lotnyk2019,Varga2020,Heikkinen2020}), one can form a nanoscale cavity of typically hundreds of nm in height, and tens of $\mu$m in width defining the boundaries of a picogram or femtogram scale superfluid acoustic resonator~\cite{Rojas2014,Rojas2015,Souris2017}. Such superfluid acoustic resonator could be formed by means of a microsale hollow volume within a glass or silicon substrate. However, filling this volume with superfluid helium requires an opened filling channel, which can lead to spurious acoustic modes and loss channels. We propose to solve this technical problem by confining our superfluid acoustic modes in phononic nanostructures.

\begin{figure}[h]
\centering
\includegraphics[width=8.6cm]{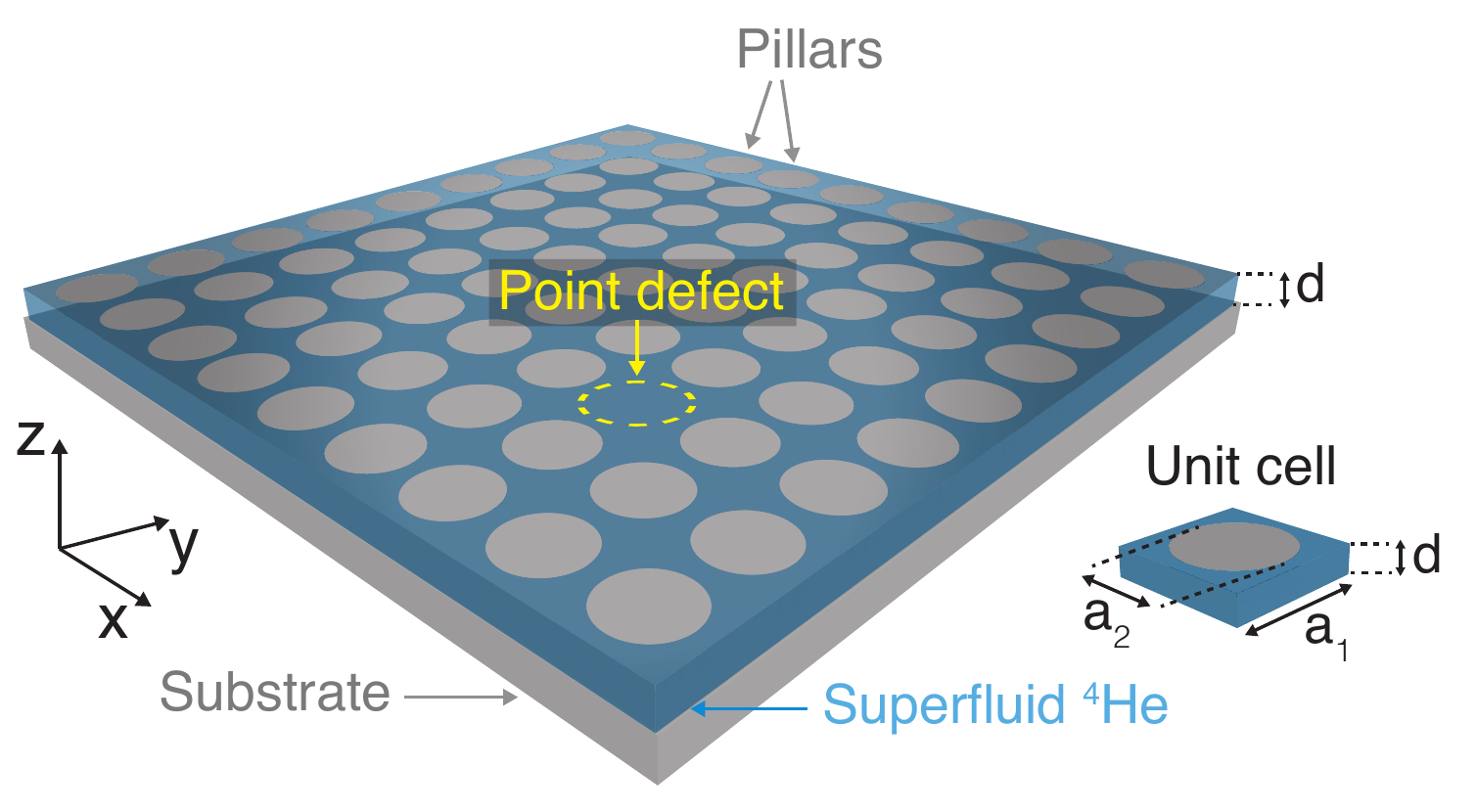}
\caption{2D sonic crystal composed of an array of cylindrical pillars patterned in a substrate (gray regions). Hollow volumes in-between pillars are filled with superfluid \he4 (blue regions). The square unit cell is shown on the right-hand side with its dimensions: side length ($a_1\sim100$ $\mu$m) and pillar diameter ( $a_2\sim80$ $\mu$m). A point defect in the lattice will host the acoustic mode of the superfluid optomechanical system. The nanofluidic geometry is enclosed by bonding another substrate on top (not shown here for clarity), confining acoustic propagation within a thin superfluid \he4 slab ($d\sim100$ nm).}
\label{PhononicCrystal_perspective}
\end{figure}

Using a methodology borrowed from previous work on phononic crystal slabs~\cite{Safavi-Naeini2010}, and macroscopic sonic crystals~\cite{LaudeBook,Martinez-Sala1995}, we designed nanofluidic 2-dimensional superfluid sonic crystals, which consist of hollow nanostructures composed of channels and cavities formed in a solid substrate, and filled with superfluid helium (see Fig.~\ref{PhononicCrystal_perspective}). The fluid is confined in a thin slab containing hundreds of cylindrical pillars arranged in a periodic lattice structure forming a 2D artificial crystal. Sound waves propagating in the superfluid embedded within this structure must satisfy Bragg scattering conditions, which leads to a phononic band structure defining the possible propagating modes. This system forms a sonic crystal (or phononic crystal). By removing one pillar from the periodic lattice, we create a point defect in the artificial crystal. This defect can host an acoustic mode, which has a frequency located at the centre of the sonic crystal's bandgap. Because in the sonic crystal, acoustic propagation is forbidden for frequencies within a bandgap, radiation loss out of the acoustic mode is strongly suppressed, greatly enhancing the acoustic mode's quality factor.

The highly confined acoustic mode is coupled (via electrostriction) to an electric field generated by a nanoscale capacitor located at the point defect of the sonic crystal. Enclosed within the nanofluidic geometry, the capacitor is terminated by large antennas, which couple the microwave field of a 3D cavity mode to the capacitor. This forms a cavity optomechanical system in which the mechanical mode is a superfluid acoustic mode confined at the point defect of the sonic crystal, and the optical cavity mode is defined by the microwave cavity coupled to the nanoscale capacitor. We find that the proposed superfluid optomechanical system should have a relatively large optomechanical coupling strength ($g_0\sim10^{-2}$ Hz), which is about 6 orders of magnitude larger than in previous superfluid optomechanics work with microwave fields~\cite{DeLorenzo2014,DeLorenzo2017}. This potential improvement is due to the high confinement of the acoustic mode, and the strong mode overlap that the nanofluidic environment can provide. 

In section~\ref{sec:2}, we give a short background on the superfluid \he4 two-fluid model, and collective excitations. In section~\ref{sec:3}, we describe the characteristic properties of superfluid \he4 as an acoustic medium, including the different sources of internal attenuation identified as relevant to quantum optomechanics. In section~\ref{sec:4}, we introduce our proposed nanofluidic system, the sound propagation inside, and the different sources of acoustic loss in these geometries. In section~\ref{sec:5}, we present a possible implementation of the proposed phononic nanostructures in a cavity optomechanical setup, and the optomechanical coupling.

\section{\label{sec:2} Superfluid \he4: Background}
\subsection{Two-fluid model}
Liquid \he4 is a system of strongly correlated bosons, a quantum Bose liquid, which behaves like an ordinary fluid (He I) down to a critical temperature $T_\lambda = 2.17$ K. Below this critical temperature, liquid \he4 transits into a superfluid phase (He II), which exhibits peculiar properties, such as a vanishing viscosity for fluid flow in thin capillaries~\cite{Kapitza1938}. Most of the properties of the superfluid phase are well described by a two-fluid model originally introduced by Tisza~\cite{Tisza1938}, and then rigorously reformulated by Landau~\cite{Landau1941}. A discussion of the discovery of superfluidity in \he4 can be found at Ref.~\cite{Balibar2017}.

In the two-fluid model, the liquid is divided into two components. These two components are not a real division of the liquid but rather a mathematical abstraction, which works well to describe the properties of the superfluid phase. In this construction, the superfluid component of mass density $\rho_s$ and velocity $\bm{v}_s$, carries neither entropy nor viscosity. The superfluid component represents a collective phenomenon, in which particles move together to preserve the macroscopic occupation of a single quantum state of the Bose liquid. The second component, the normal fluid component of mass density $\rho_n$ and velocity $\bm{v}_n$, represents the gas of elementary excitations (or quasi-particles) of the Bose liquid, which is discussed in the next section. The total mass density $\rho$ and momentum density $\bm{j}$ of the fluid are given by
\beqn
\rho & = & \rho_n + \rho_s, \\
\bm{j} & =  & \rho_n \bm{v}_n + \rho_s \bm{v}_s.
\eeqn
At the transition temperature ($T=T_\lambda$), the superfluid component density vanishes ($\rho_s=0$, $\rho_n=\rho$). In the zero temperature limit, the normal fluid component density vanishes ($\rho_n=0$, $\rho_s=\rho$), and with it, the entropy of the fluid $\rho S=\rho_n S_n $, where $S_n$ is the entropy of the normal fluid component. Since the superfluid component describes a single macroscopic quantum state, there is no viscosity associated with the superfluid flow ($\eta_s=0$). The viscosity of the liquid comes from the gas of excitations, identified as the normal fluid component. These excitations scatter randomly against the atomically rough walls of the container transferring momentum from the fluid to the walls, hence providing a viscous friction with a coefficient of viscosity $\eta=\eta_n$ comparable to that of the normal liquid phase (He I). The gas of excitations is also responsible for sound attenuation.

\subsection{Spectrum of excitations}
\begin{figure}[h]
\centering
\includegraphics[width=8cm]{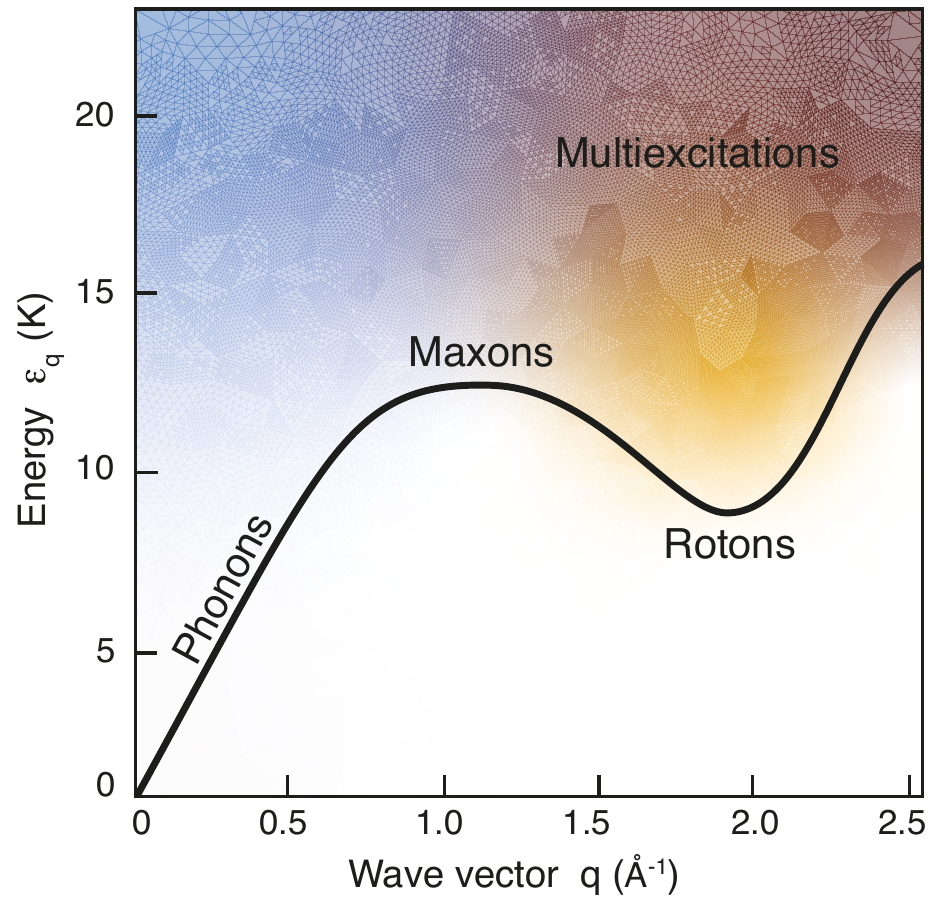}
\caption{Schematic of superfluid \he4 dispersion curve, showing the low energy branch of the excitation spectrum. Thanks to liquid \he4's isotropy, the dispersion curve does not depends on the orientation of the excitation wavevector $\mathbf{q}$.}
\label{DispersionRelation}
\end{figure}

Landau proposed that the normal fluid component can be seen as a gas of weakly interacting elementary excitations~\cite{Landau1941,Landau1949}. The concept of elementary excitations can be used when their interaction energy is small compared to their own energy~\cite{KhalatnikovBook2000}. The background through which these quasiparticles move corresponds to the superfluid component. Hence, the dynamical properties of He II at low temperature are dictated by the excitations spectrum (Fig.~\ref{DispersionRelation}), in particular the phonon-maxon-roton excitations branch of the dispersion curve $\epsilon=\epsilon(q)$, with $q$ the magnitude of the excitation wave vector. 

The linear section of the dispersion curve at low energy ($q\rightarrow0$) corresponds to phonon excitations, i.e. long wavelengths density fluctuations. We note that while phonons usually emerge as the quantized vibrational states of a crystal lattice, in superfluid \he4, phonons correspond to the gapless Goldstone modes of the Bose liquid~\cite{NozieresBook}. The low energy part of the dispersion curve can be approximated by
\beq
\epsilon_{\rm ph}(q)\simeq c q(1-\gamma q^2),
\eeq
where $c$ is the first-sound (i.e. density waves) velocity and $\gamma$ the phonon dispersion coefficient. Both parameters are pressure dependent, and importantly $\gamma$ is negative at low pressure (anomalous dispersion) and positive at high pressure (normal dispersion). The sign of $\gamma$ dictates the nature of phonon interactions, and therefore the nature of sound attenuation in superfluid \he4.

\begin{figure}[h]
\centering
\includegraphics[width=9.0cm]{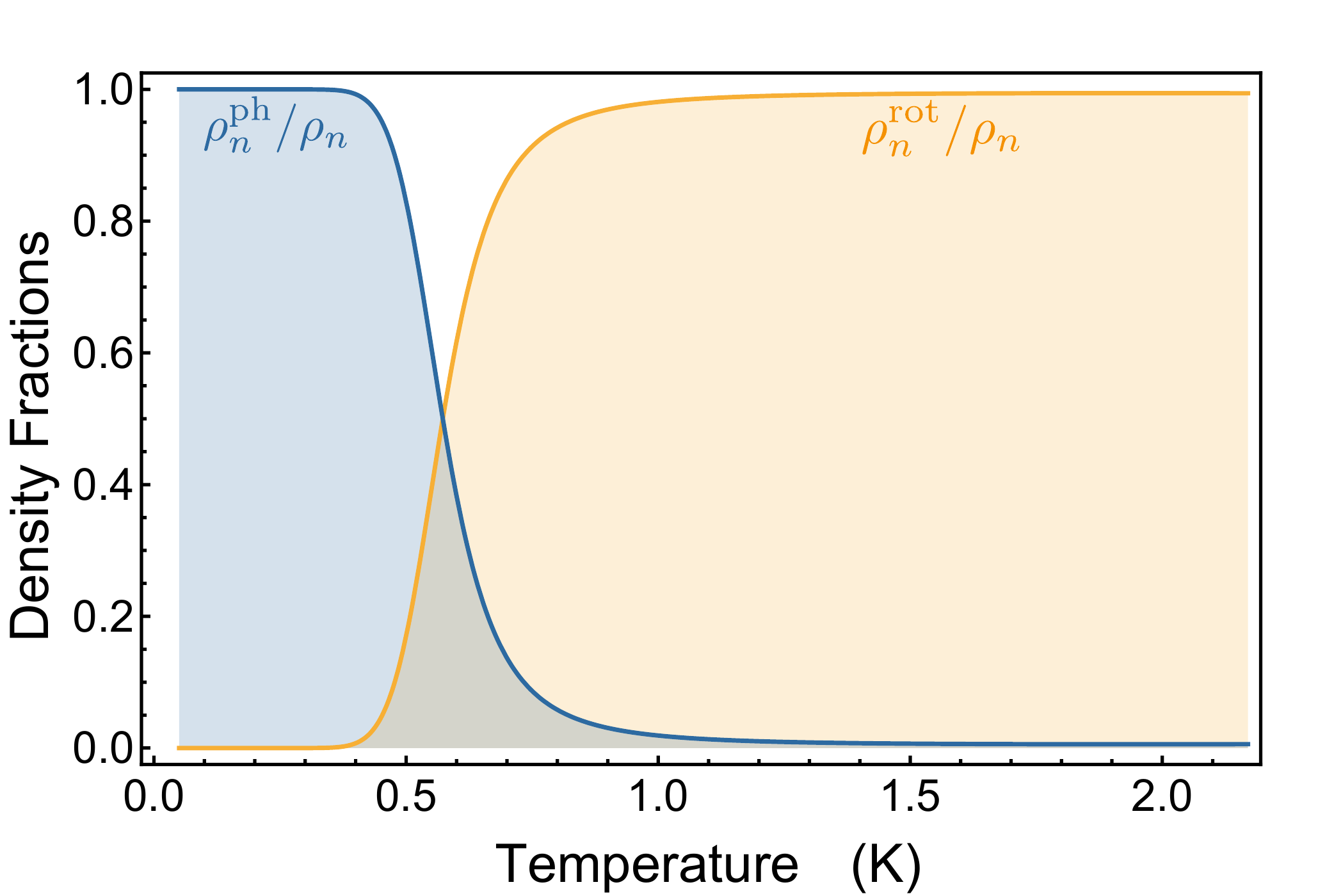}
\caption{Phonon contribution $\rho_n^{\rm ph}/\rho_n$ (blue curve) and roton contribution $\rho_n^{\rm rot}/\rho_n$ (yellow curve) to the normal fluid density fraction. These two contributions are equal at a temperature $T\simeq0.6$ K. }
\label{DensityFractions}
\end{figure}

The minimum of the dispersion curve in Fig.~\ref{DispersionRelation} corresponds to roton excitations. The true nature of roton excitations is still debated. In Feynman's picture~\cite{Feynman1955}, a roton was interpreted as a vortex ring (i.e. a toroidal vortex) with a radius close to its core radius such that only one atom can pass through the ring. In an alternative view, a roton can be seen as the backflow induced by a moving impurity atom~\cite{Miller1962}, or more recently, as the \textit{ghost of a Bragg spot}, signalling the proximity of a solidification phase transition~\cite{Nozieres2004}. The dispersion curve near the roton minimum can be approximated by
\beq
\epsilon_{\rm rot}(q)\simeq\Delta_r + \frac{\hbar^2}{2 \mu_r} (q-q_r)^2,
\eeq
where $\Delta_r/k_B=8.594$ K and $q_r=1.926\ \AA^{-1}$ are the energy and momentum coordinates of the roton minimum, and $\mu_r=(1/\hbar^2)\partial^2 \epsilon(q)/\partial q^2=0.124$ $m_4$ the roton effective mass associated with the curvature of the dispersion curve at the roton miminum. Values for these parameter have been extracted from neutron scattering measurements~\cite{Beauvois2018}, at saturated vapour pressure (SVP). The maximum of the dispersion curve corresponds to maxon excitations. We have been exhaustive in describing the low energy excitations in superfluid \he4 as they may be relevant to various types of superfluid optomechanical systems, for instance, those using optical light. In this work, however, maxons, and higher energy multi-excitations, are not relevant to describe the low temperature properties of liquid He II, and its interaction with the microwave fields of the proposed architecture. Therefore, we will ignore their contribution in the next sections. In addition, one can show that at low temperature the normal fluid density is only the sum of a phonon and roton contribution $\rho_n=\rho_n^{\rm ph} +\rho_n^{\rm rot}$, given by~\cite{Wilks1967}:
\beqn
\rho_n^{\rm ph}(T) & = & \frac{2\pi^2k_B^4}{45\hbar^3 c_1^5}T^4, \\
\rho_n^{\rm rot}(T) & = & \sqrt{\frac{\mu_r}{2\pi^3}} \frac{\hbar q_r^4}{3 \sqrt{k_B T}} e^{-\Delta_r/k_B T}.
\eeqn
These two contributions (shown Fig.~\ref{DensityFractions}) are equal at around 0.6 K. Below this temperature, the roton density vanishes exponentially with temperature, and only the phonon excitations contribute to the normal density. Therefore, in the mK temperature range, which is our temperature range of interest, we can ignore the contribution of rotons. The dynamic properties of the fluid, such as sound attenuation will be dictated by phonon dynamics only.

\subsection{\he3 impurities: quasiparticles}\label{xx}
The impurities in solution represent another loss channel for sound propagation. In superfluid \he4, the only impurities are the isotopic \he3 impurities since at low temperature all the other impurities have either been filtered or adsorbed on the walls of the container. In this work, we will consider typical natural concentrations at the part per million (ppm) level, and ultra-low concentrations at the part per billion (ppb) level or lower. In the mK temperature range, the gas of \he3 impurities will be non-degenerate, $T\gg T_F$, where $T_F$ is the Fermi temperature of \he3 atoms. At thermal equilibrium, \he3 atoms form a gas of slowly moving quasiparticles of momentum $\bm k$. In the limit of low \he3 concentration, considering the superfluid \he4 at rest, the gas of \he3 quasiparticles is described by the following energy spectrum~\cite{Landau1948,TerHaar1965} 
\beq\label{eq:he3spectrum}
\epsilon_3(k)=\epsilon_3(0)+\frac{\hbar^2 k^2}{2m^{\ast}},
\eeq
where $m_3$ is the bare mass of a \he3 atom, and $m^{\ast}\simeq 2.34\ m_3$ is the effective mass of \he3 atoms in superfluid \he4 at zero pressure~\cite{BaymBook2004}. The effective mass is larger than the bare \he3 atomic mass because a \he3 atom drags a flow of \he4 atoms with it, carrying momentum. $\epsilon_3(0)\simeq-2.785$~K depends, in principle, on both \he3 and \he4 densities, however at low concentration, the \he3-dependence is negligible and $\epsilon_3(0)$ can be identified as the chemical potential $\mu_3$ of the \he3 in \he4.

\section{\label{sec:3} Superfluid \he4: acoustic medium}
\subsection{Sound propagation in superfluid \he4}
In electromagnetism, the wave equation of the field follows directly from Maxwell equations, for which  monochromatic plane waves are rigorous solutions. In contrast, the wave equation for acoustics is only an approximate equation derived from the non-linear hydrodynamic equations, and plane waves are approximate solutions. Nevertheless, it was proposed that for pure \he4 and long wavelength excitations, an arbitrary state of the sound field can be decomposed into a linear superposition of plane waves~\cite{Landau1941,Landau1949,LondonBook1954}.

A complete description of the motion of the fluid in the superfluid phase (He II) involves solving the equations of motion for the two-fluid model, which have been derived in various textbooks~\cite{Wilks1967,PuttermanBook,NozieresBook,KhalatnikovBook2000}. Deriving superfluid hydrodynamic equations for the most general case falls outside the scope of this paper. Instead, we discuss the propagation of sound in He II by investigating the special case of small disturbances (indicated by $\delta$) from a steady state for which $\bm{v}_n=0$ and $\bm{v}_s=0$, so
\beqn
\bm{v}_n & = & \delta \bm{v}_n (\bm{r},t), \\
\bm{v}_s & = & \delta \bm{v}_s (\bm{r},t),
\eeqn
and the thermodynamic variables are close to their equilibrium value indicated by the subscript zero:
\beqn
\rho & = &  \rho_0 + \delta \rho(\bm{r},t), \\
p & = & p_0 + \delta p(\bm{r},t),  \\
T & = & T_0 + \delta T(\bm{r},t),  \\
s & = & s_0 + \delta s(\bm{r},t),
\eeqn
where $\rho$, $p$, $T$, and $s$ are respectively the mass density, pressure, temperature, and entropy fields. We assume the disturbance to be small (e.g. $\delta \rho \ll \rho_0$), and the velocity fields $\delta \bm{v}_n, \delta \bm{v}_n$ to be smaller than the sound velocity. As for classical fluids, we can linearise the equations of motion neglecting all quadratic terms in the small quantities. Hence, the linearised equations of motion without dissipative effects become
\beqn
\frac{\partial \delta\rho}{\partial t} + \bm{\nabla}\cdot \delta\bm{j} = 0 \label{eqn:EQM_1} \\
\rho_0  \frac{\partial (\delta s)}{\partial t} + s_0 \frac{\partial (\delta \rho)}{\partial t} + \rho_0 s_0 \bm{\nabla}\cdot \delta\bm{v}_n = 0 \label{eqn:EQM_2} \\
\frac{\partial \delta\bm{j}}{\partial t} + \bm{\nabla} \delta p = 0 \label{eqn:EQM_3} \\
\frac{\partial \bm{v}_s}{\partial t} + \bm{\nabla} \delta\mu = 0 \label{eqn:EQM_4} 
\eeqn
where $\delta\bm{j}=\rho_{n,0} \delta\bm{v}_n + \rho_{s,0} \delta\bm{v}_s $ and $\delta\mu=(1/\rho_0)\delta p -s_0\delta T $ are respectively the mass current and the chemical potential disturbance from equilibrium value. The two first equations above derive from conservation of mass and entropy, the other two equations are typical of the two-fluid model for which a pressure gradient drives a total mass current, and a chemical gradient drives a superfluid flow. Combining Eq.~\ref{eqn:EQM_1} and Eq.~\ref{eqn:EQM_3}, we obtain the \textit{first-sound} acoustic wave equation:
\beq\label{eq:sound-wave}
\frac{\partial^2 \delta \rho}{\partial t^2} +  \nabla^2 \delta p = 0,
\eeq
with $\nabla^2$ the Laplacian operator defined in Cartesian coordinates as $\nabla^2=\partial_x^2 + \partial_y^2 + \partial_z^2$. Considering the equation of state for the fluid $p=p(\rho, s)$, and its (isentropic) differential form
\beq
\delta p=\left.\frac{\partial  p}{\partial \rho} \right |_s \delta \rho,
\eeq
we define the isentropic sound velocity $c_1$ given by
\beq
c_1^2=\left . \frac{\partial p}{\partial \rho} \right|_{s},
\eeq
which represents the velocity at which pressure (or density) waves propagate in He II. To the first-order approximation, we obtain a constitutive equation
\beq
\delta p = c_1^2 \delta \rho,
\eeq
and can rewrite the first-sound acoustic wave equation as
\beq\label{eq:soundwave_p}
\frac{\partial^2 \delta p}{\partial t^2} - c_1^2 \nabla^2 \delta p = 0
\eeq
or
\beq\label{eq:soundwave_rho}
\frac{\partial^2 \delta \rho}{\partial t^2} - c_1^2 \nabla^2 \delta \rho = 0.
\eeq
Hence, we see that the first-sound mode in He II is analog to a classical sound mode (density waves) in ordinary fluids, for which plane waves are simple solutions given by
\beq
\delta p (\bm{r},t) = A e^{i (\bm{k}\cdot\bm{r} - \Omega t)},
\eeq
with $A$ the acoustic wave amplitude, $\Omega=2\pi f$ the angular frequency, $\bm{k}=\bm{n} k=\bm{n} (\Omega/c_1)$ the wave vector, and $\bm{n}$ the direction of propagation of the plane wave.

Combining the other two equations of motion, Eq.~\ref{eqn:EQM_2} and Eq.~\ref{eqn:EQM_4}, leads to a wave equation describing \textit{second sound} propagation, a sound mode corresponding to a temperature (or entropy) wave, which can be interpreted as compressional waves in the gas of excitations.

The first sound (density waves) and second sound (entropy waves) wave equations derive from the two-fluid model hypothesis taken in the hydrodynamic regime, for which spatial variations of density and entropy fields are slow compared to the relaxation time needed to establish thermal equilibrium in the gas of excitations (i.e. phonons)~\cite{NozieresBookCh7}. Thus, the condition of validity of the hydrodynamic limit can be written as
\beq
\Omega \tau_r \ll 1,
\eeq
where $\tau_r=1.43\times10^{-10}/T^5$ is the relaxation time required for achieving local thermodynamic equilibrium in the phonon gas~\cite{Jackle1971}. The hydrodynamic limit works well at low frequency, however, at the typical temperature ($T<0.1$ K) and frequency range ($\Omega/2\pi \sim 1$ MHz) of this work, we always have $\Omega\tau_r \gg 1$, and the hydrodynamic limit is not a valid approximation. The appropriate limit for this work is the collisionless regime ($\Omega\tau_r \gg 1$), for which the typical lifetime of excitations $\tau_r$ is much greater than the period of the sound wave $2\pi/\Omega$. In this limit, sound propagation corresponds to a \textit{quasi-particle sound} mode (also called \textit{zero sound} mode)~\cite{NozieresBookCh7}, in which the restoring force on a given particle comes from the averaged field of all other particles. This effect leads to slight modification of the isothermal first-sound velocity. At $T<0.1$ K, where the only thermal excitations of importance are phonons, calculations predict an increase in quasi-particle sound velocity compare to first-sound velocity $c(T)=c_1(T)(1+\rho_n/\rho)$~\cite{NozieresBookCh7}. However, more importantly for us, sound attenuation in the collisionless regime, which is detailed in the next section, is significantly different from attenuation in the hydrodynamic regime. 

\subsection{Sound attenuation in superfluid \he4}\label{section:SoundAttenuation}
In general, the presence of dissipative effects in ordinary fluids, such as those related to viscosity or heat conduction lead to sound attenuation. This is true of superfluid \he4 in the hydrodynamic limit. However, in the collisionless regime the gas of excitations does not reach local thermodynamic equilibrium, therefore it does not dissipate sound energy as effectively as in the hydrodynamic regime, and neither viscosity nor heat conduction are well-defined. Besides, the presence \he3 quasiparticles in solution provides another acoustic loss channel. Finding the full temperature and frequency dependence of sound attenuation in He II is a challenging problem. However, in the low temperature range ($T<0.1$ K) and for low \he3 concentration ($x_3<10^{-6}$) sound attenuation is simplified. For these conditions, sound attenuation can be separated into a phonon-phonon interaction contribution, in which energy loss is caused by acoustic phonons scattering against thermal phonons, and a phonon-\he3 interaction contribution, in which acoustic phonons scatter against \he3 quasiparticles.

\subsubsection{Phonon - phonon interaction}
Theoretical investigation of the phonon-phonon interaction contribution to sound attenuation was originally developed in refs.~\cite{Dransfeld1962,Woodruff1962,Pethick1966,Disatnik1967}, following different methods, whose main ideas are well described in ref.~\cite{PhysicalAcoustics6BookCh5}. The phonon dispersion coefficient $\gamma$ responsible for the phonon-phonon interaction, was first assumed to be positive at all pressure, which led to an underestimation of sound attenuation at low pressure. Later, as shown in ref.~\cite{Maris1973b}, this error was corrected, and the theoretical value for sound attenuation was brought in line with experiments~\cite{Chase1955,Abraham1969a,Roach1972}. It was then accepted that at low pressure the phonon-phonon interaction term was dominated by 3-phonon interaction processes, in which an acoustic phonon is absorbed by a thermal phonon to produce a third phonon (Landau process), leading to a sound attenuation coefficient given by~\cite{Isihara1989}
\beqn
\alpha_q & = & \frac{3}{2} \frac{\rho^{\rm ph}_n}{\rho_0}(u+1)^2 q \nonumber\\
&& \times [\arctan{(2\Omega_q\tau)} + \arctan{(3\gamma \bar{q}_{\rm th}^2\Omega_q\tau)}],
\eeqn
where $q$ and $\Omega_q$ are respectively the acoustic phonon wavenumber and pulsation frequency, $\tau$ the typical phonon relaxation time, $\bar{q}_{\rm th}=3k_BT/c$ is the average thermal phonon momentum, and
\beq
u = \frac{\rho_0}{c}\frac{\partial c_1}{\partial \rho},
\eeq
is the Grüneisen constant, a dimensionless parameter. Hence, in the collisionless regime ($\Omega_q\tau\gg1$), the sound attenuation can be written as
\beq\label{eq:AttenuationLandau}
\alpha_q=\frac{\pi^3}{30}\frac{(u+1)^2 k_B^4}{\hbar^3\rho_0c^6}\Omega_qT^4.
\eeq

\begin{figure}[h]
\centering
\includegraphics[width=8cm]{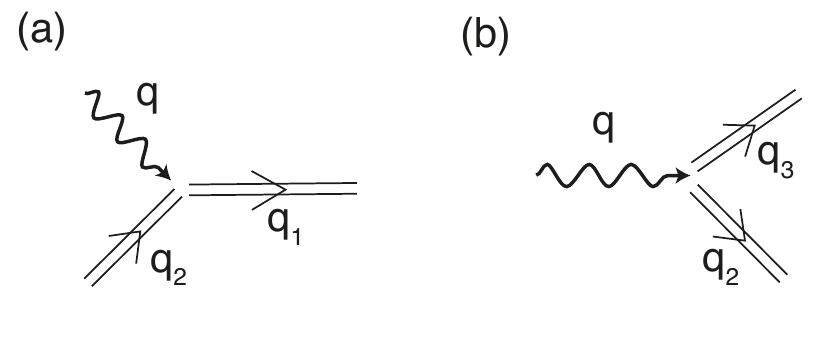}
\caption{Beliaev-Landau processes where acoustic phonons are represented by wavy lines and thermal phonons by double lines. (a) Landau process ($ q + q_2 \rightarrow q_1 $) where an acoustic phonon $q$ interacts with a thermal phonon $q_2$ to give a thermal phonon $q_1$, (b) Beliaev process ($q \rightarrow q_2 + q_3$) where an acoustic phonon $q$ decays into 2 thermal phonons $q_2$ and $q_3$.}
\label{Beliaev-Landau}
\end{figure}

Note that this expression is only valid for an acoustic phonon frequency much smaller than the typical thermal phonon frequency, that is for $\hbar \Omega_q\ll k_B T$. More recently, it was shown by Kurkjian~\etal\cite{Kurkjian2017} that at higher frequency Beliaev processes, for which an acoustic phonon can decay into two thermal phonons, become significant. Schematics of Beliaev-Landau processes are shown Fig.~\ref{Beliaev-Landau}.

Including Beliaev processes leads to a more complex expression of the damping rate, which was computed for quantum gases by Kurkjian~\etal\cite{Kurkjian2017}. In the low temperature limit, we can derive a simple expression of the damping rate given by
\beq\label{eq:BLDamping}
\Gamma^{\rm Bel/Lan}_q \underset{T\to0}{\sim} \frac{(u+1)^2}{8\pi}\frac{(k_B T)^5}{\rho_0\hbar^4c^5}\ \tilde{\Gamma}^{\rm Bel/Lan}(\tilde{q}),
\eeq
where we introduced the dimensionless wave number
\beq
\tilde{q}=\frac{\hbar c q}{k_BT},
\eeq
rescaled by the typical thermal phonon wave number $k_B T/\hbar c$. The functions $\tilde{\Gamma}^{\rm Bel/Lan}(\tilde{q})$ are universal functions of $\tilde{q}$ given by~\cite{Kurkjian2017}
\beqn
\tilde{\Gamma}^{\rm Bel}(\tilde{q})& =& \frac{\tilde{q}^5}{30}-\frac{4\pi^4}{15}\tilde{q}+48\left[\zeta{(5}) - g_5(e^{-\tilde{q}}) \right] \nonumber \\
& & - 24\tilde{q}\ g_4(e^{-\tilde{q}}) + 4\tilde{q}^2 \left[\zeta(3)-g_3(e^{-\tilde{q}})\right] \label{eq:GammaBel} \\
\tilde{\Gamma}^{\rm Lan}(\tilde{q}) & =& \tilde{\Gamma}^{\rm Bel}(\tilde{q})-\frac{\tilde{q}^5}{30}+\frac{8\pi^4}{15}\tilde{q}, \label{eq:GammaLan}
\eeqn
where the Bose functions $g_\alpha(z)$, also called polylogarithms $\mathrm{Li}_\alpha(z)$, are given by
\beq
g_\alpha(z)=\sum^{+\infty}_{n=1}\frac{z^n}{n^\alpha},
\eeq
and the Riemann zeta functions are given by $\zeta(\alpha)=g_\alpha(1)$.

In the limit $\tilde{q}\to0$, which corresponds to acoustic phonons in the mode $\bm{q}$ having an energy $\hbar\omega_q$ much lower than the typical thermal phonons energy $k_B T$, we obtain
\beq
\tilde{\Gamma}^{\rm Lan}(\tilde{q}) \underset{\tilde{q}\to0}{\sim} \frac{8\pi^4}{15}\tilde{q},
\eeq
which leads to 
\beq
\Gamma^{\rm Lan} _q \underset{\tilde{q}\to0}{\underset{T\to0}{\sim}} \frac{\pi^3}{15} \frac{(u+1)^2 k_B^4}{\hbar^3 \rho_0 c^5} \Omega_q T^4,
\eeq
which corresponds to the commonly used expression, given at Eq.~\ref{eq:AttenuationLandau}, for the acoustic attenuation in superfluid \he4 due to 3-phonon interaction in the collisionless regime ($\omega_q \tau\gg1$), and the low temperature and low frequency limit ($\hbar \Omega_q \ll k_B$). By including Beliaev processes, one can compute the total damping rate $\Gamma_q=\Gamma^{\rm Lan}_q+\Gamma^{\rm Bel}_q$ numerically, which gives the acoustic damping and the attenuation at larger mode frequency.

\begin{figure}[h]
\centering
\includegraphics[width=9cm]{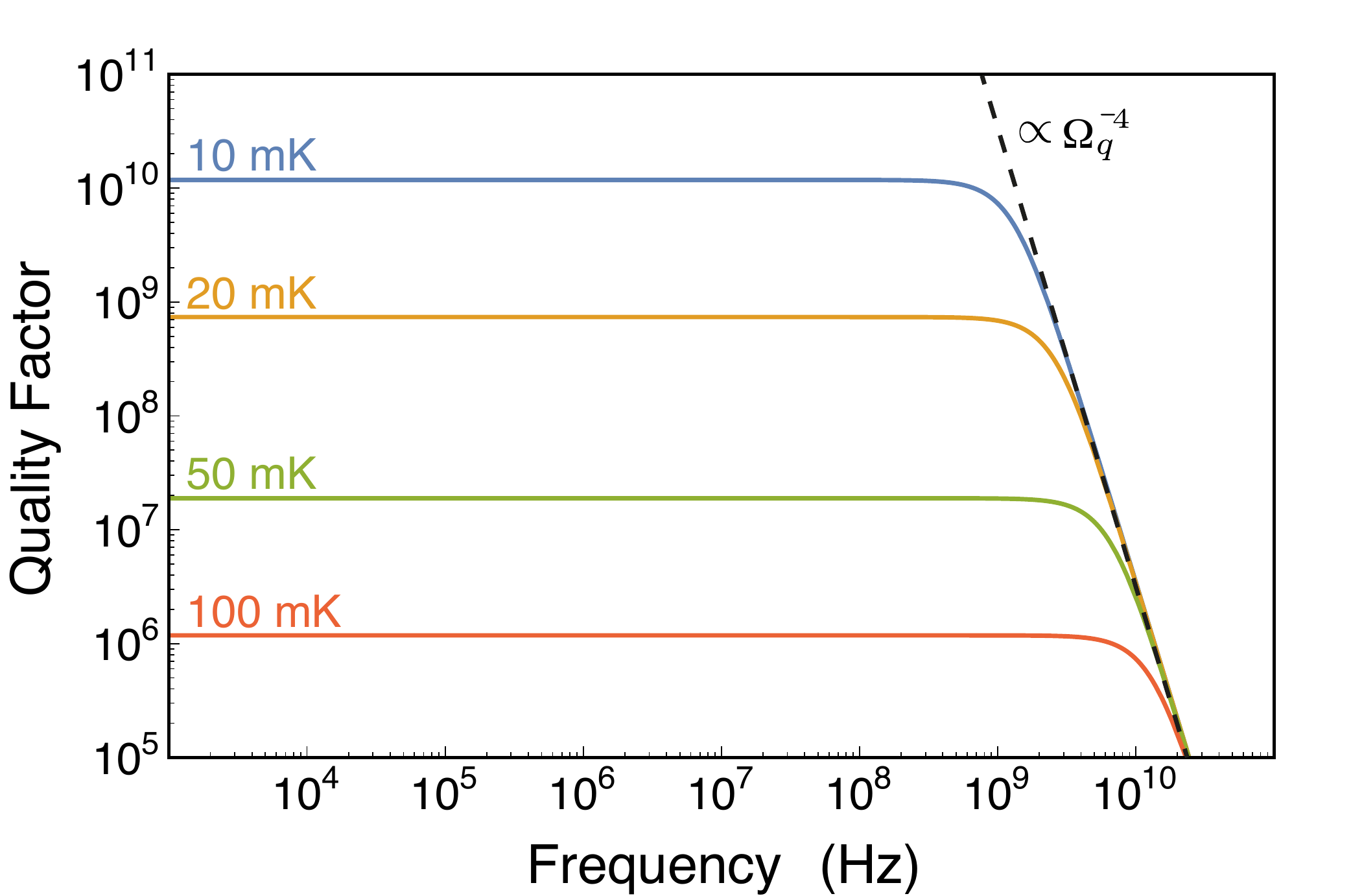}
\caption{Quality factor of an acoustic mode $\bm q$, computed using the expression of the damping rate given at Eq.~\ref{eq:BLDamping} for 3-phonon interaction, as a function of the mode frequency $\Omega_q/2\pi$ at temperatures 10 (blue line), 20 (orange line), 50 (green line) and 100 mK (red line). The black dashed line indicates the $\Omega_q^{-4}$ frequency dependence at frequencies higher than $k_B T/\hbar$.}
\label{Q3phonon}
\end{figure}

Fig.~\ref{Q3phonon} shows the frequency dependence of the quality factor for a resonator of acoustic mode $\bm q$ given by $Q_a=\Omega_q / \Gamma_q$, where $\Gamma_q=2c\alpha_q$ represents the mode damping rate for both Beliaev and Landau processes.  Hence, we observe that at low pressure and at the typical base temperature of a dilution refrigerator ($T\sim10$ mK), the quality factor is independent of frequency up to a cut-off frequency of the order of 1 GHz, above which the quality factor decreases following a $1/\Omega_q^4$ frequency dependence.

Recent high-precision neutron measurements from Beauvois~\etal\cite{Beauvois2018} compared to theory~\cite{Beauvois2019} confirm that the dispersion coefficient, negative at low pressure ($\gamma<0$), changes sign and becomes positive ($\gamma>0$) at a pressure $P\simeq20$ bar ($\rho_0=169$ kg/m$^3$). This indicates a concave dispersion relation at high pressure and low wave numbers. The 3-phonon interaction processes are forbidden at high pressure as these processes cannot satisfy the conservation of both energy and momentum. This leads to a regime where 4 phonon processes ($2\leftrightarrow2$) are the dominant damping factor.

\begin{figure}[h]
\centering
\includegraphics[width=9cm]{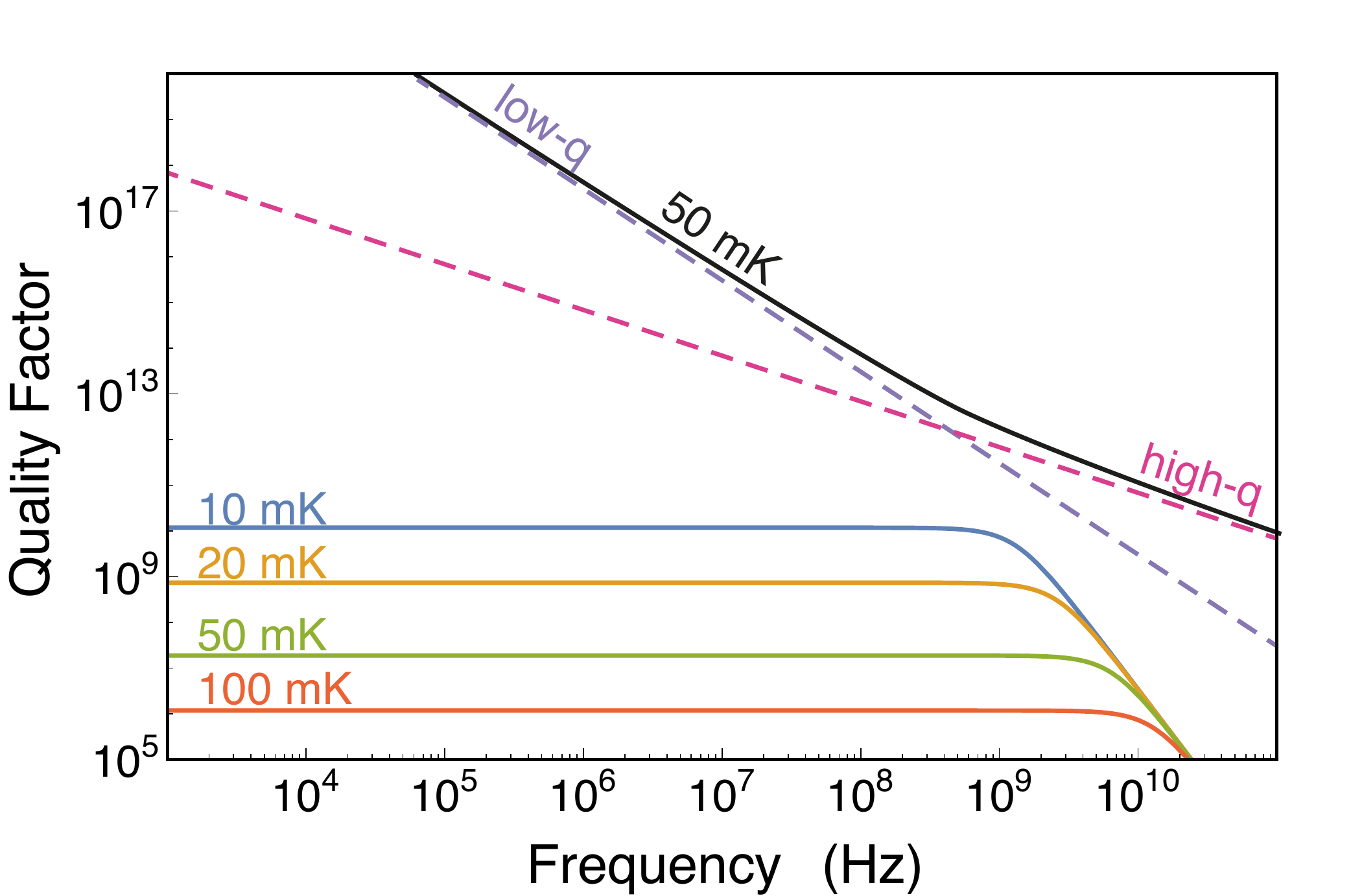}
\caption{Quality factor of an acoustic mode $\bm q$, computed using the expression of the damping rate given at Eq.~\ref{eq:BLDamping} for 3-phonon interaction as a function of the mode frequency $\Omega_q/2\pi$. The different lines show the quality factor at different temperatures 10 (blue line), 20 (orange line), 50 (green line) and 100 mK (red line). The purple dashed lines shows the asymptotic limits of the damping rate for the 4-phonon interaction (Eq.~\ref{eq:4PhononsAttenuation}), taken at 50 mK, at low-$\tilde{q}$ (purple dashed line), and at high-$\tilde{q}$ (pink dashed line). The superimposed black line is guide to the eye to show the trend of the quality factor. The exact frequency dependence has a more complex structure around $\tilde{q}\sim1$, which could be obtained by computing the full $\tilde{q}$ dependence of the 4-phonon damping rate.}
\label{Q4phonon}
\end{figure}

The damping rate due to 4-phonon interaction was originally calculated, in quantum hydrodynamics, by Landau and Khalatnikov~\cite{Landau1949}. As for crystal lattices, to introduce transitions between different phonon states (i.e. off-diagonal terms), one needs to consider the anharmonic contributions (i.e. third order and higher perturbation terms) in the interaction Hamiltonian. To fully account for all the 4-phonon processes, one has to consider both the quartic term of the interaction Hamiltonian $\mathcal{V}_4$ at the first order in perturbation, but also the cubic terms of the interaction Hamiltonian $\mathcal{V}_3$ at the second order in perturbation. However, as pointed out in reference~\cite{Tucker1992} certain 4-phonon processes, neglected by Landau and Khalatnikov, turn out to be of the same order of magnitude as the kept one. In a more recent work, Kurkjian~\etal\cite{Kurkjian2017} included all the important terms and found a damping rate that is smaller than the original one given by Landau and Khalatnikov. The damping rate for 4-phonon processes is given in the low temperature limit by
\beq\label{eq:4PhononsAttenuation}
\Gamma^{2\leftrightarrow2}_q\underset{T\to0}{\sim} \frac{(u+1)^4}{128 \pi^4}\frac{(k_B T)^7}{\rho_0|\gamma|c^8\hbar^5}\ \tilde{\Gamma}^{2\leftrightarrow2}(\tilde{q}),
\eeq
where $\tilde{\Gamma}^{2\leftrightarrow2}(\tilde{q})$ is a quadruple integral whose exact form is given in ref.~\cite{Kurkjian2017}. It is interesting to consider the asymptotic limits at low-$\tilde{q}$  given by
\beq
\tilde{\Gamma}^{2\leftrightarrow2}(\tilde{q})\underset{\tilde{q}\to0}{\sim} \frac{16\pi^5}{135}\tilde{q}^3,
\eeq
and at high-$\tilde{q}$ given by 
\beq
\tilde{\Gamma}^{2\leftrightarrow2}(\tilde{q})\underset{\tilde{q}\to\infty}{\sim} \frac{16\pi \zeta(5)}{3}\tilde{q}^2,
\eeq
where $\zeta(5)=1.03693$ is the Riemann zeta function $\zeta(\alpha)$ taken at $\alpha=5$. Using these asymptotic forms, we can compute the damping rate in the low temperature limit and compare it with the damping rate originating from 3-phonon interactions. We show Fig.~\ref{Q4phonon} the quality factor we obtain from the 4-phonon interaction terms. We clearly see that the 4-phonon damping at 50 mK represents a negligible contribution compared to 3-phonon damping. At lower temperature, we expect an even lower contribution from the 4-phonon damping. Therefore, it is interesting to consider the possibility to study superfluid acoustic resonators at high pressure where sound attenuation will be only limited by 4-phonon damping, and the quality factor greatly enhanced. We should note that as the dominant terms in the sound attenuation becomes exceptionally small, one may need to consider other loss channels previously ignored, including the phonon-roton interaction.

\subsubsection{Phonon - \he3 interaction}
In addition to phonon-phonon scattering, the other important intrinsic acoustic loss source in superfluid \he4 is due to \he 3 impurities. The natural concentration of standard commercially available helium 4 gas varies depending on the location of the helium plant, from a ratio \he3/\he4 of the order of 0.1 ppm to a few ppb as measured by Souris~\etal\cite{Souris2014}. Beyond this, the purification technique developed by Hendry and McClintock~\cite{Hendry1987} can produce extremely pure samples with concentrations as low as $x_3=n_3/(n_4+n_3)<10^{-13}$, with $n_3$ and $n_4$ the quantity of \he3 and \he4 atoms respectively. In this work, we are interested in describing sound attenuation in the limit of ultralow concentrations at the ppb level or lower, thus limiting \he3 dissipation.

Most of the work on sound attenuation in dilute mixtures of \he3 in \he4 has been concerned with relatively large concentrations, with \he3/\he4 ratio at the percent level ($x_3\sim10^{-2}$). Much of this work is based on Bardeen, Baym and Pines (BBP) theory~\cite{Bardeen1966, Bardeen1967}. In the low temperature limit ($T\leq0.2$ K) and for large concentrations, sound attenuation caused by \he3-phonon scattering has been described solving the \he3 Boltzmann equation, including the \he3-\he3 collision integral, in the relaxation time approximation~\cite{Baym1967a, Baym1967b}. This topic is thoroughly described in~\cite{BaymBook2004}. For these conditions of concentration and temperature, the \he3 atoms were described as a Fermi gas, which is not valid at ultralow concentrations where it behaves as a classical gas since $T\gg T_F$. Typically, the Fermi temperature given by $k_BT_F=\hbar^2k_F^2/2m^{\ast}$, where $k_F=(3\pi^2x_3\rho_0/m_4)^{1/3}$ is the Fermi wave number, is of the order of 0.3 mK at \he3 concentration of 1 ppm, and 3 $\mu$K at 1 ppb.

More importantly, the primary mechanism for sound attenuation at relatively large concentrations relies on the viscosity of the \he3 gas, caused by rapid \he3-\he3 collisions maintaining a local thermodynamic equilibrium among themselves. For the viscous attenuation by the \he3 gas to be significant, it requires $\omega\tau_\eta\ll1$ where $\tau_\eta$ is the \he3-\he3 scattering relaxation time appropriate to viscosity~\cite{Baym1968}. This relaxation time should be of the same order or greater than the \he3-\he3 collision time $\tau_{33}$ given by
\beq
\tau_{33}=\frac{l_{33}}{\bar{v}_3},
\eeq
where $l_{33}\simeq 8.66\times10^{-10}/x_3$ m is the mean free path of an unpolarized \he3 gas~\cite{Baym2015a}, and $\bar{v}_3=\sqrt{3k_BT/m^{\ast}}$ is the root mean square thermal velocity of the \he3 gas. Typically, at 10 mK, and for a concentration of 1 ppb, $\tau_{33}\sim0.1$ s, which means that the condition $\omega\tau_\eta\ll1$ would only be satisfied at frequencies of the order of 1 Hz or less, which is far below the high MHz frequency range discussed in this work. Hence, the concept of viscosity for the \he3 gas is not well defined at the frequency range of our experiments, and so extending the theoretical results for sound attenuation to the case of ultralow \he3 concentrations is not trivial.

Currently, there is no reported theoretical work on sound attenuation in the case of ultralow \he3 concentrations, low temperature, and high sound frequency (i.e. collisionless regime). In the work by De Lorenzo~\etal\cite{DeLorenzo2017,DeLorenzoThesis}, sound attenuation caused by the very dilute \he3 impurity gas is derived using a classical viscous gas approximation in the ballistic regime~\cite{Kerscher2001}. For concentrations at the ppm level, and sound frequencies in the kHz range and below, it should be a relatively satisfying approximation. However, for concentrations at the ppb level, and sound frequencies in the MHz frequency range such as used here (i.e. in the collisionless regime), this approximation must fail. One could however use the results of the theoretical work on transport properties (e.g. thermal conductivity) in very dilute solutions~\cite{Baym2013,Baym2015a,Baym2015b} to estimate the sound attenuation for this case. Since phonon absorption/emission processes are forbidden by energy and momentum conservation conditions, sound attenuation is caused solely by phonon scattering.

Using the result of ref.~\cite{Baym2015a}, one can calculate the rate of scattering of a phonon in the mode $\bm q$ against a \he3 atom, given by
\beq\label{eq:dampingrate_ph-3}
\Gamma_q^{ph-3}=\frac{x_3 m_4 c q^4}{4\pi\rho_0} J,
\eeq
where $J$ is an angular integral defined by
\beqn
J&=&\frac{1}{2}\int_{0}^{\pi} (A+B\cos{\theta})^2 \sin{\theta}\rm{d}\theta \nonumber\\
&=&A^2+\frac{B^2}{3}\simeq1.6,
\eeqn
where the parameters $A=-1.2\pm0.2$ and $B=0.70\pm0.035$ are dimensionless quantities~\cite{Boghosian1967,Watson1969,Abraham1969b,BaymBook2004} .

\begin{figure}[h]
\centering
\includegraphics[width=9cm]{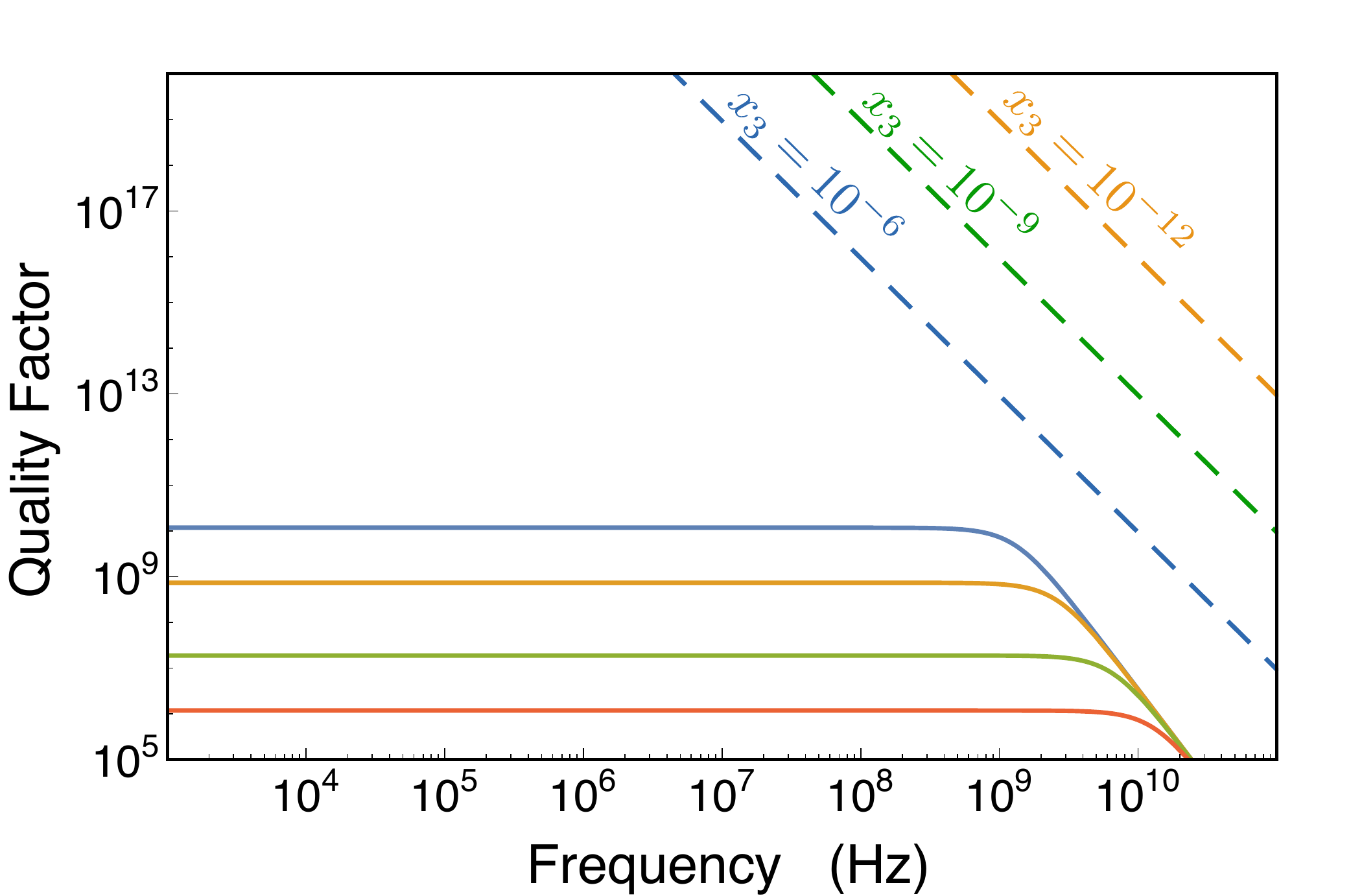}
\caption{Quality factor of an acoustic mode $\bm q$, computed using the expression of the damping rate given at Eq.~\ref{eq:BLDamping} for 3-phonon interaction, as a function of the mode frequency $\Omega_q/2\pi$ at temperatures 10 (blue line), 20 (orange line), 50 (green line) and 100 mK (red line). The quality factor is computed from Eq.~\ref{eq:dampingrate_ph-3} for phonon-\he3 interaction processes is shown for different \he3 concentrations: $x_3=10^{-6}$ (blue dashed line), $x_3=10^{-9}$ (green dashed line) and $x_3=10^{-12}$ (orange dashed line).}
\label{Qph-3n}
\end{figure}

We show Fig.~\ref{Qph-3n} the quality factor calculated for different \he3 concentrations using the phonon scattering expression given above in Eq.~\ref{eq:dampingrate_ph-3}. We observe that for low \he3 concentrations this damping mechanism should give a negligible effect at all frequencies compared to the 3-phonon interaction contribution. Note that this results gives us an estimation of sound attenuation when we only consider phonon-\he3 scattering. A complete theory would need to consider contributions from phonon absorption and emission by \he3 quasiparticles.

\subsubsection{Boundary scattering}
In previous sections, we considered phonon-phonon and phonon-\he3 interactions in the case where the thermal phonon mean free path is limited by collisions between phonons, and where the \he3 quasiparticle mean free path is limited by collisions between \he3 quasiparticles; this is a valid approximation in bulk. In the nanofluidic geometries proposed in this work, the confinement is considerable, such that the boundaries of the geometries could limit the phonon or \he3 quasiparticle mean free paths. In principle, if the walls of the nanofluidic geometries were rough they would significantly limit the mean free paths; as originally described by Casimir~\cite{Casimir1938}. In the special case of extremely smooth walls, possible to achieve with quantum nanofluidics, the boundary scattering of phonon and \he3 quasiparticle becomes less important. As described in ref.\cite{ZimanBook1960Ch11}, one can define a roughness-dependent mean free path for boundary scattering defined by:
\beq
l_B=l_B^0 \left( \frac{1+S}{1-S} \right),
\eeq
where $l_B^0$ is the boundary mean free path defined by the characteristic size of the confinement geometry, and $S$, quantifies the roughness of the wall, representing the fraction of the incident phonons (or \he3 quasiparticles) that are specularly reflected from the walls. The remaining phonons are diffusively scattered in all directions. The parameter $S$ describes the whole range of cases from $S=0$ (perfectly rough) to $S=1$ (perfectly smooth). In nanofluidic geometries, atomically smooth silicon surfaces can be used to reach a specularity close to its maximum value~\cite{Parpia1991}. Additionally, the specularity of the walls in these geometries could be tuned in-situ with the amount of \he4 atoms covering the surface~\cite{Heikkinen2020}. The maximum specularity ($S=0.98$) was obtained with a coverage of 3-4 atomic layers of \he4, which is enough to form a superfluid \he4 layer on the surface. The specularity reported in these works corresponded to the fraction of \he3 cooper pairs being specularly reflected from the walls, and is then related to the typical wavelength of these cooper pairs, approximately given by the Fermi wavelength $\lambda_F\sim1 \AA$. Thus, the same surface will appear even smoother to the longer wavelengths of phonons and \he3 quasiparticles. We would then expect minimal boundary effects in nanofluidic geometries made of atomically smooth substrate materials (e.g. silicon or glass wafers).

\section{\label{sec:4}Phononic Nanostructures}

\subsection{Nanofluidic channels}
Exploiting  the recent development in quantum nanofluidics~\cite{Duh2012,Thomson2014,Zhelev2018} (i.e. nanoscale confinement geometries for quantum fluids), we aim to fabricate precisely defined phononic nanostructures to confined a superfluid acoustic mode at the nanoscale.

\begin{figure}[h]
\centering
\includegraphics[width=8cm]{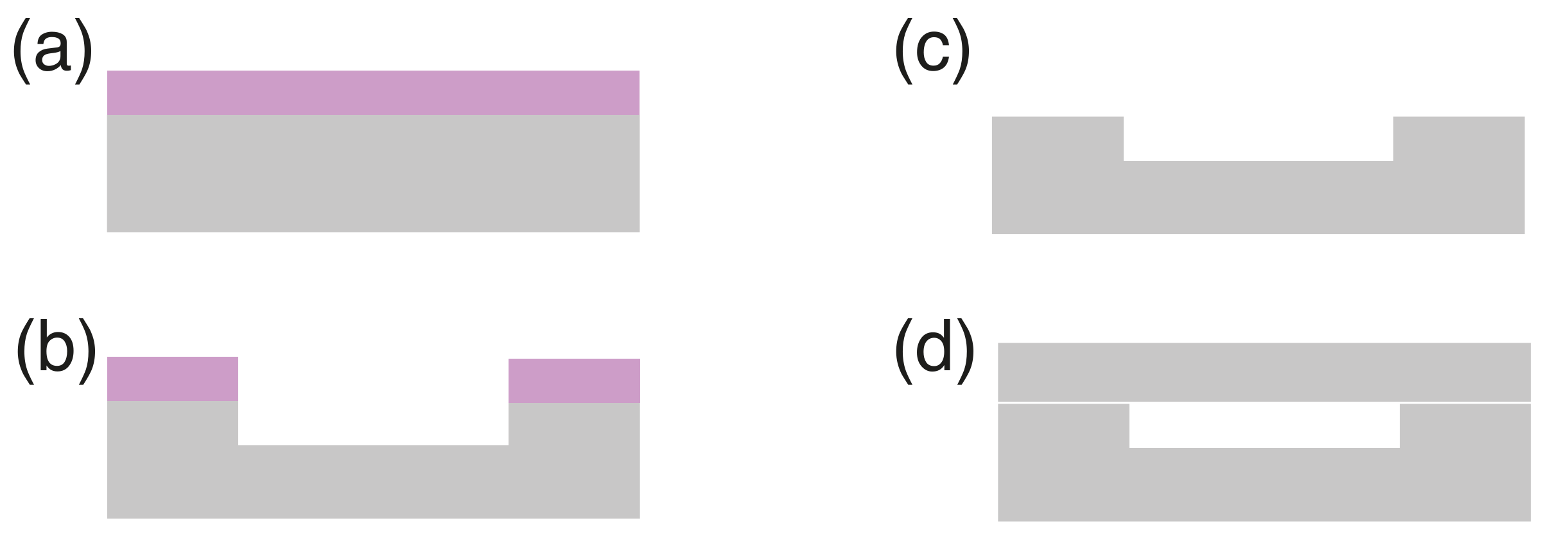}
\caption{Schematic of the nanofabrication process steps of nanofluidic geometries. a) Substrate (grey color) is covered with a masking layer (pink color) prior to the lithography step, b) lithography followed by substrate etching, c) substrate cleaning, d) direct wafer bonding.}
\label{NanofabricationProcess}
\end{figure}

A schematic of the nanofabrication process steps is shown in Fig.~\ref{NanofabricationProcess}, more details of this type of fabrication process can be found at references~\cite{Duh2012,Rojas2014,Rojas2015,Souris2017}. Using optical lithography techniques, we can easily pattern micron sized planar geometries on a substrate (e.g. silicon, glass). By etching these patterns with nanometer scale height, we form shallow 3D structures. This process is followed by a direct wafer bonding technique to enclose the nanofluidic geometry. This fabrication technique allows us to create duct channels, cavities, and more complex hollow structures, which can then be filled with superfluid \he4.  

\subsection{Sound propagation in nanofluidic channels}
In this proposal, we aim to study superfluid \he4, at mK temperatures, deep into the superfluid phase ($\rho_s\rightarrow\rho$). In this regime, superfluid \he4 can be described as an ideal fluid with no viscosity, with dissipative effects treated separately. For this case, superfluid hydrodynamics is governed by Euler equation~\cite{PuttermanBook}, leading to the classical wave equation for the scalar pressure field:
\beq
\frac{\partial^2 p}{\partial t ^2} - c_h^2 \nabla^2 p= 0,
\eeq
where $p=p(\bm{r},t)$ is the scalar pressure field in liquid helium at position $\bm{r}$ and time $t$, and $c_h$ is the sound velocity in liquid \he4 at zero temperature and saturated vapour pressure~\cite{Donnelly1998}. Sound propagation from liquid helium into a solid substrate is a difficult problem to treat exactly; as the scalar pressure field of the liquid must be matched with the longitudinal and transverse modes of the solid substrate. A simpler problem to solve is to consider the solid material as a simple isotropic acoustic media of density $\rho_{\rm sub}$, and sound velocity $c_{\rm sub}$, associated to the longitudinal velocity of the solid (i.e. pressure waves), given by
\beq\label{eq:YoungMod}
c_{\rm sub}=\sqrt{\frac{E(1-\nu)}{\rho_{\rm sub}(1+\nu)(1-2\nu)}},
\eeq
where $E$ is the Young's modulus, and $\nu$ is the Poisson ratio of the solid material. Assuming continuity of the pressure field at the liquid/solid interface, one obtains the normal incidence reflection coefficient for the acoustic wave amplitude at the liquid/solid interface:
\beq
r=\left | \frac{Z_h-Z_{\rm sub}}{Z_h+Z_{\rm sub}} \right |,
\eeq
where $Z_h=\rho_h c_h$ and $Z_{\rm sub}=\rho_{\rm sub} c_{\rm sub}$ are respectively the characteristic acoustic impedances of liquid \he4, and the solid substrate. The characteristic acoustic impedance of the different substrate materials for the sonic crystal are reported in Table~\ref{tab:AcousticImpedance}. 
\begin{table}[h]
\caption{\label{tab:AcousticImpedance}Table of acoustic properties of the substrate materials composing sonic crystals. The characteristic acoustic impedance is given in \textit{rayl} (Pa.s/m) unit. The sound velocity for solid materials was derived from the literature using Eq.~\ref{eq:YoungMod}.}
\begin{ruledtabular}
\begin{tabular}{lrrr}
  Material& $\rho$ (kg/m$^3$) & $c$ (m/s) & $Z$ (rayl)\\
  \hline
  Liquid \he4~\footnote{At saturated vapour pressure and T= 0 K, values from~\cite{Donnelly1998}.} & 145.1 & 229.5 & $3.33\times10^{4}$\\
  Fused silica~\cite{Semiwafer}& 2200& 5900 & $1.30\times10^7$\\
  Crystal quartz~\footnote{AT cut crystal quartz values from~\cite{Zhang2006}.} & 2650 & 7000 & $1.86\times10^7$  \\
  Borosilicate glass~\cite{Semiwafer}   & 2230& 5460 & $1.22\times10^7$ \\
  Crystal Silicon~\footnote{[100] orientation, values from~\cite{Wang2004}.} & 2330 & 8560 & $1.99\times10^7$   \\
  Sapphire~\cite{Zhang2006} & 3910 & 9940  & $3.89\times10^7$\\
\end{tabular}
\end{ruledtabular}
\end{table}

Typically, due to the strong acoustic impedance mismatch between liquid \he4 and any other solid material, the reflection coefficient, as estimated using the normal incidence reflection coefficient, will be greater than 99.5\%. We exploit this strong acoustic impedance mismatch and the high reflection coefficient to confine the acoustic mode within the superfluid. Similarly to optical fibre waveguides, in which light is guided by total internal reflection, we can use nanofluidic channels as waveguides for acoustic waves propagating in superfluid helium.
\begin{figure}[h]
\centering
\includegraphics[width=8cm]{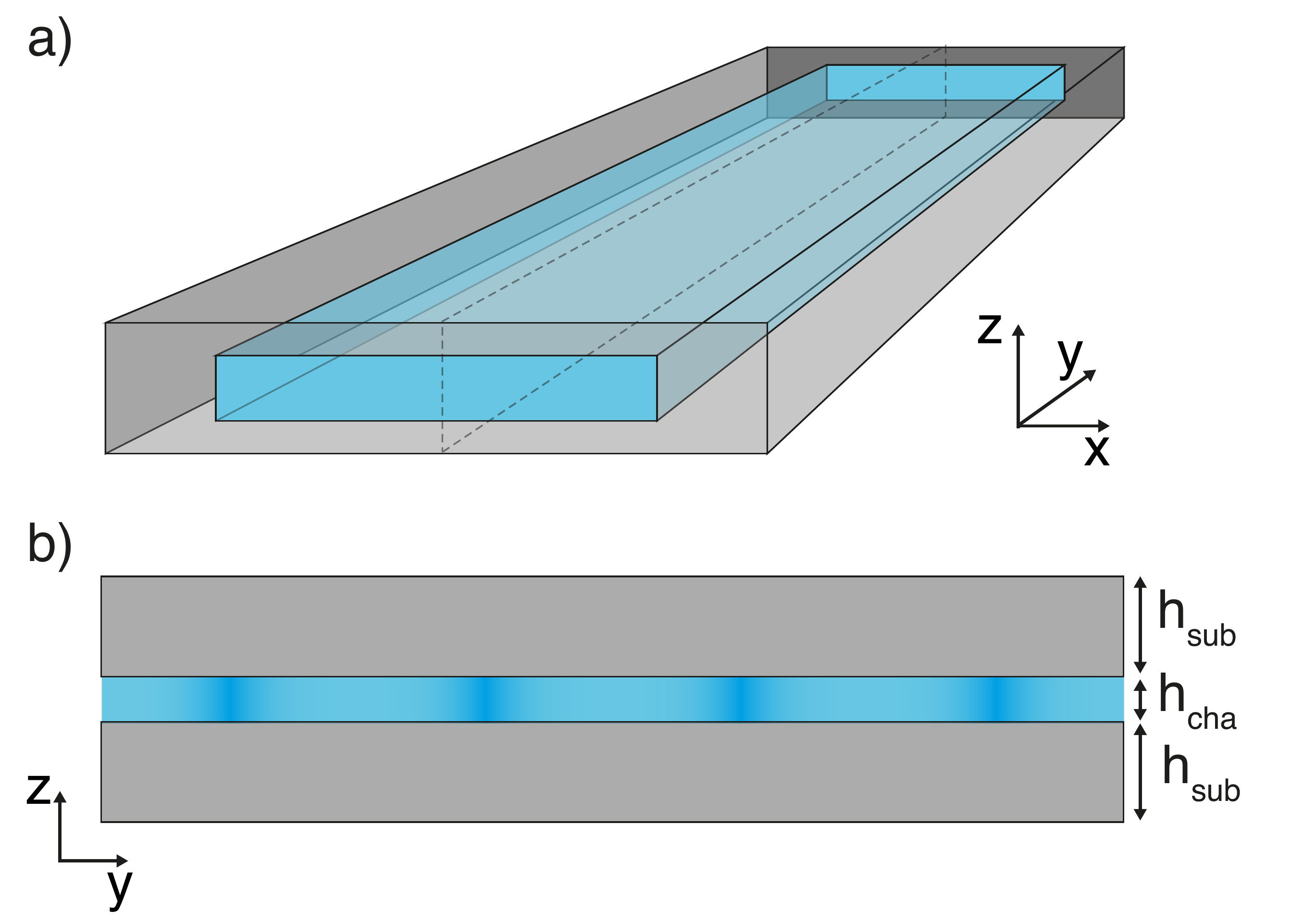}
\caption{a) Schematic of a rectangular nanofluidic channel  made in a solid substrate (grey color), and filled with superfluid \he4 (blue color) b) Cross-section of the nanofluidic channel taken along the black dashed line rectangle drawn in (a) showing an acoustic wave propagating in the $y$-direction. In the range of parameters used in this proposal, such nanofluidic channel acts like an acoustic waveguide.}
\label{DuctChannelWaveGuide}
\end{figure}

Because of the high reflection coefficient, we can approximate the boundary conditions for acoustic propagation within the channel by the sound-hard boundary conditions given by
\beq\label{eq:soundhard}
\bm{n}\cdot\bm{\nabla}{p}=0,
\eeq
where $\bm{n}$ is a unit vector normal to the surface of the wall. This Neumann boundary condition means that the normal derivative of pressure vanishes on the wall, expressing the fact that a sound wave is perfectly reflected at the liquid/solid interface. We show in the next section how deviations from this ideal model lead to radiation losses, which can be taken into account using numerical simulations. In the proposed nanofluidic geometries, the height of the geometry is much smaller than the lateral dimensions. An example of a typical rectangular nanofluidic channel is shown Fig.~\ref{DuctChannelWaveGuide}. The height of the nanofluidic channel defined along the $z$-axis is typically $h_{\rm cha}\sim100$ nm, while its lateral dimensions are, at least, of the order of tens of microns. In the ideal case of perfectly reflected sound waves, we can define the quantized $z$-component of the wave vector as
\beq
k_z^n = \frac{n \pi}{h_{\rm cha}},
\eeq
where $n=0,1,2,...$ is an integer. Therefore, the cutoff frequency is
\beq
\frac{\omega_c}{2\pi} = \frac{c_h}{2 h_{\rm cha}}\sim 1\ \mathrm{GHz},
\eeq
at a channel height of the order of 100 nm. Modes with $k_z\neq0$ and a frequency below the cut-off frequency are evanescent and do not propagate in the nanofluidic geometry. In the proposed experiment, the sound frequency will be kept well below the cut-off frequency so that the sound propagation inside the nanofluidic geometry with $k_z=0$ can be considered 2D to a good approximation.

\subsection{2D sonic crystals}
Using standard optical lithography techniques, we can pattern microscale cylindrical solid pillars into a wafer substrate. Following a direct wafer bonding step, the pillars on the bottom wafer are enclosed by a second substrate on top, creating a nanofluidic environment.
A lattice of these pillars forms an artificial crystal. Filling this structure with a fluid (i.e. superfluid helium) creates a 2D \textit{sonic crystal}, a type of artificial phononic crystal for pressure waves in the fluid. Fig.~\ref{PhononicCrystal_perspective} shows a 2D sonic crystal made of a square-lattice of cylindrical pillars, without showing the top substrate enclosing the geometry for clarity. The point defect in the sonic crystal structure traps the acoustic mode of interest.

Bulk superfluid helium is an isotropic acoustic medium with a quasi-linear dispersion relation $\Omega\simeq c_h k$ at low wave vectors, which is a valid approximation at all the sound frequencies used here. Meaning that the acoustic eigenmodes (i.e. phonons) in \text{bulk} superfluid helium are well-defined plane waves. In our nanofluidic geometries, the confinement along the $z$-axis, causes the sound propagation to be guided in 2-dimensions. Also, the boundary conditions imposed by the lattice of pillars defining the sonic crystal, generate Bragg scattering. Hence, propagating modes in the sonic crystal are Bloch waves of the form
\beq
p(\bm{r})=e^{-i\bm{k}\cdot\bm{r}} \tilde{p}(\bm{r}),
\eeq
with $\bm{k}$ the 2D wave vector. These Bloch waves are plane waves modulated by a periodic function $\tilde{p}(\bm{r})$ of the same periodicity as the sonic crystal. This derives from the classical Bloch's theorem of solid state physics here applied to acoustic waves in an artificial crystal~\cite{LaudeBook}. Sound propagation within the sonic crystal is defined by a phononic band structure, analogous to an electronic band structure but for acoustic waves.

\begin{figure}[h]
\centering
\includegraphics[width=8.6cm]{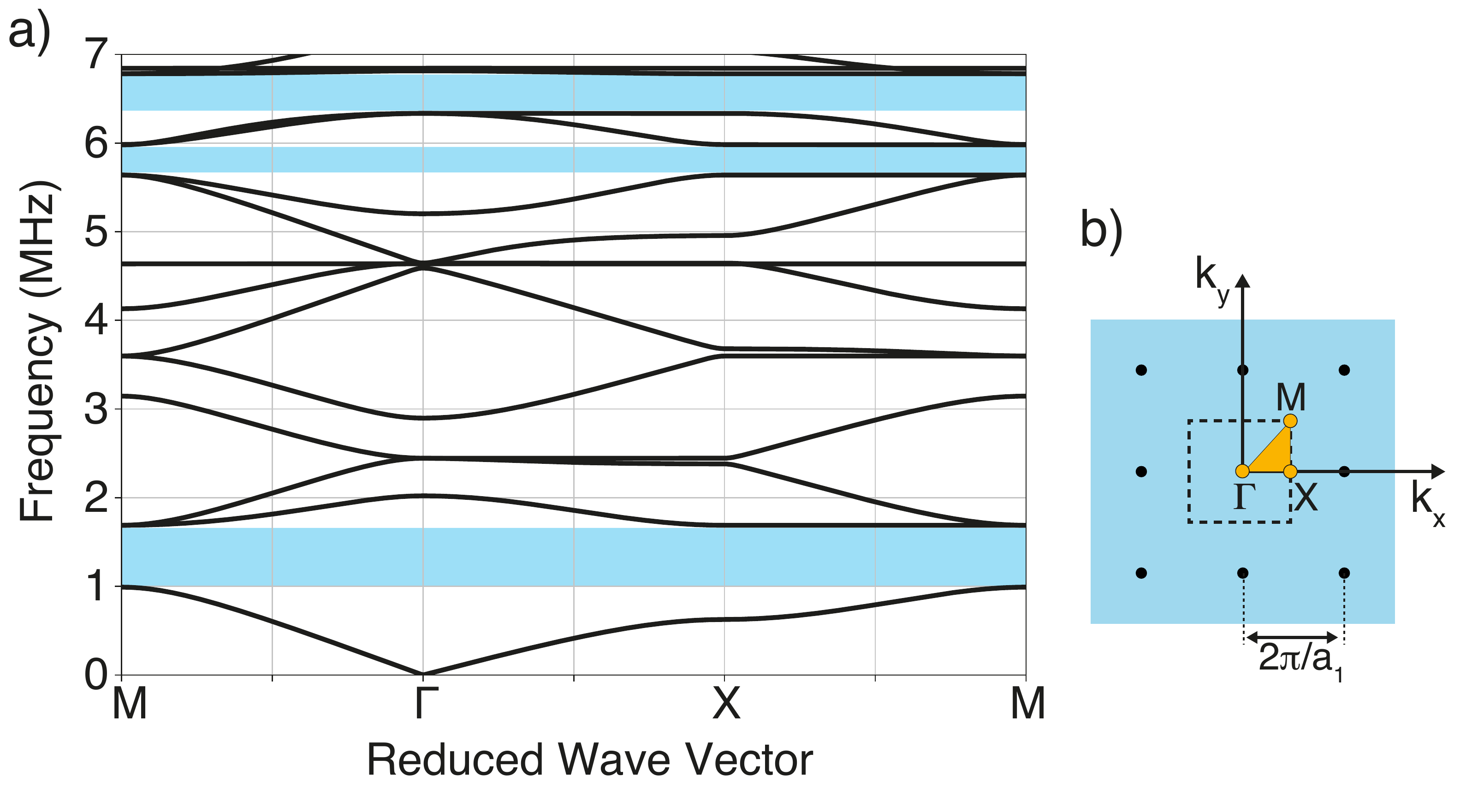}
\caption{a) Calculated phononic band structure for a square lattice of cylindrical pillars in a superfluid \he4. The lattice spacing is $a_1=100$ $\mu$m and the pillar diameter is $a_2=90$ $\mu$m. Three bandgaps are visible (blue bands) in this structures. b) The reciprocal lattice showing the irreducible Brillouin zone (orange triangle), and the characteristic points: $\Gamma$, $M$ and $X$.}
\label{Three_Bandgaps90}
\end{figure}

Phononic band structures plotted as a function of the reduced wave vector in the irreducible Brillouin zone (IBZ), such as the one shown Fig.~\ref{Three_Bandgaps90}, were found using numerical calculations based on Finite-Element Method (FEM) with COMSOL Multiphysics. To calculate phononic band structures, we find the eigenmodes of a square shape unit cell of size $a_1$ with a disk of diameter $a_2$, using Floquet periodic boundary conditions on the edges of the unit cell, and sound-hard boundary conditions at edge of the disk.

The phononic band structure (Fig.~\ref{Three_Bandgaps90}) shows 3 complete bandgaps. Phononic bandgaps originates from Bragg interferences in the crystal, caused by the important contrast between solid and liquid acoustic impedances. The bandgaps are \textit{complete bandgaps}, meaning that they are opened for all values of the wavevector $\bm k$ in the IBZ. Thus, there is an entire frequency band, for which sound propagation is forbidden. Sound waves at frequencies within the bandgap are evanescent waves, which do not propagate more than a characteristic length inside the sonic crystal. These evanescent waves are waves whose wave number is imaginary, therefore their amplitude decays exponentially within the sonic crystal (see Fig.~\ref{ExpDecay}).

\begin{figure}[h]
\centering
\includegraphics[width=8.6cm]{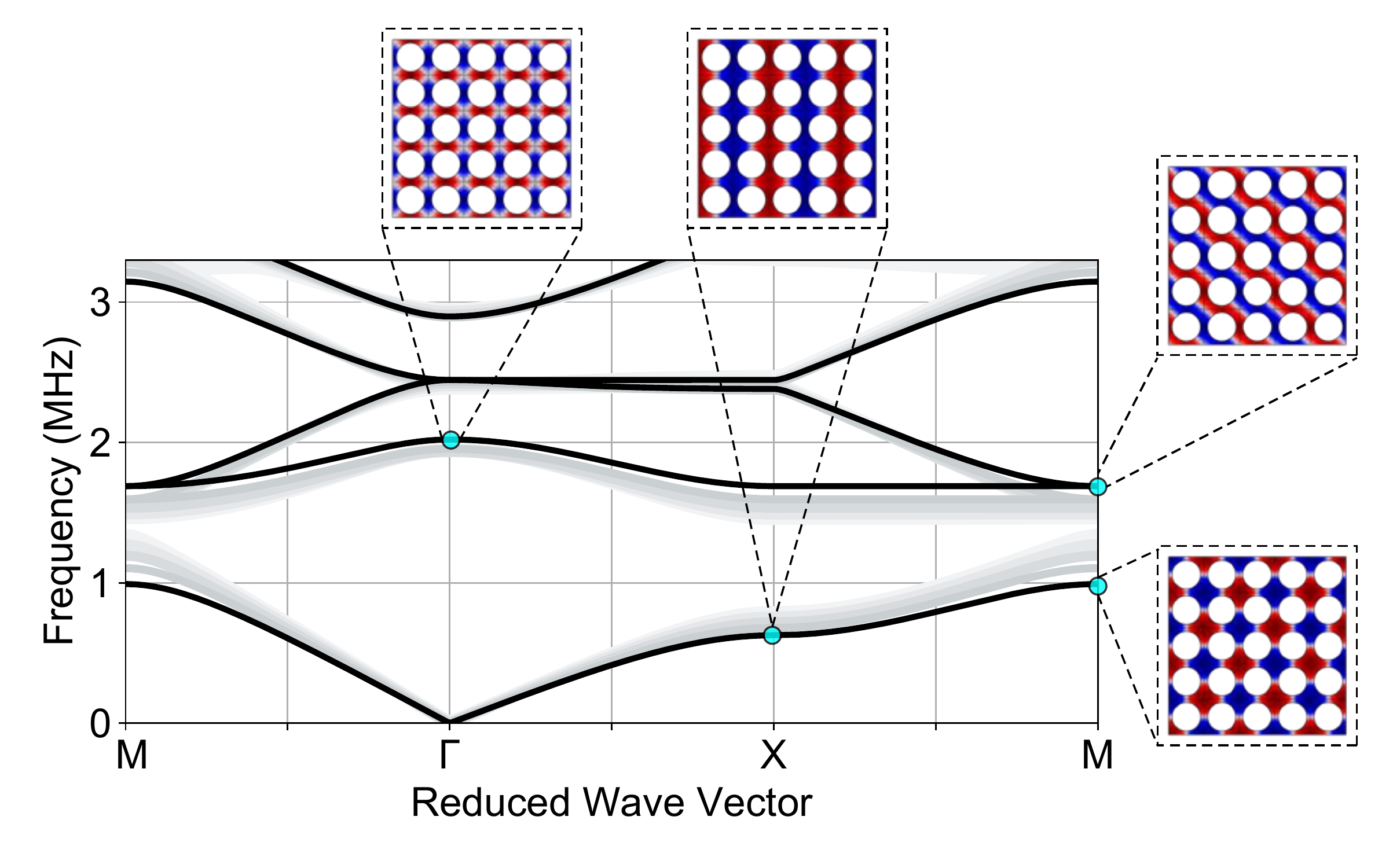}
\caption{Phononic band structure zoomed around the lowest frequency bandgap calculated for a lattice spacing $a_1=100$ $\mu$m, at different values of the pillar diameter $a_2$, from $90$ $\mu$m (black line) to $70$ $\mu$m by steps of 5 $\mu$m (grey lines). The mode shape of the pressure field is shown at different characteristic points (color plot).}
\label{1stBandgap_a2values}
\end{figure}

The position and width of the bandgap are functions of the geometric parameters of the lattice. For this sonic crystal geometry, the frequency of the center of the bandgap is governed by the distance between pillars ($a_1$), while the bandgap width is also governed by the ratio pillar diameter ($a_2$). For instance, increasing the $a_2/a_1$ ratio towards unity, increases the bandgap width. Results of numerical simulations show (see Fig.~\ref{Three_Bandgaps90}) that the bandgaps are centred at $1.34$, $5.81$, and $6.56$ MHz, with a width of $0.7$, $0.34$, and $0.45$ MHz respectively. In Fig.~\ref{1stBandgap_a2values} the closing of the lowest frequency bandgap at the characteristic point $M$ of the reciprocal space for smaller values of $a_2$. The bandgap closes at around $a_2\simeq65$ $\mu$m in our geometry, that is for a ratio $a_2/a_1\simeq0.65$.

\begin{figure}[h]
\centering
\includegraphics[width=8cm]{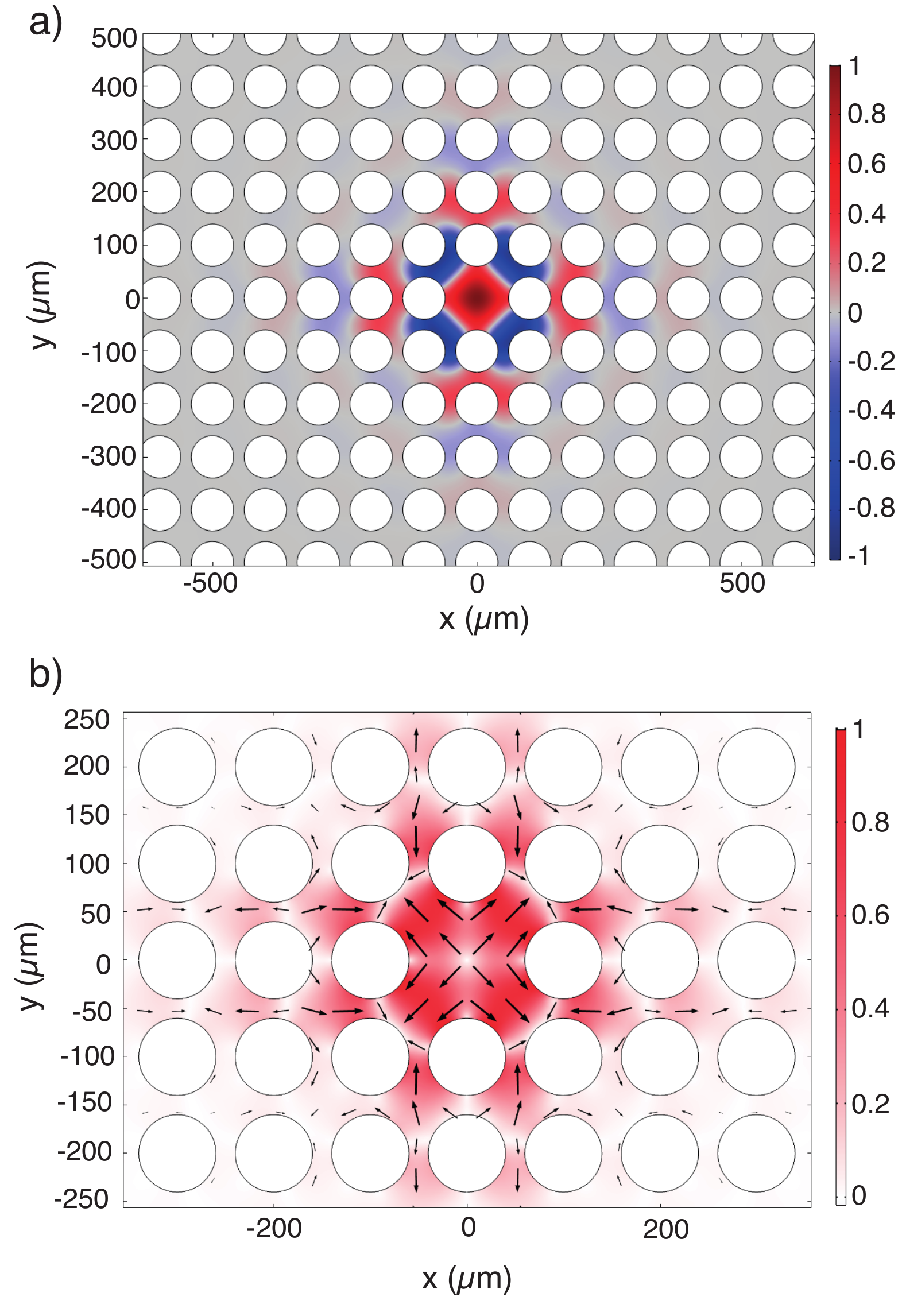}
\caption{Mode shape of a point defect acoustic mode ($\Omega_a/2\pi$=1.34 MHz) at the center of a 2D sonic crystal of parameters $a_1=100$ $\mu$m and $a_2=80$ $\mu$m. (a) Acoustic pressure field mode shape, (b) velocity field mode shape indicating the velocity magnitude (color scale) and direction (black arrows).}
\label{DefectModeShape}
\end{figure}

By removing a single pillar from the 2D square-lattice structure, we define a point defect, with an acoustic mode at $\Omega_a/2\pi=1.34$ MHz, right at the centre of the lowest frequency bandgap of the sonic crystal. Fig.~\ref{DefectModeShape} shows the mode shape of the point defect acoustic mode.
Since the point defect is surrounded by a 2D sonic crystal, acoustic waves at the mode frequency ($\Omega_a/2\pi=1.34$ MHz) cannot propagate outside the defect, completely confining the mode. While this is in principle true for an infinite sonic crystal, for a real sonic crystal, there will be finite size effects limiting the quality factor. This leads to an acoustic mode quality factor dependent on the details of the sonic crystal parameters. We investigate in the next section the different sources of radiation losses for a point defect acoustic mode.

\subsection{Extrinsic losses mechanisms}
The acoustic mode of interest is localised at the point defect of a 2D sonic crystal embedded in a nanofluidic device. The nanofluidic device will be immersed in liquid helium contained within a 3D superconducting microwave cavity as described in section~\ref{sec:5}. The extrinsic sources of loss are associated with mechanisms by which acoustic energy is radiated out of the sonic crystal's point defect into, both the substrate and the surrounding liquid helium. 

The acoustic mode defined at the point defect of a sonic crystal has a quality factor defined by
\beq\label{eq:Q_a}
\frac{1}{Q_a}=\frac{1}{Q_a^{\rm int}} + \frac{1}{Q_a^{\rm ext}},
\eeq
where $Q_a^{\rm int}$ corresponds to the intrinsic quality factor due to internal dissipation mechanisms in superfluid \he4, which are discussed at section~\ref{sec:4}. $Q_a^{\rm ext}$ on the other hand corresponds to the extrinsic sources of loss from acoustic radiation out of the point defect mode, and is given by  
\beq
\frac{1}{Q_a^{\rm ext}}=\frac{1}{Q_a^{\rm sc}} + \frac{1}{Q_a^{\rm sub}},
\eeq 
where $Q_a^{\rm sc}$ is the quality factor related to the planar acoustic radiation loss out of the sonic crystal caused by the finite size of the sonic crystal, and $Q_a^{\rm sub}$ is the quality factor related to substrate internal loss. These two quantities, $Q_a^{\rm sc}$ and $Q_a^{\rm sub}$, constitute the main contributions to the external quality factor.

\subsubsection{Planar radiation loss: $1/Q_a^{\rm sc}$}
The planar radiation loss out of the sonic crystal ($1/Q_a^{\rm sc}$) contribution is obtained by assuming that the substrate enclosing the nanofluidic geometry behaves as infinitely rigid walls, leading to perfect acoustic wave reflection at the liquid/solid interfaces. This condition leads to a sound-hard wall boundary condition as given by Eq.~\ref{eq:soundhard}, and here the substrate loss contribution is ignored.

In this case, and because the acoustic pressure field is uniform in the $z$-direction inside the nanofluidic geometry, we can ignore the 3\up{rd} dimension, and only consider acoustic propagation in the $xy$-plane. In theory, the band structure calculated for a 2D sonic crystal, shown in Fig.~\ref{Three_Bandgaps90}, leads to a set of complete bandgaps. In these bandgaps, acoustic propagation is forbidden and the quality factor associated to radiation losses out of the sonic crystal should diverge. In practice, however, because the sonic crystal has a finite size, the phononic band structure does not describe the entire picture, and acoustic energy can actually radiate out of the sonic crystal. 
\begin{figure}[h]
\centering
\includegraphics[width=8.6cm]{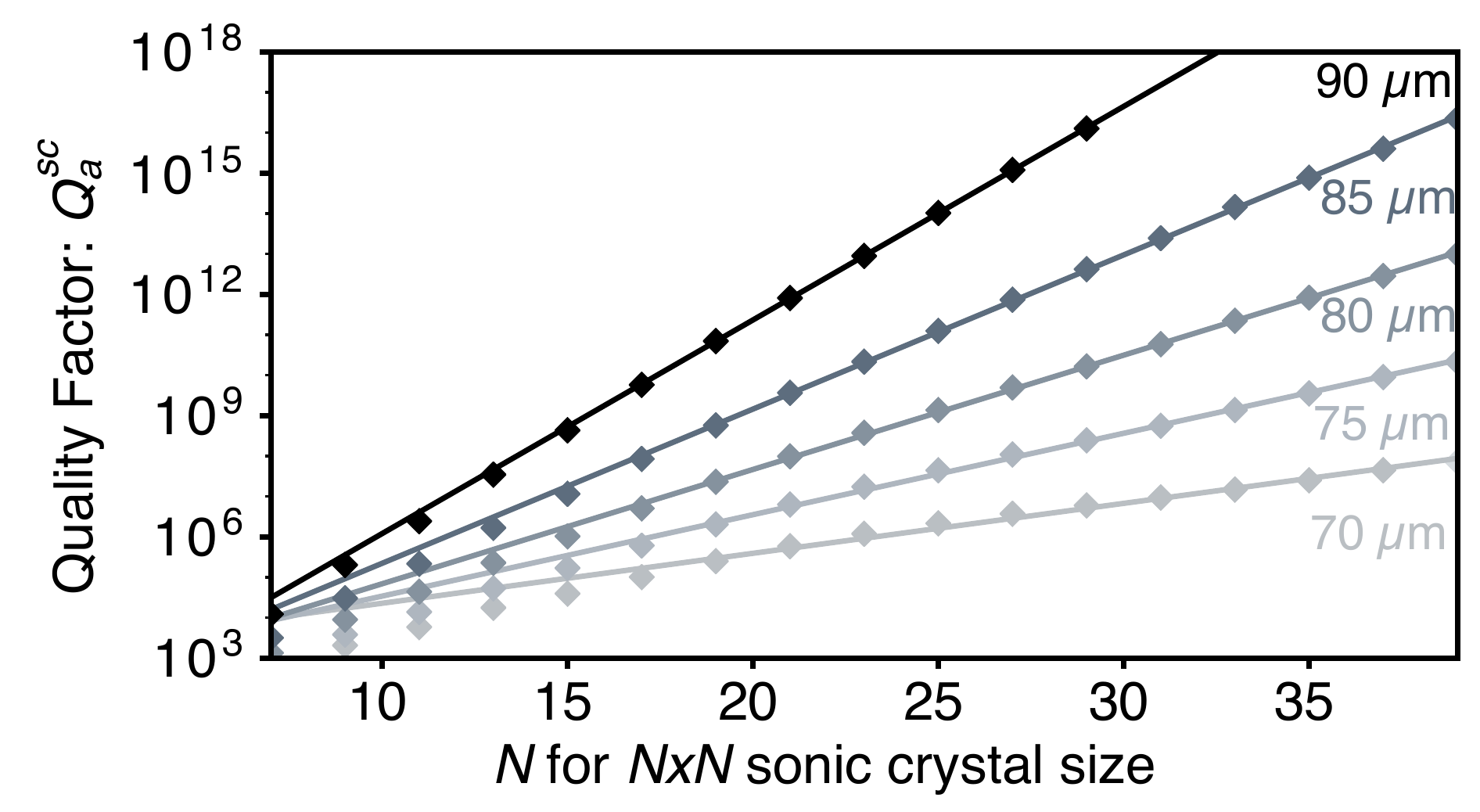}
\caption{Quality factor $Q_a^{\rm sc}$ of a point defect mode obtained considering planar radiation loss in helium caused by the finite size $N$ of the sonic crystal for different values of $a_2$ from $90$ $\mu$m (black dots) to 70 $\mu$m (grey dots), by step of 5 $\mu$m, as indicated on the figure. The lines are linear fits of the datapoint in this log scale calculated for crystal of size $N>17$.}
\label{Qrad_FiniteSizeEffects} 
\end{figure}

\begin{figure}[h]
\centering
\includegraphics[width=8cm]{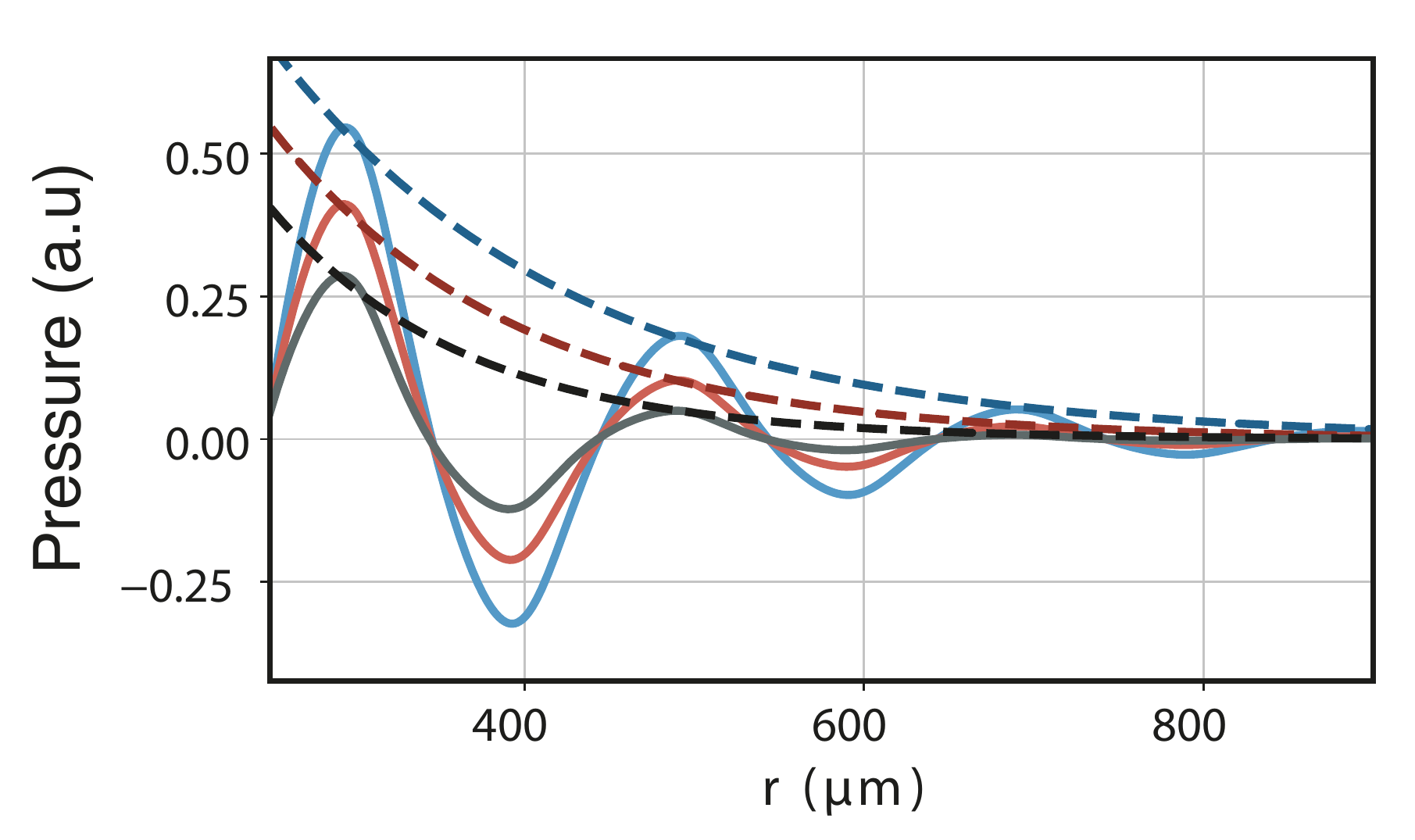}
\caption{Calculated pressure field outside of the defect at the mode frequency ($\Omega_a/2\pi=1.34$ MHz) as a function of the distance to the centre of the point defect for 3 sonic crystals with parameters $a_1=100$ $\mu$m, $a_2=75$ $\mu$m (blue line), 80 $\mu$m (red line), and 85 $\mu$m (black line). The pressure field is normalized by the maximum of the field at the centre of the point defect. The envelope of the normalized pressure field is fitted with an exponential function $p(r)=\exp(-r/l_c)$ where $l_c$ is the characteristic decay length extracted from the fit and given by $l_c=176$ $\mu$m (dashed blue line), $l_c=143$ $\mu$m (dashed red line), $l_c=115$ $\mu$m (dashed black line).}
\label{ExpDecay}
\end{figure}

In an attempt to quantify this effect, we run numerical simulations for sonic crystals of different sizes. Fig.~\ref{Qrad_FiniteSizeEffects} shows the quality factor $Q_a^{\rm sc}$ defined by considering radiation losses out of the sonic crystal as a function of $N$ where $N\times N$ is the number of unit cells in the sonic crystal. As expected, larger sonic crystals provide a higher confinement of the acoustic field within the defect, and therefore higher quality factors. The data also shows that the quality factor is increasing more rapidly with crystal size at larger $a_2/a_1$ ratios. This effect can be understood by looking at the exponential decay of the acoustic pressure field radiated outside of the crystal's defect, which vanishes more rapidly at a larger $a_2/a_1$ ratio (Fig.~\ref{ExpDecay}). Hence, wider frequency bandgaps, caused by larger $a_2/a_1$ ratios, are preferable for limiting radiation losses out of the sonic crystal.

For the largest ratio shown ($a_2/a_1=0.9$) and for a crystal of size $N=17$ (i.e. few mm$^2$), we obtain a quality factor $Q_a^{\rm sc}\sim10^{10}$, which is comparable to the internal quality factor caused by 3-phonon processes at 10 mK (see section~\ref{section:SoundAttenuation}). Hence, if radiation loss out of the sonic crystal was the only external loss factor, obtaining an ultra-high quality factor where internal losses would be the limiting factor, is within reach considering standard nanofabrication techniques. However, radiation loss is not the only external loss factor, which is discussed in the next section.

\subsubsection{Substrate losses: $1/Q_a^{\rm sub}$}
Modelling substrate losses is not trivial, and required a full 3D FEM numerical simulation of the nanofluidic device's components. We identified two main channels of loss originating from the substrate. First, out-of-plane acoustic radiation in the $z$-direction, propagating through the substrate into the surrounding bulk helium. Secondly, internal loss caused by the substrate's participation in the acoustic mode. 

\begin{figure}[h]
\centering
\includegraphics[width=8.6cm]{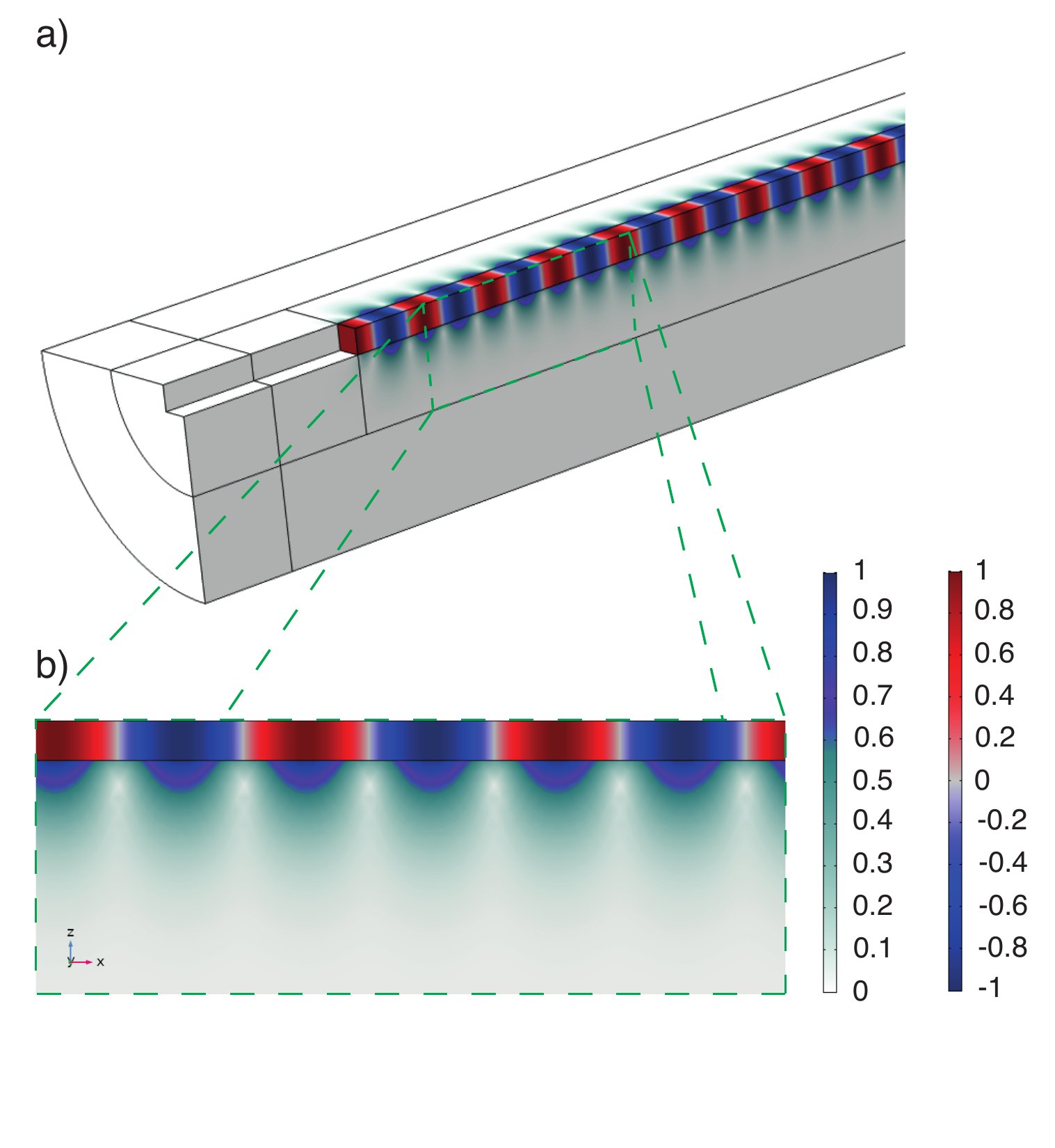}
\caption{3D numerical simulation showing both the acoustic field (blue to red color scale) in helium and the displacement field (white to purple color scale) in the substrate taken at the mode frequency ($\Omega_a=1.34$ MHz). The fields are normalized to their maximum value. a) represents a 3D view, and b) is a zoomed view of the cross section of the rectangular channel.}
\label{Rectangular_Waveguide_FEM}
\end{figure}

To account for out-of-plane acoustic radiation, we studied the acoustic propagation in a simpler acoustic waveguide system made from a nanofluidic rectangular channel inside a thick substrate. The cross section dimensions of the rectangular channel (100 $\mu$m $\times250$ nm) are mimicking acoustic propagation in a particular direction of the $xy$-plane of the sonic crystal. Fig.~\ref{Rectangular_Waveguide_FEM} shows both the acoustic field within the rectangular channel and the displacement field within the solid substrate.

The displacement field in the substrate decays exponentially as expected for evanescent waves. Like for an optical fibre waveguide, because of the wave impedance mismatch, and the fact that the wavevector is confined in the $x$-direction of the channel, or the $xy$-plane in the sonic crystal, it leads to total internal reflections at liquid/solid interfaces. Since evanescent waves do not carry acoustic energy, one would expect zero acoustic radiation through the substrate. However, evanescent waves would only be the correct solutions of the wave equation for an infinitely thick substrate, which in practice, is not the case. Hence, numerical simulations provides a useful tool to quantify deviations from evanescent waves, and estimate the associated loss channel. The results of our numerical simulations, nevertheless, indicates that this loss channel is negligible compare to the second loss channel discussed below. 

The main source of loss in the substrate comes from the substrate's internal loss via its participation in the acoustic mode. To account for this loss channel, we add a loss coefficient for superfluid \he4 and the substrate in our FEM simulations. For the fabrication of the sonic crystal, an appropriate substrate material must be selected. Substrate loss coefficient ($\alpha$) values were calculated from the quality factor values given in the literature for resonators made of these materials, a reasonable approximation provided that internal losses were the limiting factor in these experiments. This proposal aims to provide an architecture in which we can preserve the intrinsic quality factor of superfluid \he4, which is expected to be large at low temperature. Hence, in these simulations, we chose a relatively large value of the internal quality factor for superfluid \he4 ($Q_a^{\rm int}=10^{10}$), the one given by 3 phonons processes at 10 mK.

\begin{figure}[h]
\centering
\includegraphics[width=8.6cm]{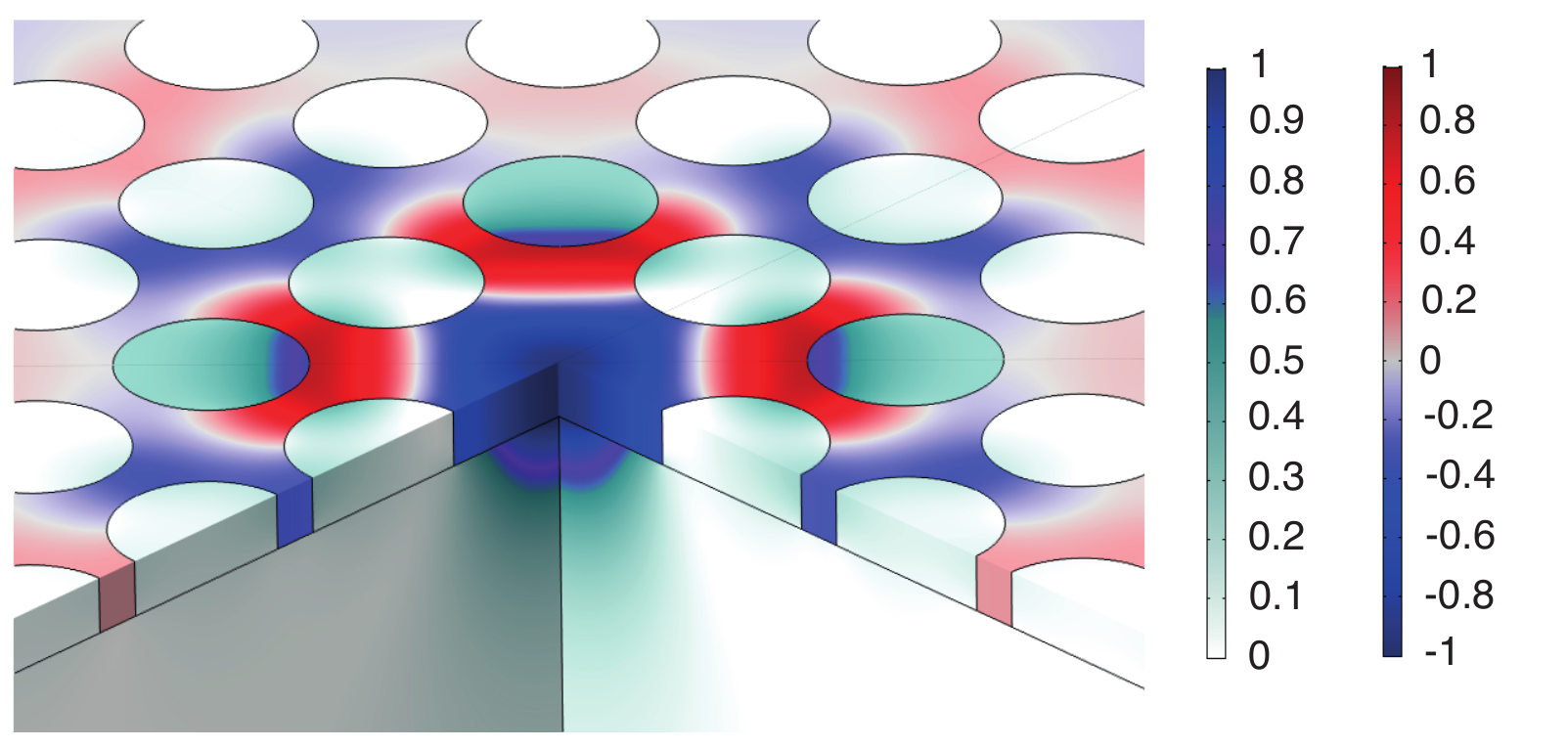}
\caption{3D numerical simulation showing the acoustic field in the vicinity of the sonic crystal defect, and its coupling to elastic deformation in the substrate taken at the mode frequency ($\Omega_a=1.34$ MHz). Both the acoustic field (blue to red color scale) in helium and the displacement field (white to purple color scale) in the substrate are normalized to their maximum value.}
\label{SonicCrystalPie}
\end{figure}

From Eq.~\ref{eq:Q_a}, one can define the reduction coefficient $\eta$ corresponding to the reduction of the intrinsic quality factor of superfluid \he4 caused by substrate loss as
\beq\label{eq:ReductionCoefficient}
\eta=1-\frac{Q_a}{Q_a^{int}}.
\eeq
The numerical simulations allow us to calculate this quality factor reduction coefficient for different type of substrate materials. We run these simulations for the simple rectangular acoustic waveguide geometry and the sonic crystal geometry. In the sonic crystal simulation, we observed that the contribution of the solid substrate to the acoustic mode is localised in the vicinity of the point defect. See Fig.~\ref{SonicCrystalPie}, where both the acoustic field in helium and the displacement field in the substrate are shown.

The results of FEM simulations for different substrate materials are summarized in table~\ref{tab:substrate_loss}, where the reduction coefficient is for the waveguide and the sonic crystal agree within 0.5\%. These results allow us to the identify the most suitable substrates for this application but also a way to quantify the expected reduction coefficient for different geometries.

\begin{table}[hbtp]
\caption{\label{tab:substrate_loss} Table of the quality factor reduction coefficient ($\eta$) induced by substrate loss for different materials, obtained by FEM numerical simulations. Typical materials properties values: density ($\rho$), Young's modulus ($E$), and Poisson's ratio ($\nu$) have been extracted from an online database~\cite{Matweb}. The isotropic loss factor was computed from the quality factor values reported in the cited references ($\alpha=1/Q$).}
\begin{ruledtabular}
 \begin{tabular}{lccccccc}
 Material & $\rho$& $E$ & $\nu$ & $\alpha$ & $\eta$ \\
 & (kg/m$^3$) & (GPa) & & & & \\
\hline
 Silicon~\cite{McGuigan1978} & 2330 & 170 & 0.06 & $5\times10^{-10}$ & 4.3\%\\
 Quartz~\cite{Goryachev2012}  & 2649& 97 & 0.08 & $5\times10^{-10}$& 7.6\%\\
 Sapphire~\cite{Wang2004} & 3910 & 330 & 0.24 & $1.7\times10^{-9}$& 8.5\% \\
 Fused Silica~\cite{Penn2001} & 2203 & 70.4 & 0.15 & $1.7\times10^{-8}$ & 82.2\%\\
 Borofloat~\cite{Senkal2015} & 2230 & 64 & 0.2 & $1\times10^{-6}$ & 96.8\%\\
 [0.5ex]
\end{tabular}
\end{ruledtabular}
\end{table}

\section{\label{sec:5} Cavity optomechanical system}
We can form a cavity optomechanical system by coupling the sonic crystal's point defect mode described in the previous section to a microwave cavity mode. The architecture we propose aims to exploit the high quality factors offer by 3D superconducting microwave cavities in combination with the high degree of mode confinement provided by nanofluidic geometries.

\subsection{Microwave cavity mode}
\begin{figure}[h]
\centering
\includegraphics[width=8.6cm]{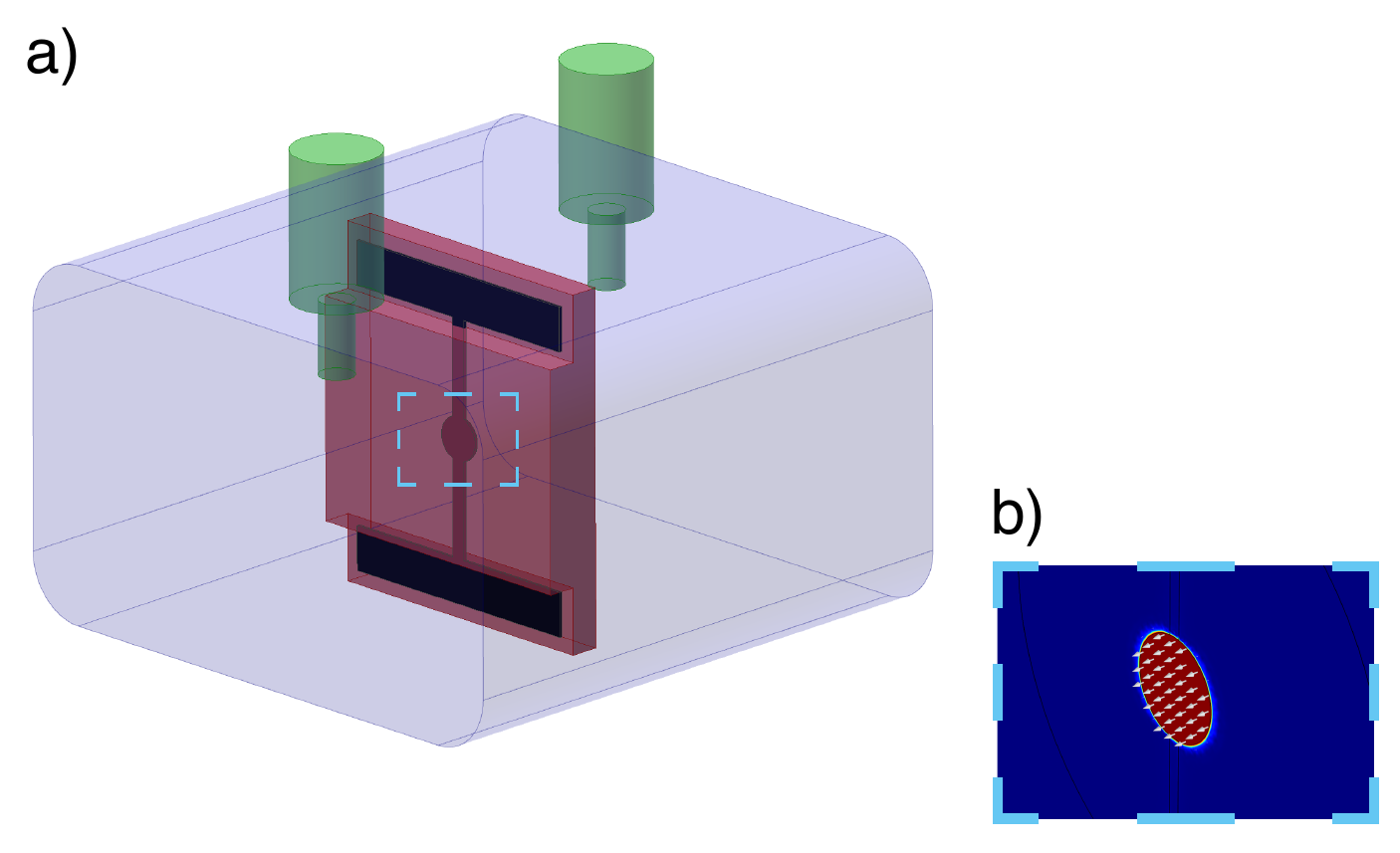}
\caption{a) Schematic of a nanofluidic device (light red area) composed of a nanoscale capacitor (dark red area) terminated by two large antennas (black area), embedded in a 3D superconducting microwave cavity (blue area), which can be readout with a pair of pin couplers (green areas). b) Numerical simulation of the electric field mode shape in this type of resonator showing the focusing of the electric field inside the nanoscale capacitor with arb. unit scale (red is for high intensity and blue for low intensity).}
\label{MicrowaveCavityAndChip}
\end{figure}

A nanoscale parallel plate capacitor can be formed inside the nanofluidic geometry by depositing two superconducting electrodes, on the top and bottom inner surfaces of the sonic's crystal point defect. The typical capacitance will be given by the standard formula $C_{\rm nano}=A\epsilon_0\epsilon_h / d$, where $d$ is the gap between the two parallel electrodes, $A$ is the effective surface area of the electrodes, $\epsilon_0=8.85\times10^{-12}$ pF/m is the vacuum permittivity, and $\epsilon_h=1.057$ the relative permittivity of liquid \he4~\cite{Donnelly1998}. With the values used to define the phononic crystal nanostructure in the previous sections ($d\sim$ 100 nm, $A\sim 0.01$ mm$^2$), we obtain a typical capacitance for the nanoscale capacitor that lies in the pF range. Therefore, to form a microwave resonator with a mode frequency at $\omega_0/2\pi=5$ GHz, which is appropriate for most RF experiments at low temperature, we need to connect to the nanoscale capacitor to an effective inductance ($L=1/(\omega_0^2 C)$) in the nH range. This type of architecture can be realized with good flexibility, by embedding the nanoscale capacitor in a 3D superconducting microwave cavity (schematic shown Fig.~\ref{MicrowaveCavityAndChip})

\begin{figure}[h]
\centering
\includegraphics[width=8cm]{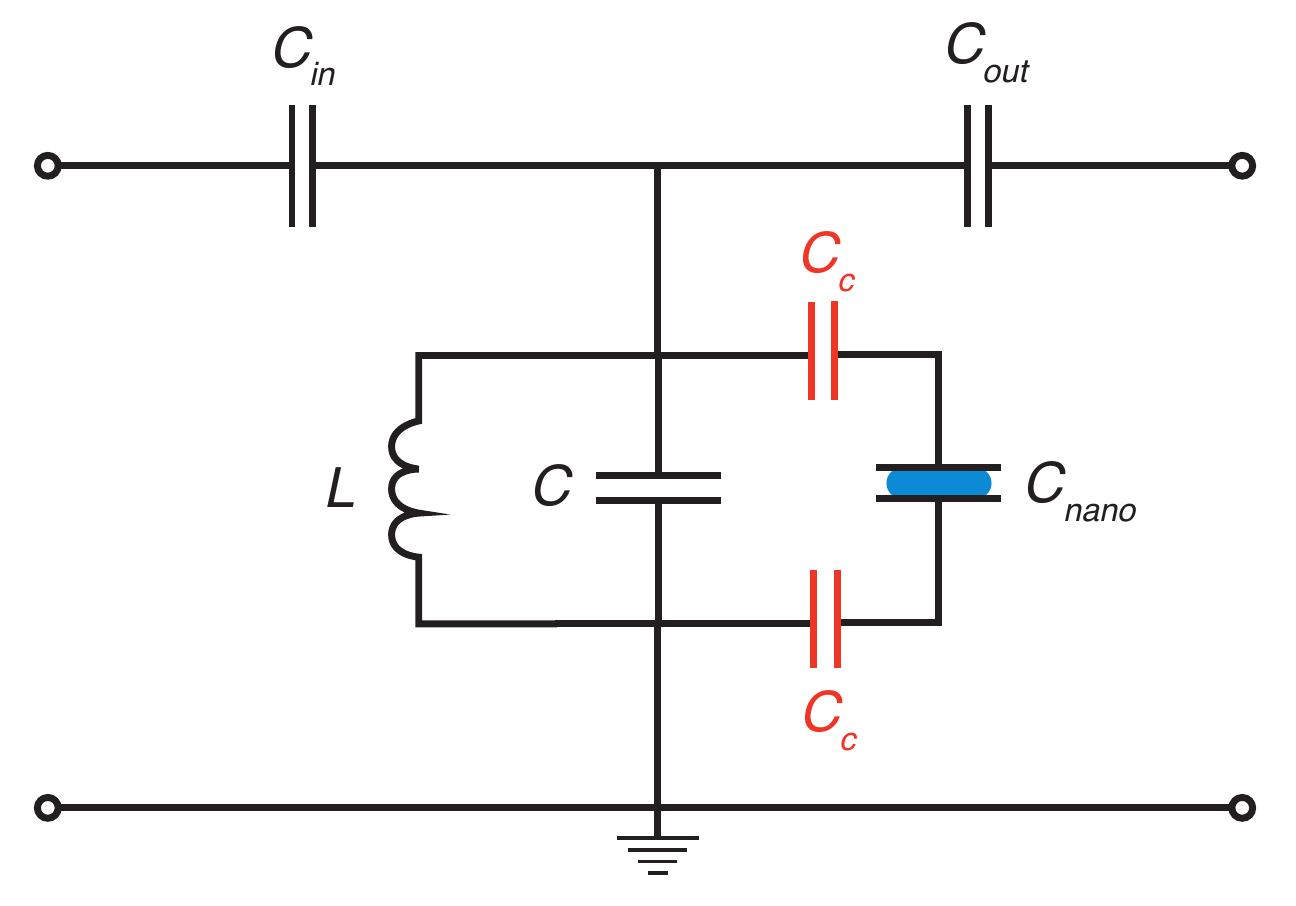}
\caption{Circuit diagram showing the coupling between the nanoscale capacitor ($C_{\rm nano}$) and the 3D superconducting microwave cavity with effective inductance $L$ and capacitance $C$. The coupling capacitors ($C_{\rm c}$) couple the in-cavity field to the nanoscale capacitor. The in and out coupling capacitors ($C_{\rm in}$  and $C_{\rm out}$) represent the pin coupling to the measurement transmission line.}
\label{MicrowaveCavitySchematic}
\end{figure}

To engineer the coupling between the nanofluidic capacitor and the cavity in this type of architecture, different strategies have been used, and the most appropriate for our application would be to either use the capacitive coupling strategy~\cite{Yuan2015} or the galvanic coupling strategy~\cite{Noguchi2016}. In the case of the capacitive coupling strategy, the nanoscale capacitor ($C_{\rm nano}$) is terminated by two large antennas coupling the in-cavity field to the nanoscale capacitor. These two antennas can be represented by coupling capacitors $C_{\rm c}$ between the cavity's inductance and the nanofluidic capacitor as shown in the circuit diagram Fig.~\ref{MicrowaveCavitySchematic}. In the case of a galvanic coupling, the two electrodes of the nanoscale capacitor are grounded to the cavity with electrical contacts, and the two coupling capacitors $C_{\rm c}$ can be removed from the circuit diagram. Finally, to couple the field in and out of the microwave cavity, the most commonly used strategy is referred to as ``pin coupling", which can be represented by two coupling capacitors ($C_{\rm in}$  and $C_{\rm out}$) in the circuit diagram. The pin coupling strategy is thoroughly described in the literature~\cite{ReagorThesis2015}.

\subsection{Electrostrictive coupling}
Using the standard cavity optomechanics framework~\cite{Aspelmeyer2014}, the microwave cavity mode ($\omega_0$) and the acoustic mode ($\Omega_a$) can be represented by two harmonic oscillators, leading to the following bare Hamiltonian,
\beq
\mathcal{H}_0=\hbar\omega_0 a^{\dagger}a + \hbar\Omega_a b^{\dagger}b,
\eeq
where $a^{\dagger}$ are the creation operator for cavity mode photons, and $b^{\dagger}$ for acoustic mode phonons. The coupling between the cavity mode and the acoustic mode is parametric, such that the cavity resonance frequency is modulated by density fluctuations ($\delta\rho$) of the acoustic mode, and can be expanded as:
\beq
\omega_0(\delta\rho)=\omega_0 + \delta\rho \frac{\partial \omega_0}{\partial \rho} + ...\ \ .
\eeq
Considering small density fluctuations, it is sufficient to keep the linear term only. We define the frequency shift by density as $G=-\partial \omega_0/\partial\rho$. Expanding to leading order in the density, we obtain for the optomechanical interaction part of the Hamiltonian,
\beq
\mathcal{H}_{\rm int}=-\hbar g_0 a^{\dagger} a (b^{\dagger} + b),
\eeq
where $g_0=G\delta\rho_{\rm zpf}$ is the single-photon optomechanical coupling strength, expressed as a rate, quantifying the interaction between a single photon and a single phonon.

In the proposed implementation, the microwave cavity mode frequency is modulated by density fluctuations of the acoustic mode, via the liquid \he4 dielectric constant's dependence on density $\epsilon_h(\rho_h)$. This dependence is expressed by Clausius-Mossotti relation given by
\beq
\frac{\epsilon_h - 1}{\epsilon_h -2}=\rho_h \frac{4\pi}{3}\frac{\alpha_M}{M},
\eeq
where $M=4.0026$ g/mol is \he4 molar mass, $\alpha_M=0.123$ cm$^3$/mol is \he4 molar polarizability, which remains constant from room temperature to low temperature~\cite{Harris-Lowe1970}. Differentiating this expression with respect to the density, we find
\beq
\delta \epsilon_h =\frac{(\epsilon_h+2)(\epsilon_h-1)}{3\rho_h} \delta\rho_h,
\eeq
which gives the linear part of the dielectric constant's dependence on density fluctuations. Assuming that the frequency modulation is small compared to the resonance frequency, one can use the result of the theory of cavity perturbations~\cite{PozarBook}, which gives the expression of the relative cavity frequency shift,
\beq\label{eq:CavityPerturbation}
\frac{\delta\omega_0}{\omega_0}=-\frac{\displaystyle\int d^3r  \left( \delta \mu_h |\bm{H}_0(\bm{r})|^2+ \delta \epsilon_h |\bm{E}_0(\bm{r})|^2\right)}{\displaystyle\int d^3r  \left( \mu_h |\bm{H}_0(\bm{r})|^2+ \epsilon_h |\bm{E}_0(\bm{r})|^2\right)},
\eeq
caused by a change in dielectric constant $\delta\epsilon_h$, and a change in magnetic constant $\delta\mu_h$, where $\mu_h$ is the magnetic constant, $|\bm{E}_0(\bm{r})|$ ($|\bm{H}_0(\bm{r})|$) is the electric field (magnetic field) amplitude mode shape, and the integral is taken over the entire volume of the cavity mode. We observe that the terms are related to the electromagnetic stored energy, so that a decrease in the resonance frequency can be related to the increase in the stored energy of the perturbed cavity. Since the magnetic susceptibility of liquid helium is negligible ($\chi_m\sim-10^{-7}$)~\cite{Bruch2000}, compared to its electric susceptibility ($\chi_e\sim0.057$)~\cite{Niemela1995}, we can neglect the change in magnetic constant.
The two terms of the denominator in the right-hand side of Eq.~\ref{eq:CavityPerturbation} are equal and correspond to half the total stored energy of the unperturbed cavity,
\beq
\frac{W_e}{2}=\displaystyle\int d^3r\  \mu_h |\bm{H}_0(\bm{r})|^2 =  \displaystyle\int d^3r\  \epsilon_h |\bm{E}_0(\bm{r})|^2,
\eeq
and so the relative frequency shift can be written as
\beq
\frac{\delta\omega_0}{\omega_0}=-\frac{(\epsilon_h+2)(\epsilon_h-1)}{6\rho_h}\frac{\displaystyle\int d^3r\ \delta\rho(\bm{r}) |\bm{E}_0(\bm{r})|^2 }{\epsilon_h\displaystyle\int d^3r\  |\bm{E}_0(\bm{r})|^2 }.
\eeq

To find the single photon coupling strength, we now write the relative frequency shift associated with zero-point fluctuations of the acoustic mode. The density variation is concentrated in the sonic crystal point defect resonator, and can be written as
\beq
\delta\rho(\bm{r})=\delta\rho_{\rm zpf}\ f(\bm{r}).
\eeq
where $f(\bm{r})$ is a dimensionless, normalised function, representing the mode shape function associated to the acoustic mode. The total acoustic energy stored in this mode is given by the usual expression of the potential energy for an acoustic resonator,
\beq
W_a= \frac{1}{2} \frac{1}{K_h} \int d^3r\ p(\bm{r})^2 ,
\eeq
where $p(\bm{r})=K_h \delta\rho(\bm{r}) /\rho$ is the pressure variation caused by density variations, and $K_h$ is helium's bulk modulus (i.e. inverse of compressibility) defined as
\beq
K_h=-V \frac{\partial P}{\partial V}=\rho_h \frac{\partial P}{\partial \rho} = \rho_h c_h^2.
\eeq
We can then write the total acoustic energy as
\beq
W_a= \frac{c_h^2 \delta\rho_{\rm zpf}^2}{\rho_h} \int d^3\bm{r}f(\bm{r})^2, 
\eeq
which we equate to the zero-point energy ($\hbar\Omega_a/2$) to find the zero-point density fluctuation amplitude:
\beq
\delta\rho_{zpf} = \sqrt{\frac{\rho_h\hbar\Omega_a}{2c_h^2 V_{\rm eff}}}.
\eeq
where $V_{\rm eff}=\int d^3 r f(\bm{r})^2$ is the acoustic mode effective volume. The typical zero-point fractional density change of our acoustic resonators, equivalent to a strain, is given by
\beq
\frac{\delta\rho_{\rm zpf}}{\rho_h}=\sqrt{\frac{\hbar\Omega_a}{2\rho_hc_h^2 V_{\rm eff}}}\sim10^{-10},
\eeq
where $\Omega_a/2\pi\sim1.5$ MHz is the typical frequency of the acoustic mode, $V_{\rm eff}=\alpha V_{\rm def}$ is the effective volume of the acoustic mode with $V_{\rm def}\sim10^{-15}$ m$^3$ the volume of the sonic crystal's point defect, and $\alpha$ a numerical constant that depends on the mode shape. Numerical simulations give $\alpha\sim0.3$ for a sonic crystal defined by $a_2/a_1=0.8$, and $\alpha\sim0.2$ for $a_2/a_1=0.9$. 

\begin{figure}[h]
\centering
\includegraphics[width=8cm]{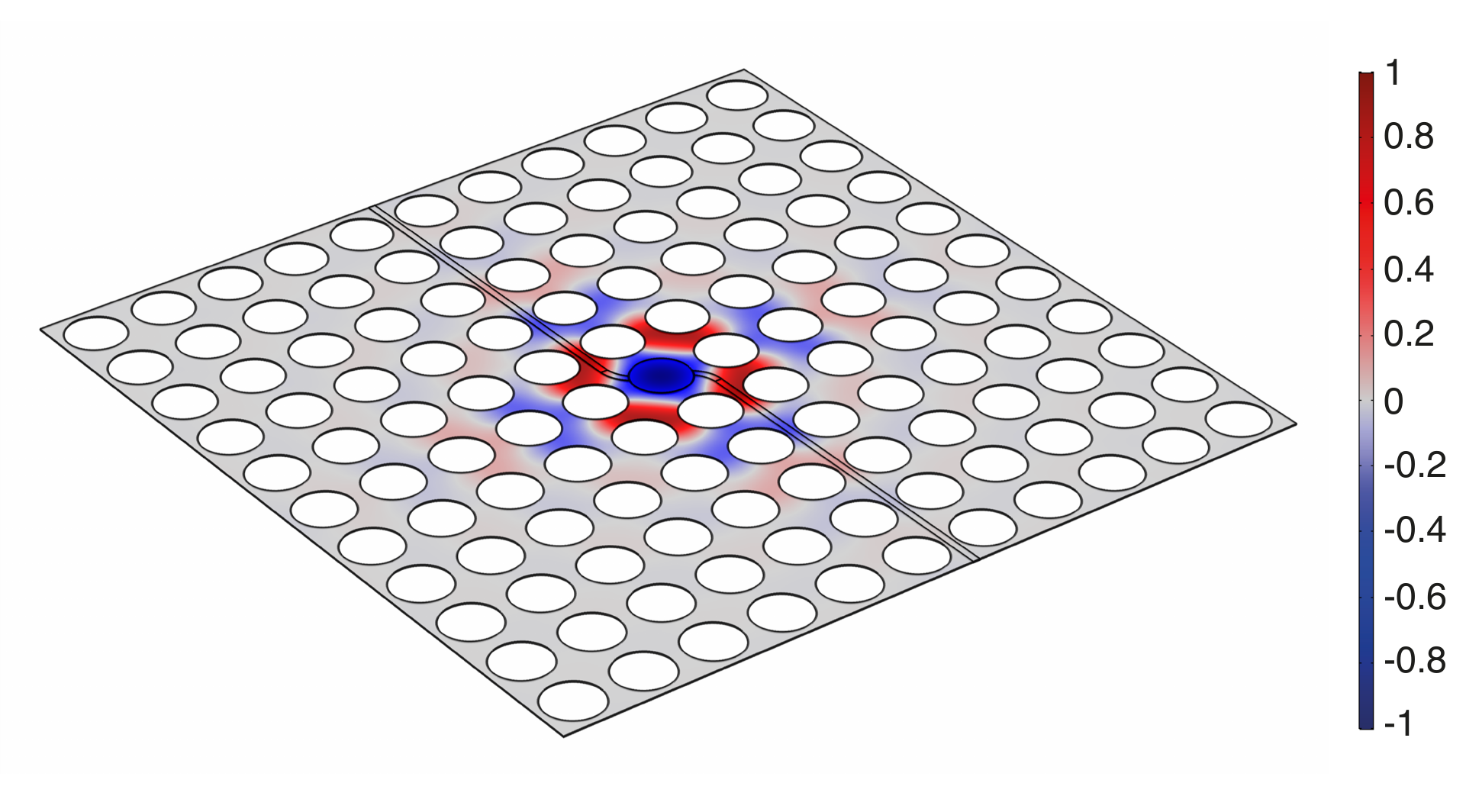}
\caption{Illustration of the overlap between the pressure field (normalized color scale) of the sonic crystal's defect mode and the electric field confined in the nanoscale capacitor located at the center of defect where the pressure field is maximum leading to a mode overlap function close to unity.}
\label{ModeOverlap}
\end{figure}

We can now express the single photon optomechanical coupling strength:
\beqn\label{eq:g0}
g_0 & = & \omega_0 \left ( \frac{\delta\omega_0}{\omega_0} \right)_{\rm zpf} \nonumber\\
& = &\omega_0 \frac{(\epsilon_h+2)(\epsilon_h-1)}{6\epsilon_h}\frac{\delta\rho_{\rm zpf}}{\rho_h}\displaystyle\int d^3r\ f(\bm{r}) g(\bm{r})^2,
\eeqn
where $g(\bm{r})$ is a dimensionless, square-normalised function representing the mode shape of the cavity mode electric field. In order to maximize the mode coupling integral in Eq.~\ref{eq:g0}, the electrodes forming the nanofluidic capacitor can be positioned so that they the electric field overlaps with the maximum of the pressure field (Fig.~\ref{ModeOverlap}).

In an ideal scenario, where the mode coupling integral is of the order of unity ($g(\bm{r})\sim1$), it leads to a single-photon optomechanical coupling strength $g_0\sim0.03\omega_0 (\delta\rho_{\rm zpf} / \rho_h)$, which gives $g_0/2\pi\sim0.015$ Hz for a cavity mode resonance frequency $\omega_0/2\pi=5$ GHz. We show how the optomechanical parameters of the proposed system compare to other recent superfluid optomechanical systems in Table~\ref{tab:SuperfluidOptomechanics}. We note that with the miniaturization of the superfluid acoustic resonator into a nanofluidic geometry, the optomechanical coupling strength is increased by 6 orders of magnitude compared to the other architectures made using a microwave cavity optomechanics.
\begin{table}[h]
\caption{\label{tab:SuperfluidOptomechanics}Table of the figure of merit for superfluid optomechanical systems reported in the literature and this proposal.}
\begin{ruledtabular}
\begin{tabular}{lrrr}
  Systems & $\Omega_a/2\pi$ & $\omega_0/2\pi$ & $g_0/2\pi$\\
  \hline
  This proposal & 1.34 MHz & 5 GHz &  0.015 Hz\\
  De Lorenzo~\etal\cite{DeLorenzo2017} & 8.1 kHz & 10.6 GHz & $4\times10^{-8}$ Hz\\
  Kashkanova~\etal\cite{Kashkanova2016} & 317.5 MHz & $194.5$ THz & $\sim3$ kHz\\
  Shkarin~\etal\cite{Shkarin2019} & 319 MHz & $196$ THz & $3.6$ kHz\\
  Childress~\etal\cite{Childress2017} & 0.02 -1 kHz & $300$ THz & 0.2-10 kHz\\
  He~\etal\cite{He2020} & 1 - 10 MHz & $193$ THz & 133 kHz\\
\end{tabular}
\end{ruledtabular}
\end{table}

\section{\label{sec:6} Conclusion}
Superfluid optomechanics offers great prospects for quantum sensing and quantum technology applications. In this work, we presented the characteristic properties of superfluid \he4 as an acoustic medium that are relevant for these applications. Of particular interest, is superfluid \he4's extremely low sound attenuation in the low temperature limit, and naturally high acoustic impedance mismatch with most solid materials. 

We described the different phonon processes responsible for sound attenuation in superfluid \he4, and how its properties can be tuned with pressure. We highlighted, for instance, the interesting regime at high pressure where sound attenuation is dominated by 4-phonons processes only, allowing to reach stratospheric acoustic quality factors at low temperature. Such regime however has not yet been investigated in part because external factors introduce larger losses preventing its observation. We described how in practice sound attenuation is also affected by the unavoidable presence of \he3 impurities and container boundaries, and how to limit the additional losses that these may introduce.

Furthermore, the exquisite properties of superfluid \he4 can easily be spoiled by the design of the acoustic cavity, which can cause radiation loss and heating. Our proposal focuses on a novel type of devices based on phononic nanostructures enabling the confinement of superfluid acoustic modes at the nanoscale leading to higher frequency modes and the enhancement of optomechanical coupling strengths. We identified the different loss mechanisms for our proposed geometry. Our results, based on FEM numerical simulations, show that sonic crystals provide a useful method to significantly limit radiations losses out of the acoustic mode. Additional external losses arising from the substrate participation to the acoustic mode have been identified, and could be limited with the appropriate choice of substrate materials (e.g. silicon, glass, quartz) and geometries. Finally, an implementation of these phononic nanostructures in a cavity optomechanical setup based on superconducting microwave cavities was proposed. The system parameters have been calculated and compared to the literature. We note that this proposal provides a six order of magnitude improvement on the optomechanical coupling strength compared to previous microwave optomechanical systems. 

Future prospects include the possibility of reducing the mode volume to enhance the optomechanical coupling strength, coupling multiple acoustic modes to create optomechanical arrays and metamaterials, and forming hybrid system by coupling these acoustic modes to other nanomechanical structures (e.g. compliant membranes). 

\begin{acknowledgments}
We acknowledge insightful discussions with Keith C. Schwab, John P. Davis and Fabien Souris. This research was supported by the Royal Society (grants No.UF150140, No. RGF\bs EA\bs 180099, No. RGF\bs R1\bs 180059, and No. RGF\bs EA\bs 201047, No. RPG\bs2016\bs186), the EPSRC (grant No. EP/R04533X/1), and the Royal Holloway Strategy Fund ("FEM simulation for designing nanofabricated structures").

\end{acknowledgments}

\bibliography{Spence_PrApplied_2020}

\providecommand{\noopsort}[1]{}\providecommand{\singleletter}[1]{#1}%
\begin{thebibliography}{110}%
\makeatletter
\providecommand \@ifxundefined [1]{%
 \@ifx{#1\undefined}
}%
\providecommand \@ifnum [1]{%
 \ifnum #1\expandafter \@firstoftwo
 \else \expandafter \@secondoftwo
 \fi
}%
\providecommand \@ifx [1]{%
 \ifx #1\expandafter \@firstoftwo
 \else \expandafter \@secondoftwo
 \fi
}%
\providecommand \natexlab [1]{#1}%
\providecommand \enquote  [1]{``#1''}%
\providecommand \bibnamefont  [1]{#1}%
\providecommand \bibfnamefont [1]{#1}%
\providecommand \citenamefont [1]{#1}%
\providecommand \href@noop [0]{\@secondoftwo}%
\providecommand \href [0]{\begingroup \@sanitize@url \@href}%
\providecommand \@href[1]{\@@startlink{#1}\@@href}%
\providecommand \@@href[1]{\endgroup#1\@@endlink}%
\providecommand \@sanitize@url [0]{\catcode `\\12\catcode `\$12\catcode
  `\&12\catcode `\#12\catcode `\^12\catcode `\_12\catcode `\%12\relax}%
\providecommand \@@startlink[1]{}%
\providecommand \@@endlink[0]{}%
\providecommand \url  [0]{\begingroup\@sanitize@url \@url }%
\providecommand \@url [1]{\endgroup\@href {#1}{\urlprefix }}%
\providecommand \urlprefix  [0]{URL }%
\providecommand \Eprint [0]{\href }%
\providecommand \doibase [0]{https://doi.org/}%
\providecommand \selectlanguage [0]{\@gobble}%
\providecommand \bibinfo  [0]{\@secondoftwo}%
\providecommand \bibfield  [0]{\@secondoftwo}%
\providecommand \translation [1]{[#1]}%
\providecommand \BibitemOpen [0]{}%
\providecommand \bibitemStop [0]{}%
\providecommand \bibitemNoStop [0]{.\EOS\space}%
\providecommand \EOS [0]{\spacefactor3000\relax}%
\providecommand \BibitemShut  [1]{\csname bibitem#1\endcsname}%
\let\auto@bib@innerbib\@empty
\bibitem [{\citenamefont {Aspelmeyer}\ \emph {et~al.}(2014)\citenamefont
  {Aspelmeyer}, \citenamefont {Kippenberg},\ and\ \citenamefont
  {Marquardt}}]{Aspelmeyer2014}%
  \BibitemOpen
  \bibfield  {author} {\bibinfo {author} {\bibfnamefont {M.}~\bibnamefont
  {Aspelmeyer}}, \bibinfo {author} {\bibfnamefont {T.}~\bibnamefont
  {Kippenberg}},\ and\ \bibinfo {author} {\bibfnamefont {F.}~\bibnamefont
  {Marquardt}},\ }\bibfield  {title} {\bibinfo {title} {Cavity optomechanics},\
  }\href {https://doi.org/10.1103/RevModPhys.86.1391} {\bibfield  {journal}
  {\bibinfo  {journal} {Rev. Mod. Phys}\ }\textbf {\bibinfo {volume} {86}},\
  \bibinfo {pages} {1391} (\bibinfo {year} {2014})}\BibitemShut {NoStop}%
\bibitem [{\citenamefont {Bowen}\ and\ \citenamefont
  {Milburn}(2016)}]{BowenBook}%
  \BibitemOpen
  \bibfield  {author} {\bibinfo {author} {\bibfnamefont {W.}~\bibnamefont
  {Bowen}}\ and\ \bibinfo {author} {\bibfnamefont {G.}~\bibnamefont
  {Milburn}},\ }\href@noop {} {\emph {\bibinfo {title} {Quantum
  Optomechanics}}}\ (\bibinfo  {publisher} {CRC Press},\ \bibinfo {year}
  {2016})\BibitemShut {NoStop}%
\bibitem [{\citenamefont {Teufel}\ \emph {et~al.}(2009)\citenamefont {Teufel},
  \citenamefont {Donner}, \citenamefont {Castellanos-Beltran}, \citenamefont
  {Harlow},\ and\ \citenamefont {Lehnert}}]{Teufel2009}%
  \BibitemOpen
  \bibfield  {author} {\bibinfo {author} {\bibfnamefont {J.~D.}\ \bibnamefont
  {Teufel}}, \bibinfo {author} {\bibfnamefont {T.}~\bibnamefont {Donner}},
  \bibinfo {author} {\bibfnamefont {M.~A.}\ \bibnamefont
  {Castellanos-Beltran}}, \bibinfo {author} {\bibfnamefont {J.}~\bibnamefont
  {Harlow}},\ and\ \bibinfo {author} {\bibfnamefont {K.}~\bibnamefont
  {Lehnert}},\ }\bibfield  {title} {\bibinfo {title} {Nanomechanical motion
  measured with an imprecision below that at the standard quantum limit.},\
  }\href {https://doi.org/10.1038/nnano.2009.343} {\bibfield  {journal}
  {\bibinfo  {journal} {Nat. Nanotechnol.}\ }\textbf {\bibinfo {volume} {4}},\
  \bibinfo {pages} {820} (\bibinfo {year} {2009})}\BibitemShut {NoStop}%
\bibitem [{\citenamefont {Anetsberger}\ \emph {et~al.}(2010)\citenamefont
  {Anetsberger}, \citenamefont {Gavartin}, \citenamefont {Arcizet},
  \citenamefont {Unterreithmeier}, \citenamefont {Weig}, \citenamefont
  {Gorodetsky}, \citenamefont {Kotthaus},\ and\ \citenamefont
  {Kippenberg}}]{Anetsberger2010}%
  \BibitemOpen
  \bibfield  {author} {\bibinfo {author} {\bibfnamefont {G.}~\bibnamefont
  {Anetsberger}}, \bibinfo {author} {\bibfnamefont {E.}~\bibnamefont
  {Gavartin}}, \bibinfo {author} {\bibfnamefont {O.}~\bibnamefont {Arcizet}},
  \bibinfo {author} {\bibfnamefont {Q.~P.}\ \bibnamefont {Unterreithmeier}},
  \bibinfo {author} {\bibfnamefont {E.~M.}\ \bibnamefont {Weig}}, \bibinfo
  {author} {\bibfnamefont {M.~L.}\ \bibnamefont {Gorodetsky}}, \bibinfo
  {author} {\bibfnamefont {J.~P.}\ \bibnamefont {Kotthaus}},\ and\ \bibinfo
  {author} {\bibfnamefont {T.~J.}\ \bibnamefont {Kippenberg}},\ }\bibfield
  {title} {\bibinfo {title} {Measuring nanomechanical motion with an
  imprecision below the standard quantum limit},\ }\href
  {https://doi.org/10.1103/PhysRevA.82.061804} {\bibfield  {journal} {\bibinfo
  {journal} {Phys. Rev. A}\ }\textbf {\bibinfo {volume} {82}},\ \bibinfo
  {pages} {061804} (\bibinfo {year} {2010})}\BibitemShut {NoStop}%
\bibitem [{\citenamefont {Teufel}\ \emph {et~al.}(2011)\citenamefont {Teufel},
  \citenamefont {Donner}, \citenamefont {Li}, \citenamefont {Harlow},
  \citenamefont {Allman}, \citenamefont {Cicak}, \citenamefont {Sirois},
  \citenamefont {Whittaker}, \citenamefont {Lehnert},\ and\ \citenamefont
  {Simmonds}}]{Teufel2011}%
  \BibitemOpen
  \bibfield  {author} {\bibinfo {author} {\bibfnamefont {J.}~\bibnamefont
  {Teufel}}, \bibinfo {author} {\bibfnamefont {T.}~\bibnamefont {Donner}},
  \bibinfo {author} {\bibfnamefont {D.}~\bibnamefont {Li}}, \bibinfo {author}
  {\bibfnamefont {J.~W.}\ \bibnamefont {Harlow}}, \bibinfo {author}
  {\bibfnamefont {M.~S.}\ \bibnamefont {Allman}}, \bibinfo {author}
  {\bibfnamefont {K.}~\bibnamefont {Cicak}}, \bibinfo {author} {\bibfnamefont
  {A.~J.}\ \bibnamefont {Sirois}}, \bibinfo {author} {\bibfnamefont {J.~D.}\
  \bibnamefont {Whittaker}}, \bibinfo {author} {\bibfnamefont {K.~W.}\
  \bibnamefont {Lehnert}},\ and\ \bibinfo {author} {\bibfnamefont {R.~W.}\
  \bibnamefont {Simmonds}},\ }\bibfield  {title} {\bibinfo {title} {Sideband
  cooling of micromechanical motion to the quantum ground state},\ }\href
  {https://doi.org/10.1038/nature10261} {\bibfield  {journal} {\bibinfo
  {journal} {Nature}\ }\textbf {\bibinfo {volume} {475}},\ \bibinfo {pages}
  {359} (\bibinfo {year} {2011})}\BibitemShut {NoStop}%
\bibitem [{\citenamefont {Chan}\ \emph {et~al.}(2011)\citenamefont {Chan},
  \citenamefont {Alegre}, \citenamefont {Safavi-Naeini}, \citenamefont {Hill},
  \citenamefont {Krause}, \citenamefont {Gr{\"o}blacher}, \citenamefont
  {Aspelmeyer},\ and\ \citenamefont {Painter}}]{Chan2011}%
  \BibitemOpen
  \bibfield  {author} {\bibinfo {author} {\bibfnamefont {J.}~\bibnamefont
  {Chan}}, \bibinfo {author} {\bibfnamefont {T.~P.~M.}\ \bibnamefont {Alegre}},
  \bibinfo {author} {\bibfnamefont {A.~H.}\ \bibnamefont {Safavi-Naeini}},
  \bibinfo {author} {\bibfnamefont {J.}~\bibnamefont {Hill}}, \bibinfo {author}
  {\bibfnamefont {A.}~\bibnamefont {Krause}}, \bibinfo {author} {\bibfnamefont
  {S.}~\bibnamefont {Gr{\"o}blacher}}, \bibinfo {author} {\bibfnamefont
  {M.}~\bibnamefont {Aspelmeyer}},\ and\ \bibinfo {author} {\bibfnamefont
  {O.}~\bibnamefont {Painter}},\ }\bibfield  {title} {\bibinfo {title} {Laser
  cooling of a nanomechanical oscillator into its quantum ground state},\
  }\bibfield  {journal} {\bibinfo  {journal} {Nature}\ }\textbf {\bibinfo
  {volume} {478}},\ \href {https://doi.org/10.1038/nature10461}
  {10.1038/nature10461} (\bibinfo {year} {2011})\BibitemShut {NoStop}%
\bibitem [{\citenamefont {Palomaki}\ \emph
  {et~al.}(2013{\natexlab{a}})\citenamefont {Palomaki}, \citenamefont {Harlow},
  \citenamefont {Teufel}, \citenamefont {Simmonds},\ and\ \citenamefont
  {Lehnert}}]{Palomaki2013a}%
  \BibitemOpen
  \bibfield  {author} {\bibinfo {author} {\bibfnamefont {T.}~\bibnamefont
  {Palomaki}}, \bibinfo {author} {\bibfnamefont {J.}~\bibnamefont {Harlow}},
  \bibinfo {author} {\bibfnamefont {J.}~\bibnamefont {Teufel}}, \bibinfo
  {author} {\bibfnamefont {R.}~\bibnamefont {Simmonds}},\ and\ \bibinfo
  {author} {\bibfnamefont {K.}~\bibnamefont {Lehnert}},\ }\bibfield  {title}
  {\bibinfo {title} {Coherent state transfer between itinerant microwave fields
  and a mechanical oscillator},\ }\href {https://doi.org/10.1038/nature11915}
  {\bibfield  {journal} {\bibinfo  {journal} {Nature}\ }\textbf {\bibinfo
  {volume} {495}},\ \bibinfo {pages} {210} (\bibinfo {year}
  {2013}{\natexlab{a}})}\BibitemShut {NoStop}%
\bibitem [{\citenamefont {Reed}\ \emph {et~al.}(2017)\citenamefont {Reed},
  \citenamefont {Mayer}, \citenamefont {Teufel}, \citenamefont {Burkhart},
  \citenamefont {Pfaff}, \citenamefont {Reagor}, \citenamefont {Sletten},
  \citenamefont {Ma}, \citenamefont {Schoelkopf}, \citenamefont {Knill},\ and\
  \citenamefont {Lehnert}}]{Reed2017}%
  \BibitemOpen
  \bibfield  {author} {\bibinfo {author} {\bibfnamefont {A.}~\bibnamefont
  {Reed}}, \bibinfo {author} {\bibfnamefont {K.}~\bibnamefont {Mayer}},
  \bibinfo {author} {\bibfnamefont {J.}~\bibnamefont {Teufel}}, \bibinfo
  {author} {\bibfnamefont {L.}~\bibnamefont {Burkhart}}, \bibinfo {author}
  {\bibfnamefont {W.}~\bibnamefont {Pfaff}}, \bibinfo {author} {\bibfnamefont
  {M.}~\bibnamefont {Reagor}}, \bibinfo {author} {\bibfnamefont
  {L.}~\bibnamefont {Sletten}}, \bibinfo {author} {\bibfnamefont
  {X.}~\bibnamefont {Ma}}, \bibinfo {author} {\bibfnamefont {R.}~\bibnamefont
  {Schoelkopf}}, \bibinfo {author} {\bibfnamefont {E.}~\bibnamefont {Knill}},\
  and\ \bibinfo {author} {\bibfnamefont {K.}~\bibnamefont {Lehnert}},\
  }\bibfield  {title} {\bibinfo {title} {Faithful conversation of propagating
  quantum information to mechanical motion},\ }\href
  {https://doi.org/10.1038/nphys4251} {\bibfield  {journal} {\bibinfo
  {journal} {Nat. Phys}\ }\textbf {\bibinfo {volume} {13}},\ \bibinfo {pages}
  {1163} (\bibinfo {year} {2017})}\BibitemShut {NoStop}%
\bibitem [{\citenamefont {Palomaki}\ \emph
  {et~al.}(2013{\natexlab{b}})\citenamefont {Palomaki}, \citenamefont {Teufel},
  \citenamefont {Simmonds},\ and\ \citenamefont {Lehnert}}]{Palomaki2013b}%
  \BibitemOpen
  \bibfield  {author} {\bibinfo {author} {\bibfnamefont {T.}~\bibnamefont
  {Palomaki}}, \bibinfo {author} {\bibfnamefont {J.}~\bibnamefont {Teufel}},
  \bibinfo {author} {\bibfnamefont {R.}~\bibnamefont {Simmonds}},\ and\
  \bibinfo {author} {\bibfnamefont {K.}~\bibnamefont {Lehnert}},\ }\bibfield
  {title} {\bibinfo {title} {Entangling mechanical motion with microwave
  fields},\ }\href {https://doi.org/10.1126/science.1244563} {\bibfield
  {journal} {\bibinfo  {journal} {Science}\ }\textbf {\bibinfo {volume}
  {342}},\ \bibinfo {pages} {710} (\bibinfo {year}
  {2013}{\natexlab{b}})}\BibitemShut {NoStop}%
\bibitem [{\citenamefont {Riedinger}\ \emph {et~al.}(2018)\citenamefont
  {Riedinger}, \citenamefont {Wallucks}, \citenamefont {Marinkovi{\'c}},
  \citenamefont {L{\"o}schnauer}, \citenamefont {Aspelmeyer}, \citenamefont
  {Hong},\ and\ \citenamefont {Gr{\"o}blacher}}]{Riedinger2018}%
  \BibitemOpen
  \bibfield  {author} {\bibinfo {author} {\bibfnamefont {R.}~\bibnamefont
  {Riedinger}}, \bibinfo {author} {\bibfnamefont {A.}~\bibnamefont {Wallucks}},
  \bibinfo {author} {\bibfnamefont {I.}~\bibnamefont {Marinkovi{\'c}}},
  \bibinfo {author} {\bibfnamefont {C.}~\bibnamefont {L{\"o}schnauer}},
  \bibinfo {author} {\bibfnamefont {M.}~\bibnamefont {Aspelmeyer}}, \bibinfo
  {author} {\bibfnamefont {S.}~\bibnamefont {Hong}},\ and\ \bibinfo {author}
  {\bibfnamefont {S.}~\bibnamefont {Gr{\"o}blacher}},\ }\bibfield  {title}
  {\bibinfo {title} {Remote quantum entanglement between two micromechanical
  oscillators},\ }\href {https://doi.org/10.1038/s41586-018-0036-z} {\bibfield
  {journal} {\bibinfo  {journal} {Nature}\ }\textbf {\bibinfo {volume} {556}},\
  \bibinfo {pages} {473} (\bibinfo {year} {2018})}\BibitemShut {NoStop}%
\bibitem [{\citenamefont {Suh}\ \emph {et~al.}(2014)\citenamefont {Suh},
  \citenamefont {Weinstein}, \citenamefont {Lei}, \citenamefont {Wollman},
  \citenamefont {Steinke}, \citenamefont {Meystre3}, \citenamefont {Clerk4},\
  and\ \citenamefont {Schwab}}]{Suh2014}%
  \BibitemOpen
  \bibfield  {author} {\bibinfo {author} {\bibfnamefont {J.}~\bibnamefont
  {Suh}}, \bibinfo {author} {\bibfnamefont {A.~J.}\ \bibnamefont {Weinstein}},
  \bibinfo {author} {\bibfnamefont {C.~U.}\ \bibnamefont {Lei}}, \bibinfo
  {author} {\bibfnamefont {E.~E.}\ \bibnamefont {Wollman}}, \bibinfo {author}
  {\bibfnamefont {S.~K.}\ \bibnamefont {Steinke}}, \bibinfo {author}
  {\bibfnamefont {P.}~\bibnamefont {Meystre3}}, \bibinfo {author}
  {\bibfnamefont {A.~A.}\ \bibnamefont {Clerk4}},\ and\ \bibinfo {author}
  {\bibfnamefont {K.~C.}\ \bibnamefont {Schwab}},\ }\bibfield  {title}
  {\bibinfo {title} {Mechanically detecting and avoiding the quantum
  fluctuations of a microwave field},\ }\href
  {https://doi.org/10.1126/science.1253258} {\bibfield  {journal} {\bibinfo
  {journal} {Science}\ }\textbf {\bibinfo {volume} {344}},\ \bibinfo {pages}
  {1262} (\bibinfo {year} {2014})}\BibitemShut {NoStop}%
\bibitem [{\citenamefont {Wollman}\ \emph {et~al.}(2015)\citenamefont
  {Wollman}, \citenamefont {Lei}, \citenamefont {Weinstein}, \citenamefont
  {Suh}, \citenamefont {Kronwald}, \citenamefont {Marquardt}, \citenamefont
  {Clerk},\ and\ \citenamefont {Schwab}}]{Wollman2015}%
  \BibitemOpen
  \bibfield  {author} {\bibinfo {author} {\bibfnamefont {E.}~\bibnamefont
  {Wollman}}, \bibinfo {author} {\bibfnamefont {C.}~\bibnamefont {Lei}},
  \bibinfo {author} {\bibfnamefont {A.}~\bibnamefont {Weinstein}}, \bibinfo
  {author} {\bibfnamefont {J.}~\bibnamefont {Suh}}, \bibinfo {author}
  {\bibfnamefont {A.}~\bibnamefont {Kronwald}}, \bibinfo {author}
  {\bibfnamefont {F.}~\bibnamefont {Marquardt}}, \bibinfo {author}
  {\bibfnamefont {A.}~\bibnamefont {Clerk}},\ and\ \bibinfo {author}
  {\bibfnamefont {K.}~\bibnamefont {Schwab}},\ }\bibfield  {title} {\bibinfo
  {title} {Quantum squeezing of motion in a mechanical resonator},\ }\href
  {https://doi.org/10.1126/science.aac5138} {\bibfield  {journal} {\bibinfo
  {journal} {Science}\ }\textbf {\bibinfo {volume} {349}},\ \bibinfo {pages}
  {952} (\bibinfo {year} {2015})}\BibitemShut {NoStop}%
\bibitem [{\citenamefont {Pirkkalainen}\ \emph {et~al.}(2015)\citenamefont
  {Pirkkalainen}, \citenamefont {Damsk{\"a}gg}, \citenamefont {Brandt},
  \citenamefont {Massel},\ and\ \citenamefont
  {Sillanp{\"a}{\"a}}}]{Pirkkalainen2015}%
  \BibitemOpen
  \bibfield  {author} {\bibinfo {author} {\bibfnamefont {J.-M.}\ \bibnamefont
  {Pirkkalainen}}, \bibinfo {author} {\bibfnamefont {E.}~\bibnamefont
  {Damsk{\"a}gg}}, \bibinfo {author} {\bibfnamefont {M.}~\bibnamefont
  {Brandt}}, \bibinfo {author} {\bibfnamefont {F.}~\bibnamefont {Massel}},\
  and\ \bibinfo {author} {\bibfnamefont {M.}~\bibnamefont
  {Sillanp{\"a}{\"a}}},\ }\bibfield  {title} {\bibinfo {title} {Squeezing of
  quantum noise of motion in a micromechanical resonator},\ }\href
  {https://doi.org/10.1103/PhysRevLett.115.243601} {\bibfield  {journal}
  {\bibinfo  {journal} {Phys. Rev. Lett.}\ }\textbf {\bibinfo {volume} {115}},\
  \bibinfo {pages} {243601} (\bibinfo {year} {2015})}\BibitemShut {NoStop}%
\bibitem [{\citenamefont {Lecocq}\ \emph {et~al.}(2015)\citenamefont {Lecocq},
  \citenamefont {Clark}, \citenamefont {Simmonds}, \citenamefont {Aumentado},\
  and\ \citenamefont {Teufel}}]{Lecoq2015}%
  \BibitemOpen
  \bibfield  {author} {\bibinfo {author} {\bibfnamefont {F.}~\bibnamefont
  {Lecocq}}, \bibinfo {author} {\bibfnamefont {J.}~\bibnamefont {Clark}},
  \bibinfo {author} {\bibfnamefont {R.}~\bibnamefont {Simmonds}}, \bibinfo
  {author} {\bibfnamefont {J.}~\bibnamefont {Aumentado}},\ and\ \bibinfo
  {author} {\bibfnamefont {J.}~\bibnamefont {Teufel}},\ }\bibfield  {title}
  {\bibinfo {title} {Quantum nondemolition measurement of a nonclassical state
  of a massive object},\ }\href {https://doi.org/10.1103/PhysRevX.5.041037}
  {\bibfield  {journal} {\bibinfo  {journal} {Phys. Rev. X}\ }\textbf {\bibinfo
  {volume} {5}},\ \bibinfo {pages} {041037} (\bibinfo {year}
  {2015})}\BibitemShut {NoStop}%
\bibitem [{\citenamefont {De~Lorenzo}\ and\ \citenamefont
  {Schwab}(2014)}]{DeLorenzo2014}%
  \BibitemOpen
  \bibfield  {author} {\bibinfo {author} {\bibfnamefont {L.~A.}\ \bibnamefont
  {De~Lorenzo}}\ and\ \bibinfo {author} {\bibfnamefont {K.~C.}\ \bibnamefont
  {Schwab}},\ }\bibfield  {title} {\bibinfo {title} {Superfluid optomechanics:
  coupling of a superfluid to a superconducting condensate},\ }\href
  {https://doi.org/10.1088/1367-2630/16/11/113020} {\bibfield  {journal}
  {\bibinfo  {journal} {New J. Phys.}\ }\textbf {\bibinfo {volume} {16}},\
  \bibinfo {pages} {113020} (\bibinfo {year} {2014})}\BibitemShut {NoStop}%
\bibitem [{\citenamefont {Hartung}\ \emph {et~al.}(2006)\citenamefont
  {Hartung}, \citenamefont {Bierwagen}, \citenamefont {Bricker}, \citenamefont
  {Compton}, \citenamefont {Grimm}, \citenamefont {Johnson}, \citenamefont
  {Meidlinger}, \citenamefont {Pendell}, \citenamefont {Popielarski},
  \citenamefont {Saxton},\ and\ \citenamefont {York}}]{Hartung2006}%
  \BibitemOpen
  \bibfield  {author} {\bibinfo {author} {\bibfnamefont {W.}~\bibnamefont
  {Hartung}}, \bibinfo {author} {\bibfnamefont {J.}~\bibnamefont {Bierwagen}},
  \bibinfo {author} {\bibfnamefont {S.}~\bibnamefont {Bricker}}, \bibinfo
  {author} {\bibfnamefont {C.}~\bibnamefont {Compton}}, \bibinfo {author}
  {\bibfnamefont {T.}~\bibnamefont {Grimm}}, \bibinfo {author} {\bibfnamefont
  {M.}~\bibnamefont {Johnson}}, \bibinfo {author} {\bibfnamefont
  {D.}~\bibnamefont {Meidlinger}}, \bibinfo {author} {\bibfnamefont
  {D.}~\bibnamefont {Pendell}}, \bibinfo {author} {\bibfnamefont
  {J.}~\bibnamefont {Popielarski}}, \bibinfo {author} {\bibfnamefont
  {L.}~\bibnamefont {Saxton}},\ and\ \bibinfo {author} {\bibfnamefont {R.~C.}\
  \bibnamefont {York}},\ }\bibfield  {title} {\bibinfo {title} {Rf performance
  of a superconducting s-band cavityfilled with liquid helium},\ }\href@noop {}
  {\bibfield  {journal} {\bibinfo  {journal} {Proc. of LINAC}\ ,\ \bibinfo
  {pages} {755}} (\bibinfo {year} {2006})}\BibitemShut {NoStop}%
\bibitem [{\citenamefont {De~Lorenzo}\ and\ \citenamefont
  {Schwab}(2017)}]{DeLorenzo2017}%
  \BibitemOpen
  \bibfield  {author} {\bibinfo {author} {\bibfnamefont {L.~A.}\ \bibnamefont
  {De~Lorenzo}}\ and\ \bibinfo {author} {\bibfnamefont {K.~C.}\ \bibnamefont
  {Schwab}},\ }\bibfield  {title} {\bibinfo {title} {Ultra-high q acoustic
  resonance in superfluid $^{4}\mathrm{He}$},\ }\href
  {https://doi.org/10.1007/s10909-016-1674-x} {\bibfield  {journal} {\bibinfo
  {journal} {J. Low Temp. Phys}\ }\textbf {\bibinfo {volume} {186}},\ \bibinfo
  {pages} {233} (\bibinfo {year} {2017})}\BibitemShut {NoStop}%
\bibitem [{\citenamefont {Singh}\ \emph {et~al.}(2017)\citenamefont {Singh},
  \citenamefont {Lorenzo}, \citenamefont {Pikovski},\ and\ \citenamefont
  {Schwab}}]{Singh2017}%
  \BibitemOpen
  \bibfield  {author} {\bibinfo {author} {\bibfnamefont {S.}~\bibnamefont
  {Singh}}, \bibinfo {author} {\bibfnamefont {L.~A.~D.}\ \bibnamefont
  {Lorenzo}}, \bibinfo {author} {\bibfnamefont {I.}~\bibnamefont {Pikovski}},\
  and\ \bibinfo {author} {\bibfnamefont {K.}~\bibnamefont {Schwab}},\
  }\bibfield  {title} {\bibinfo {title} {Detecting continuous gravitational
  waves with superfluid 4he},\ }\href
  {https://doi.org/10.1088/1367-2630/aa78cb} {\bibfield  {journal} {\bibinfo
  {journal} {New J. Phys.}\ }\textbf {\bibinfo {volume} {19}},\ \bibinfo
  {pages} {073023} (\bibinfo {year} {2017})}\BibitemShut {NoStop}%
\bibitem [{\citenamefont {Kashkanova}\ \emph {et~al.}(2016)\citenamefont
  {Kashkanova}, \citenamefont {Shkarin}, \citenamefont {Brown}, \citenamefont
  {Flowers-Jacobs}, \citenamefont {Childress},\ and\ \citenamefont
  {Hoch}}]{Kashkanova2016}%
  \BibitemOpen
  \bibfield  {author} {\bibinfo {author} {\bibfnamefont {A.}~\bibnamefont
  {Kashkanova}}, \bibinfo {author} {\bibfnamefont {A.}~\bibnamefont {Shkarin}},
  \bibinfo {author} {\bibfnamefont {C.}~\bibnamefont {Brown}}, \bibinfo
  {author} {\bibfnamefont {N.}~\bibnamefont {Flowers-Jacobs}}, \bibinfo
  {author} {\bibfnamefont {L.}~\bibnamefont {Childress}},\ and\ \bibinfo
  {author} {\bibfnamefont {S.}~\bibnamefont {Hoch}},\ }\bibfield  {title}
  {\bibinfo {title} {Superfluid brillouin optomechanics},\ }\href
  {https://doi.org/10.1038/nphys3900} {\bibfield  {journal} {\bibinfo
  {journal} {Nat. Phys}\ }\textbf {\bibinfo {volume} {13}},\ \bibinfo {pages}
  {74} (\bibinfo {year} {2016})}\BibitemShut {NoStop}%
\bibitem [{\citenamefont {Shkarin}\ \emph {et~al.}(2019)\citenamefont
  {Shkarin}, \citenamefont {Kashkanova}, \citenamefont {Brown}, \citenamefont
  {Garcia}, \citenamefont {Ott}, \citenamefont {Reichel},\ and\ \citenamefont
  {Harris}}]{Shkarin2019}%
  \BibitemOpen
  \bibfield  {author} {\bibinfo {author} {\bibfnamefont {A.~B.}\ \bibnamefont
  {Shkarin}}, \bibinfo {author} {\bibfnamefont {A.~D.}\ \bibnamefont
  {Kashkanova}}, \bibinfo {author} {\bibfnamefont {C.~D.}\ \bibnamefont
  {Brown}}, \bibinfo {author} {\bibfnamefont {S.}~\bibnamefont {Garcia}},
  \bibinfo {author} {\bibfnamefont {K.}~\bibnamefont {Ott}}, \bibinfo {author}
  {\bibfnamefont {J.}~\bibnamefont {Reichel}},\ and\ \bibinfo {author}
  {\bibfnamefont {J.~G.~E.}\ \bibnamefont {Harris}},\ }\bibfield  {title}
  {\bibinfo {title} {Quantum optomechanics in a liquid},\ }\href
  {https://doi.org/10.1103/PhysRevLett.122.153601} {\bibfield  {journal}
  {\bibinfo  {journal} {Phys. Rev. Lett.}\ }\textbf {\bibinfo {volume} {122}},\
  \bibinfo {pages} {153601} (\bibinfo {year} {2019})}\BibitemShut {NoStop}%
\bibitem [{\citenamefont {Forstner}\ \emph {et~al.}(2019)\citenamefont
  {Forstner}, \citenamefont {Sachkou}, \citenamefont {Woolley}, \citenamefont
  {Harris},\ and\ \citenamefont {Baker}}]{Forstner2019}%
  \BibitemOpen
  \bibfield  {author} {\bibinfo {author} {\bibfnamefont {S.}~\bibnamefont
  {Forstner}}, \bibinfo {author} {\bibfnamefont {Y.}~\bibnamefont {Sachkou}},
  \bibinfo {author} {\bibfnamefont {M.}~\bibnamefont {Woolley}}, \bibinfo
  {author} {\bibfnamefont {G.}~\bibnamefont {Harris}},\ and\ \bibinfo {author}
  {\bibfnamefont {C.}~\bibnamefont {Baker}},\ }\bibfield  {title} {\bibinfo
  {title} {Modelling of vorticity, sound and their interaction in recent
  citations- coherent vortex dynamics in a stronglytwo-dimensional
  superfluids},\ }\href {https://doi.org/10.1088/1367-2630/ab1bb5} {\bibfield
  {journal} {\bibinfo  {journal} {New J. Phys.}\ }\textbf {\bibinfo {volume}
  {21}},\ \bibinfo {pages} {053029} (\bibinfo {year} {2019})}\BibitemShut
  {NoStop}%
\bibitem [{\citenamefont {Sachkou}\ \emph {et~al.}(2019)\citenamefont
  {Sachkou}, \citenamefont {Baker}, \citenamefont {Harris}, \citenamefont
  {Stockdale}, \citenamefont {Forstner}, \citenamefont {Reeves}, \citenamefont
  {He}, \citenamefont {McAuslan}, \citenamefont {Bradley}, \citenamefont
  {Davis},\ and\ \citenamefont {Bowen}}]{Sachkou2019}%
  \BibitemOpen
  \bibfield  {author} {\bibinfo {author} {\bibfnamefont {Y.}~\bibnamefont
  {Sachkou}}, \bibinfo {author} {\bibfnamefont {C.}~\bibnamefont {Baker}},
  \bibinfo {author} {\bibfnamefont {G.}~\bibnamefont {Harris}}, \bibinfo
  {author} {\bibfnamefont {O.}~\bibnamefont {Stockdale}}, \bibinfo {author}
  {\bibfnamefont {S.}~\bibnamefont {Forstner}}, \bibinfo {author}
  {\bibfnamefont {M.}~\bibnamefont {Reeves}}, \bibinfo {author} {\bibfnamefont
  {X.}~\bibnamefont {He}}, \bibinfo {author} {\bibfnamefont {D.~L.}\
  \bibnamefont {McAuslan}}, \bibinfo {author} {\bibfnamefont {A.}~\bibnamefont
  {Bradley}}, \bibinfo {author} {\bibfnamefont {M.}~\bibnamefont {Davis}},\
  and\ \bibinfo {author} {\bibfnamefont {W.}~\bibnamefont {Bowen}},\ }\bibfield
   {title} {\bibinfo {title} {Coherent vortex dynamics in a strongly
  interacting superfluid on a silicon chip},\ }\href
  {https://doi.org/10.1126/science.aaw9229} {\bibfield  {journal} {\bibinfo
  {journal} {Science}\ }\textbf {\bibinfo {volume} {366}},\ \bibinfo {pages}
  {1480} (\bibinfo {year} {2019})}\BibitemShut {NoStop}%
\bibitem [{\citenamefont {Childress}\ \emph {et~al.}(2017)\citenamefont
  {Childress}, \citenamefont {Schmidt}, \citenamefont {Kashkanova},
  \citenamefont {Brown}, \citenamefont {Harris}, \citenamefont {Aiello},
  \citenamefont {Marquardt},\ and\ \citenamefont {Harris}}]{Childress2017}%
  \BibitemOpen
  \bibfield  {author} {\bibinfo {author} {\bibfnamefont {L.}~\bibnamefont
  {Childress}}, \bibinfo {author} {\bibfnamefont {M.~P.}\ \bibnamefont
  {Schmidt}}, \bibinfo {author} {\bibfnamefont {A.~D.}\ \bibnamefont
  {Kashkanova}}, \bibinfo {author} {\bibfnamefont {C.~D.}\ \bibnamefont
  {Brown}}, \bibinfo {author} {\bibfnamefont {G.~I.}\ \bibnamefont {Harris}},
  \bibinfo {author} {\bibfnamefont {A.}~\bibnamefont {Aiello}}, \bibinfo
  {author} {\bibfnamefont {F.}~\bibnamefont {Marquardt}},\ and\ \bibinfo
  {author} {\bibfnamefont {J.}~\bibnamefont {Harris}},\ }\bibfield  {title}
  {\bibinfo {title} {Cavity optomechanics in a levitated helium drop},\ }\href
  {https://doi.org/10.1103/PhysRevA.96.063842} {\bibfield  {journal} {\bibinfo
  {journal} {Phys. Rev. A}\ }\textbf {\bibinfo {volume} {96}},\ \bibinfo
  {pages} {063842} (\bibinfo {year} {2017})}\BibitemShut {NoStop}%
\bibitem [{\citenamefont {He}\ \emph {et~al.}(2020)\citenamefont {He},
  \citenamefont {Harris}, \citenamefont {Baker}, \citenamefont {Sawadsky},
  \citenamefont {Sfendla}, \citenamefont {Sachkou}, \citenamefont {Forstner},\
  and\ \citenamefont {Bowen}}]{He2020}%
  \BibitemOpen
  \bibfield  {author} {\bibinfo {author} {\bibfnamefont {X.}~\bibnamefont
  {He}}, \bibinfo {author} {\bibfnamefont {G.~I.}\ \bibnamefont {Harris}},
  \bibinfo {author} {\bibfnamefont {C.}~\bibnamefont {Baker}}, \bibinfo
  {author} {\bibfnamefont {A.}~\bibnamefont {Sawadsky}}, \bibinfo {author}
  {\bibfnamefont {Y.}~\bibnamefont {Sfendla}}, \bibinfo {author} {\bibfnamefont
  {Y.}~\bibnamefont {Sachkou}}, \bibinfo {author} {\bibfnamefont
  {S.}~\bibnamefont {Forstner}},\ and\ \bibinfo {author} {\bibfnamefont
  {W.~P.}\ \bibnamefont {Bowen}},\ }\bibfield  {title} {\bibinfo {title}
  {Strong optical coupling through superfluid brillouin lasing},\ }\href
  {https://doi.org/10.1038/s41567-020-0785-0} {\bibfield  {journal} {\bibinfo
  {journal} {Nat. Phys}\ }\textbf {\bibinfo {volume} {16}},\ \bibinfo {pages}
  {417} (\bibinfo {year} {2020})}\BibitemShut {NoStop}%
\bibitem [{\citenamefont {Harris}\ \emph {et~al.}(2020)\citenamefont {Harris},
  \citenamefont {Sawadsky}, \citenamefont {Sfendla}, \citenamefont {Wasserman},
  \citenamefont {Bowen},\ and\ \citenamefont {Baker}}]{Harris2020}%
  \BibitemOpen
  \bibfield  {author} {\bibinfo {author} {\bibfnamefont {G.~I.}\ \bibnamefont
  {Harris}}, \bibinfo {author} {\bibfnamefont {A.}~\bibnamefont {Sawadsky}},
  \bibinfo {author} {\bibfnamefont {Y.}~\bibnamefont {Sfendla}}, \bibinfo
  {author} {\bibfnamefont {W.}~\bibnamefont {Wasserman}}, \bibinfo {author}
  {\bibfnamefont {W.~P.}\ \bibnamefont {Bowen}},\ and\ \bibinfo {author}
  {\bibfnamefont {C.~G.}\ \bibnamefont {Baker}},\ }\bibfield  {title} {\bibinfo
  {title} {Proposal for a quantum traveling brillouin resonator},\ }\href
  {https://arxiv.org/abs/2006.04405} {\bibfield  {journal} {\bibinfo  {journal}
  {arxiv:2006.04405}\ } (\bibinfo {year} {2020})}\BibitemShut {NoStop}%
\bibitem [{\citenamefont {Sfendla}\ \emph {et~al.}(2020)\citenamefont
  {Sfendla}, \citenamefont {Baker}, \citenamefont {Harris}, \citenamefont
  {Tian},\ and\ \citenamefont {Bowen}}]{Sfendla2020}%
  \BibitemOpen
  \bibfield  {author} {\bibinfo {author} {\bibfnamefont {Y.}~\bibnamefont
  {Sfendla}}, \bibinfo {author} {\bibfnamefont {C.}~\bibnamefont {Baker}},
  \bibinfo {author} {\bibfnamefont {G.}~\bibnamefont {Harris}}, \bibinfo
  {author} {\bibfnamefont {L.}~\bibnamefont {Tian}},\ and\ \bibinfo {author}
  {\bibfnamefont {W.~P.}\ \bibnamefont {Bowen}},\ }\bibfield  {title} {\bibinfo
  {title} {Extreme quantum nonlinearity in superfluid thin-film surface
  waves},\ }\href {https://arxiv.org/abs/2005.13919} {\bibfield  {journal}
  {\bibinfo  {journal} {arxiv:2005.13919}\ } (\bibinfo {year}
  {2020})}\BibitemShut {NoStop}%
\bibitem [{\citenamefont {Gasparini}\ \emph {et~al.}(2008)\citenamefont
  {Gasparini}, \citenamefont {Kimball}, \citenamefont {Mooney},\ and\
  \citenamefont {Diaz-Avila}}]{Gasparini2008}%
  \BibitemOpen
  \bibfield  {author} {\bibinfo {author} {\bibfnamefont {F.~M.}\ \bibnamefont
  {Gasparini}}, \bibinfo {author} {\bibfnamefont {M.~O.}\ \bibnamefont
  {Kimball}}, \bibinfo {author} {\bibfnamefont {K.~P.}\ \bibnamefont
  {Mooney}},\ and\ \bibinfo {author} {\bibfnamefont {M.}~\bibnamefont
  {Diaz-Avila}},\ }\bibfield  {title} {\bibinfo {title} {Finite-size scaling of
  4he at the superfluid transition},\ }\href
  {https://doi.org/10.1103/RevModPhys.80.1009} {\bibfield  {journal} {\bibinfo
  {journal} {Rev. Mod. Phys}\ }\textbf {\bibinfo {volume} {80}},\ \bibinfo
  {pages} {1009} (\bibinfo {year} {2008})}\BibitemShut {NoStop}%
\bibitem [{\citenamefont {Duh}\ \emph {et~al.}(2012)\citenamefont {Duh},
  \citenamefont {Suhel}, \citenamefont {Hauer}, \citenamefont {Saeedi},
  \citenamefont {Kim}, \citenamefont {Biswas},\ and\ \citenamefont
  {Davis}}]{Duh2012}%
  \BibitemOpen
  \bibfield  {author} {\bibinfo {author} {\bibfnamefont {A.}~\bibnamefont
  {Duh}}, \bibinfo {author} {\bibfnamefont {A.}~\bibnamefont {Suhel}}, \bibinfo
  {author} {\bibfnamefont {B.~D.}\ \bibnamefont {Hauer}}, \bibinfo {author}
  {\bibfnamefont {R.}~\bibnamefont {Saeedi}}, \bibinfo {author} {\bibfnamefont
  {P.~H.}\ \bibnamefont {Kim}}, \bibinfo {author} {\bibfnamefont {T.~S.}\
  \bibnamefont {Biswas}},\ and\ \bibinfo {author} {\bibfnamefont {J.~P.}\
  \bibnamefont {Davis}},\ }\bibfield  {title} {\bibinfo {title} {Microfluidic
  and nanofluidic cavities for quantum fluids experiments},\ }\href
  {https://doi.org/10.1007/s10909-012-0617-4} {\bibfield  {journal} {\bibinfo
  {journal} {J. Low Temp. Phys}\ }\textbf {\bibinfo {volume} {168}},\ \bibinfo
  {pages} {31} (\bibinfo {year} {2012})}\BibitemShut {NoStop}%
\bibitem [{\citenamefont {Levitin}\ \emph {et~al.}(2013)\citenamefont
  {Levitin}, \citenamefont {Bennett}, \citenamefont {Casey}, \citenamefont
  {Cowan}, \citenamefont {Saunders}, \citenamefont {Drung}, \citenamefont
  {Schurig},\ and\ \citenamefont {Parpia}}]{Levitin2013}%
  \BibitemOpen
  \bibfield  {author} {\bibinfo {author} {\bibfnamefont {L.~V.}\ \bibnamefont
  {Levitin}}, \bibinfo {author} {\bibfnamefont {R.~G.}\ \bibnamefont
  {Bennett}}, \bibinfo {author} {\bibfnamefont {A.}~\bibnamefont {Casey}},
  \bibinfo {author} {\bibfnamefont {B.}~\bibnamefont {Cowan}}, \bibinfo
  {author} {\bibfnamefont {J.}~\bibnamefont {Saunders}}, \bibinfo {author}
  {\bibfnamefont {D.}~\bibnamefont {Drung}}, \bibinfo {author} {\bibfnamefont
  {T.}~\bibnamefont {Schurig}},\ and\ \bibinfo {author} {\bibfnamefont {J.~M.}\
  \bibnamefont {Parpia}},\ }\bibfield  {title} {\bibinfo {title} {Phase diagram
  of the topological superfluid 3he confined in a nanoscale slab geometry},\
  }\href {https://doi.org/10.1126/science.1233621} {\bibfield  {journal}
  {\bibinfo  {journal} {Science}\ }\textbf {\bibinfo {volume} {340}},\ \bibinfo
  {pages} {6134} (\bibinfo {year} {2013})}\BibitemShut {NoStop}%
\bibitem [{\citenamefont {Rojas}\ \emph {et~al.}(2014)\citenamefont {Rojas},
  \citenamefont {Hauer}, \citenamefont {MacDonald}, \citenamefont {Saberi},
  \citenamefont {Yang},\ and\ \citenamefont {Davis}}]{Rojas2014}%
  \BibitemOpen
  \bibfield  {author} {\bibinfo {author} {\bibfnamefont {X.}~\bibnamefont
  {Rojas}}, \bibinfo {author} {\bibfnamefont {B.}~\bibnamefont {Hauer}},
  \bibinfo {author} {\bibfnamefont {A.}~\bibnamefont {MacDonald}}, \bibinfo
  {author} {\bibfnamefont {P.}~\bibnamefont {Saberi}}, \bibinfo {author}
  {\bibfnamefont {Y.}~\bibnamefont {Yang}},\ and\ \bibinfo {author}
  {\bibfnamefont {J.}~\bibnamefont {Davis}},\ }\bibfield  {title} {\bibinfo
  {title} {Ultrasonic interferometer for first-sound measurements of confined
  liquid 4he},\ }\href {https://doi.org/10.1103/PhysRevB.89.174508} {\bibfield
  {journal} {\bibinfo  {journal} {Phys. Rev. B}\ }\textbf {\bibinfo {volume}
  {89}},\ \bibinfo {pages} {174508} (\bibinfo {year} {2014})}\BibitemShut
  {NoStop}%
\bibitem [{\citenamefont {Rojas}\ and\ \citenamefont
  {Davis}(2015)}]{Rojas2015}%
  \BibitemOpen
  \bibfield  {author} {\bibinfo {author} {\bibfnamefont {X.}~\bibnamefont
  {Rojas}}\ and\ \bibinfo {author} {\bibfnamefont {J.}~\bibnamefont {Davis}},\
  }\bibfield  {title} {\bibinfo {title} {Superfluid nanomechanical resonator
  for quantum nanofluidics},\ }\href
  {https://doi.org/10.1103/PhysRevB.91.024503} {\bibfield  {journal} {\bibinfo
  {journal} {Phys. Rev. B}\ }\textbf {\bibinfo {volume} {91}},\ \bibinfo
  {pages} {024503} (\bibinfo {year} {2015})}\BibitemShut {NoStop}%
\bibitem [{\citenamefont {Souris}\ \emph {et~al.}(2017)\citenamefont {Souris},
  \citenamefont {Rojas}, \citenamefont {Kim},\ and\ \citenamefont
  {Davis}}]{Souris2017}%
  \BibitemOpen
  \bibfield  {author} {\bibinfo {author} {\bibfnamefont {F.}~\bibnamefont
  {Souris}}, \bibinfo {author} {\bibfnamefont {X.}~\bibnamefont {Rojas}},
  \bibinfo {author} {\bibfnamefont {P.}~\bibnamefont {Kim}},\ and\ \bibinfo
  {author} {\bibfnamefont {J.}~\bibnamefont {Davis}},\ }\bibfield  {title}
  {\bibinfo {title} {Ultralow-dissipation superfluid micromechanical
  resonator},\ }\href {https://doi.org/10.1103/PhysRevApplied.7.044008}
  {\bibfield  {journal} {\bibinfo  {journal} {Phys. Rev. Applied}\ }\textbf
  {\bibinfo {volume} {7}},\ \bibinfo {pages} {044008} (\bibinfo {year}
  {2017})}\BibitemShut {NoStop}%
\bibitem [{\citenamefont {Perron}\ \emph {et~al.}(2019)\citenamefont {Perron},
  \citenamefont {Kimball},\ and\ \citenamefont {Gasparini}}]{Perron2019}%
  \BibitemOpen
  \bibfield  {author} {\bibinfo {author} {\bibfnamefont {J.~K.}\ \bibnamefont
  {Perron}}, \bibinfo {author} {\bibfnamefont {M.}~\bibnamefont {Kimball}},\
  and\ \bibinfo {author} {\bibfnamefont {F.}~\bibnamefont {Gasparini}},\
  }\bibfield  {title} {\bibinfo {title} {A review of giant correlation-length
  effects via proximity and weak-links coupling in a critical system: 4he near
  the superfluid transition},\ }\href
  {https://doi.org/10.1088/1361-6633/ab3df5} {\bibfield  {journal} {\bibinfo
  {journal} {Rep. Prog. Phys.}\ }\textbf {\bibinfo {volume} {82}},\ \bibinfo
  {pages} {11} (\bibinfo {year} {2019})}\BibitemShut {NoStop}%
\bibitem [{\citenamefont {Shook}\ \emph {et~al.}(2020)\citenamefont {Shook},
  \citenamefont {Vadakkumbatt}, \citenamefont {Yapa}, \citenamefont {Doolin},
  \citenamefont {Boyack}, \citenamefont {Kim}, \citenamefont {Popowich},
  \citenamefont {Souris}, \citenamefont {Christani}, \citenamefont {Maciejko},\
  and\ \citenamefont {Davis}}]{Shook2020}%
  \BibitemOpen
  \bibfield  {author} {\bibinfo {author} {\bibfnamefont {A.~J.}\ \bibnamefont
  {Shook}}, \bibinfo {author} {\bibfnamefont {V.}~\bibnamefont {Vadakkumbatt}},
  \bibinfo {author} {\bibfnamefont {P.~S.}\ \bibnamefont {Yapa}}, \bibinfo
  {author} {\bibfnamefont {C.}~\bibnamefont {Doolin}}, \bibinfo {author}
  {\bibfnamefont {R.}~\bibnamefont {Boyack}}, \bibinfo {author} {\bibfnamefont
  {P.~H.}\ \bibnamefont {Kim}}, \bibinfo {author} {\bibfnamefont {G.~G.}\
  \bibnamefont {Popowich}}, \bibinfo {author} {\bibfnamefont {F.}~\bibnamefont
  {Souris}}, \bibinfo {author} {\bibfnamefont {H.}~\bibnamefont {Christani}},
  \bibinfo {author} {\bibfnamefont {J.}~\bibnamefont {Maciejko}},\ and\
  \bibinfo {author} {\bibfnamefont {J.~P.}\ \bibnamefont {Davis}},\ }\bibfield
  {title} {\bibinfo {title} {Stabilized pair density wave via nanoscale
  confinement of superfluid 3he},\ }\href
  {https://doi.org/10.1103/PhysRevLett.124.015301} {\bibfield  {journal}
  {\bibinfo  {journal} {Phys. Rev. Lett.}\ }\textbf {\bibinfo {volume} {124}},\
  \bibinfo {pages} {015301} (\bibinfo {year} {2020})}\BibitemShut {NoStop}%
\bibitem [{\citenamefont {Zhelev}\ \emph {et~al.}(0176)\citenamefont {Zhelev},
  \citenamefont {Abhilash}, \citenamefont {Smith}, \citenamefont {Bennett1},
  \citenamefont {Rojas}, \citenamefont {Levitin}, \citenamefont {Saunders},\
  and\ \citenamefont {Parpia}}]{Zhelev2017}%
  \BibitemOpen
  \bibfield  {author} {\bibinfo {author} {\bibfnamefont {N.}~\bibnamefont
  {Zhelev}}, \bibinfo {author} {\bibfnamefont {T.}~\bibnamefont {Abhilash}},
  \bibinfo {author} {\bibfnamefont {E.}~\bibnamefont {Smith}}, \bibinfo
  {author} {\bibfnamefont {R.}~\bibnamefont {Bennett1}}, \bibinfo {author}
  {\bibfnamefont {X.}~\bibnamefont {Rojas}}, \bibinfo {author} {\bibfnamefont
  {L.}~\bibnamefont {Levitin}}, \bibinfo {author} {\bibfnamefont
  {J.}~\bibnamefont {Saunders}},\ and\ \bibinfo {author} {\bibfnamefont
  {J.}~\bibnamefont {Parpia}},\ }\bibfield  {title} {\bibinfo {title} {The a-b
  transition in superfluid helium-3 under confinement in a thin slab
  geometry},\ }\href {https://doi.org/10.1038/ncomms15963} {\bibfield
  {journal} {\bibinfo  {journal} {Nat. Commun.}\ }\textbf {\bibinfo {volume}
  {8}},\ \bibinfo {pages} {15963} (\bibinfo {year} {20176})}\BibitemShut
  {NoStop}%
\bibitem [{\citenamefont {Zhelev}\ \emph {et~al.}(2018)\citenamefont {Zhelev},
  \citenamefont {Abhilash}, \citenamefont {Bennett}, \citenamefont {Smith},
  \citenamefont {Ilic}, \citenamefont {Parpia}, \citenamefont {Levitin},
  \citenamefont {Rojas}, \citenamefont {Casey},\ and\ \citenamefont
  {Saunders}}]{Zhelev2018}%
  \BibitemOpen
  \bibfield  {author} {\bibinfo {author} {\bibfnamefont {N.}~\bibnamefont
  {Zhelev}}, \bibinfo {author} {\bibfnamefont {T.~S.}\ \bibnamefont
  {Abhilash}}, \bibinfo {author} {\bibfnamefont {R.~G.}\ \bibnamefont
  {Bennett}}, \bibinfo {author} {\bibfnamefont {E.~N.}\ \bibnamefont {Smith}},
  \bibinfo {author} {\bibfnamefont {B.}~\bibnamefont {Ilic}}, \bibinfo {author}
  {\bibfnamefont {J.~M.}\ \bibnamefont {Parpia}}, \bibinfo {author}
  {\bibfnamefont {L.~V.}\ \bibnamefont {Levitin}}, \bibinfo {author}
  {\bibfnamefont {X.}~\bibnamefont {Rojas}}, \bibinfo {author} {\bibfnamefont
  {A.}~\bibnamefont {Casey}},\ and\ \bibinfo {author} {\bibfnamefont
  {J.}~\bibnamefont {Saunders}},\ }\bibfield  {title} {\bibinfo {title}
  {Fabrication of microfluidic cavities using si-to-glass anodic bonding},\
  }\href {https://doi.org/10.1063/1.5031837} {\bibfield  {journal} {\bibinfo
  {journal} {Rev. Sci. Instrum.}\ }\textbf {\bibinfo {volume} {89}},\ \bibinfo
  {pages} {073902} (\bibinfo {year} {2018})}\BibitemShut {NoStop}%
\bibitem [{\citenamefont {Levitin}\ \emph {et~al.}(2019)\citenamefont
  {Levitin}, \citenamefont {Yager}, \citenamefont {Sumner}, \citenamefont
  {Cowan}, \citenamefont {Casey}, \citenamefont {Saunders}, \citenamefont
  {Zhelev}, \citenamefont {Bennett},\ and\ \citenamefont
  {Parpia}}]{Levitin2019}%
  \BibitemOpen
  \bibfield  {author} {\bibinfo {author} {\bibfnamefont {L.~V.}\ \bibnamefont
  {Levitin}}, \bibinfo {author} {\bibfnamefont {B.}~\bibnamefont {Yager}},
  \bibinfo {author} {\bibfnamefont {L.}~\bibnamefont {Sumner}}, \bibinfo
  {author} {\bibfnamefont {B.}~\bibnamefont {Cowan}}, \bibinfo {author}
  {\bibfnamefont {A.~J.}\ \bibnamefont {Casey}}, \bibinfo {author}
  {\bibfnamefont {J.}~\bibnamefont {Saunders}}, \bibinfo {author}
  {\bibfnamefont {N.}~\bibnamefont {Zhelev}}, \bibinfo {author} {\bibfnamefont
  {R.~G.}\ \bibnamefont {Bennett}},\ and\ \bibinfo {author} {\bibfnamefont
  {J.~M.}\ \bibnamefont {Parpia}},\ }\bibfield  {title} {\bibinfo {title}
  {Evidence for a spatially modulated superfluid phase of 3he under
  confinement},\ }\href {https://doi.org/10.1103/PhysRevLett.122.085301}
  {\bibfield  {journal} {\bibinfo  {journal} {Phys. Rev. Lett.}\ }\textbf
  {\bibinfo {volume} {122}},\ \bibinfo {pages} {085301} (\bibinfo {year}
  {2019})}\BibitemShut {NoStop}%
\bibitem [{\citenamefont {Lotnyk}\ \emph {et~al.}(2019)\citenamefont {Lotnyk},
  \citenamefont {Eyal}, \citenamefont {Zhelev}, \citenamefont {Abhilash},
  \citenamefont {Smith}, \citenamefont {Terilli}, \citenamefont {Wilson},
  \citenamefont {Mueller}, \citenamefont {Einzel}, \citenamefont {Saunders},\
  and\ \citenamefont {Parpia}}]{Lotnyk2019}%
  \BibitemOpen
  \bibfield  {author} {\bibinfo {author} {\bibfnamefont {D.}~\bibnamefont
  {Lotnyk}}, \bibinfo {author} {\bibfnamefont {A.}~\bibnamefont {Eyal}},
  \bibinfo {author} {\bibfnamefont {N.}~\bibnamefont {Zhelev}}, \bibinfo
  {author} {\bibfnamefont {T.}~\bibnamefont {Abhilash}}, \bibinfo {author}
  {\bibfnamefont {E.}~\bibnamefont {Smith}}, \bibinfo {author} {\bibfnamefont
  {M.}~\bibnamefont {Terilli}}, \bibinfo {author} {\bibfnamefont
  {J.}~\bibnamefont {Wilson}}, \bibinfo {author} {\bibfnamefont
  {E.}~\bibnamefont {Mueller}}, \bibinfo {author} {\bibfnamefont
  {D.}~\bibnamefont {Einzel}}, \bibinfo {author} {\bibfnamefont
  {J.}~\bibnamefont {Saunders}},\ and\ \bibinfo {author} {\bibfnamefont
  {J.}~\bibnamefont {Parpia}},\ }\bibfield  {title} {\bibinfo {title} {Thermal
  transport of helium-3 in a strongly confining channel},\ }\href
  {https://arxiv.org/abs/1910.08414} {\bibfield  {journal} {\bibinfo  {journal}
  {arXiv:1910.08414}\ } (\bibinfo {year} {2019})}\BibitemShut {NoStop}%
\bibitem [{\citenamefont {Varga}\ \emph {et~al.}(2020)\citenamefont {Varga},
  \citenamefont {Vadakkumbatt}, \citenamefont {Shook}, \citenamefont {Kim},\
  and\ \citenamefont {Davis}}]{Varga2020}%
  \BibitemOpen
  \bibfield  {author} {\bibinfo {author} {\bibfnamefont {E.}~\bibnamefont
  {Varga}}, \bibinfo {author} {\bibfnamefont {V.}~\bibnamefont {Vadakkumbatt}},
  \bibinfo {author} {\bibfnamefont {A.}~\bibnamefont {Shook}}, \bibinfo
  {author} {\bibfnamefont {P.}~\bibnamefont {Kim}},\ and\ \bibinfo {author}
  {\bibfnamefont {J.}~\bibnamefont {Davis}},\ }\bibfield  {title} {\bibinfo
  {title} {Observation of bistable turbulence in quasi-two-dimensional
  superflow},\ }\href {https://doi.org/10.1103/PhysRevLett.125.025301}
  {\bibfield  {journal} {\bibinfo  {journal} {Phys. Rev. Lett.}\ }\textbf
  {\bibinfo {volume} {125}},\ \bibinfo {pages} {025301} (\bibinfo {year}
  {2020})}\BibitemShut {NoStop}%
\bibitem [{\citenamefont {Heikkinen}\ \emph {et~al.}(2019)\citenamefont
  {Heikkinen}, \citenamefont {Casey}, \citenamefont {Levitin}, \citenamefont
  {Rojas}, \citenamefont {Vorontsov}, \citenamefont {Sharma}, \citenamefont
  {Zhelev}, \citenamefont {Parpia},\ and\ \citenamefont
  {Saunders}}]{Heikkinen2020}%
  \BibitemOpen
  \bibfield  {author} {\bibinfo {author} {\bibfnamefont {P.~J.}\ \bibnamefont
  {Heikkinen}}, \bibinfo {author} {\bibfnamefont {A.}~\bibnamefont {Casey}},
  \bibinfo {author} {\bibfnamefont {L.~V.}\ \bibnamefont {Levitin}}, \bibinfo
  {author} {\bibfnamefont {X.}~\bibnamefont {Rojas}}, \bibinfo {author}
  {\bibfnamefont {A.}~\bibnamefont {Vorontsov}}, \bibinfo {author}
  {\bibfnamefont {P.}~\bibnamefont {Sharma}}, \bibinfo {author} {\bibfnamefont
  {N.}~\bibnamefont {Zhelev}}, \bibinfo {author} {\bibfnamefont {J.~M.}\
  \bibnamefont {Parpia}},\ and\ \bibinfo {author} {\bibfnamefont
  {J.}~\bibnamefont {Saunders}},\ }\bibfield  {title} {\bibinfo {title}
  {Fragility of surface states in topological superfluid 3he},\ }\href
  {https://arxiv.org/abs/1909.04210} {\bibfield  {journal} {\bibinfo  {journal}
  {arxiv:1909.04210}\ } (\bibinfo {year} {2019})}\BibitemShut {NoStop}%
\bibitem [{\citenamefont {Safavi-Naeini}\ and\ \citenamefont
  {Painter}(2010)}]{Safavi-Naeini2010}%
  \BibitemOpen
  \bibfield  {author} {\bibinfo {author} {\bibfnamefont {A.~H.}\ \bibnamefont
  {Safavi-Naeini}}\ and\ \bibinfo {author} {\bibfnamefont {O.}~\bibnamefont
  {Painter}},\ }\bibfield  {title} {\bibinfo {title} {Design of optomechanical
  cavities and waveguides on a simultaneous bandgap phononic-photonic crystal
  slab},\ }\href {https://doi.org/10.1364/OE.18.014926} {\bibfield  {journal}
  {\bibinfo  {journal} {Opt. Express}\ }\textbf {\bibinfo {volume} {18}},\
  \bibinfo {pages} {14926} (\bibinfo {year} {2010})}\BibitemShut {NoStop}%
\bibitem [{\citenamefont {Laude}(2015)}]{LaudeBook}%
  \BibitemOpen
  \bibfield  {author} {\bibinfo {author} {\bibfnamefont {V.}~\bibnamefont
  {Laude}},\ }\href@noop {} {\emph {\bibinfo {title} {Phononic Crystals}}}\
  (\bibinfo  {publisher} {De Gruyter},\ \bibinfo {year} {2015})\BibitemShut
  {NoStop}%
\bibitem [{\citenamefont {Martinez-Sala}\ \emph {et~al.}(1995)\citenamefont
  {Martinez-Sala}, \citenamefont {Sancho}, \citenamefont {Sanchez},
  \citenamefont {Gomez}, \citenamefont {Llinares},\ and\ \citenamefont
  {Meseguer}}]{Martinez-Sala1995}%
  \BibitemOpen
  \bibfield  {author} {\bibinfo {author} {\bibfnamefont {R.}~\bibnamefont
  {Martinez-Sala}}, \bibinfo {author} {\bibfnamefont {J.}~\bibnamefont
  {Sancho}}, \bibinfo {author} {\bibfnamefont {J.~V.}\ \bibnamefont {Sanchez}},
  \bibinfo {author} {\bibfnamefont {V.}~\bibnamefont {Gomez}}, \bibinfo
  {author} {\bibfnamefont {J.}~\bibnamefont {Llinares}},\ and\ \bibinfo
  {author} {\bibfnamefont {F.}~\bibnamefont {Meseguer}},\ }\bibfield  {title}
  {\bibinfo {title} {Sound attenuation by scuplture},\ }\href
  {https://doi.org/10.1038/378241a0} {\bibfield  {journal} {\bibinfo  {journal}
  {Nature}\ }\textbf {\bibinfo {volume} {378}},\ \bibinfo {pages} {241}
  (\bibinfo {year} {1995})}\BibitemShut {NoStop}%
\bibitem [{\citenamefont {Kapitza}(1938)}]{Kapitza1938}%
  \BibitemOpen
  \bibfield  {author} {\bibinfo {author} {\bibfnamefont {P.}~\bibnamefont
  {Kapitza}},\ }\bibfield  {title} {\bibinfo {title} {Viscosity of liquid
  helium below the $\lambda$-point},\ }\href {https://doi.org/10.1038/141074a0}
  {\bibfield  {journal} {\bibinfo  {journal} {Nature}\ }\textbf {\bibinfo
  {volume} {141}},\ \bibinfo {pages} {74} (\bibinfo {year} {1938})}\BibitemShut
  {NoStop}%
\bibitem [{\citenamefont {Tisza}(1938)}]{Tisza1938}%
  \BibitemOpen
  \bibfield  {author} {\bibinfo {author} {\bibfnamefont {L.}~\bibnamefont
  {Tisza}},\ }\bibfield  {title} {\bibinfo {title} {Transport phenomena in
  helium ii},\ }\href {https://doi.org/10.1038/141913a0} {\bibfield  {journal}
  {\bibinfo  {journal} {Nature}\ }\textbf {\bibinfo {volume} {141}},\ \bibinfo
  {pages} {913} (\bibinfo {year} {1938})}\BibitemShut {NoStop}%
\bibitem [{\citenamefont {Landau}(1941)}]{Landau1941}%
  \BibitemOpen
  \bibfield  {author} {\bibinfo {author} {\bibfnamefont {L.~D.}\ \bibnamefont
  {Landau}},\ }\bibfield  {title} {\bibinfo {title} {The theory of
  superfluidity helium ii},\ }\href@noop {} {\bibfield  {journal} {\bibinfo
  {journal} {JETP}\ }\textbf {\bibinfo {volume} {11}},\ \bibinfo {pages} {592}
  (\bibinfo {year} {1941})}\BibitemShut {NoStop}%
\bibitem [{\citenamefont {Balibar}(2017)}]{Balibar2017}%
  \BibitemOpen
  \bibfield  {author} {\bibinfo {author} {\bibfnamefont {S.}~\bibnamefont
  {Balibar}},\ }\bibfield  {title} {\bibinfo {title} {Laszlo tisza and the
  two-fluid model of superfluidity},\ }\href
  {https://doi.org/10.1016/j.crhy.2017.10.016} {\bibfield  {journal} {\bibinfo
  {journal} {C.R. Physique}\ }\textbf {\bibinfo {volume} {18}},\ \bibinfo
  {pages} {586} (\bibinfo {year} {2017})}\BibitemShut {NoStop}%
\bibitem [{\citenamefont {Landau}\ and\ \citenamefont
  {Khalatnikov}(1949)}]{Landau1949}%
  \BibitemOpen
  \bibfield  {author} {\bibinfo {author} {\bibfnamefont {L.}~\bibnamefont
  {Landau}}\ and\ \bibinfo {author} {\bibfnamefont {I.~M.}\ \bibnamefont
  {Khalatnikov}},\ }\bibfield  {title} {\bibinfo {title} {The theory of the
  viscosity of helium ii: I collision of elementary excitations in helium ii},\
  }\href@noop {} {\bibfield  {journal} {\bibinfo  {journal} {JETP}\ }\textbf
  {\bibinfo {volume} {19}},\ \bibinfo {pages} {637} (\bibinfo {year}
  {1949})}\BibitemShut {NoStop}%
\bibitem [{\citenamefont {Khalatnikov}(2000)}]{KhalatnikovBook2000}%
  \BibitemOpen
  \bibfield  {author} {\bibinfo {author} {\bibfnamefont {I.}~\bibnamefont
  {Khalatnikov}},\ }\href@noop {} {\emph {\bibinfo {title} {An Introduction To
  The Theory Of Superfluidity}}}\ (\bibinfo  {publisher} {CRC Press},\ \bibinfo
  {year} {2000})\BibitemShut {NoStop}%
\bibitem [{\citenamefont {Nozieres}\ and\ \citenamefont
  {Pines}(1994{\natexlab{a}})}]{NozieresBook}%
  \BibitemOpen
  \bibfield  {author} {\bibinfo {author} {\bibfnamefont {P.}~\bibnamefont
  {Nozieres}}\ and\ \bibinfo {author} {\bibfnamefont {D.}~\bibnamefont
  {Pines}},\ }\href@noop {} {\emph {\bibinfo {title} {The Theory of Quantum
  Liquids (Vol. II), Superfluid Bose Liquid}}}\ (\bibinfo  {publisher} {CRC
  Press},\ \bibinfo {year} {1994})\BibitemShut {NoStop}%
\bibitem [{\citenamefont {Feynman}(1955)}]{Feynman1955}%
  \BibitemOpen
  \bibfield  {author} {\bibinfo {author} {\bibfnamefont {R.}~\bibnamefont
  {Feynman}},\ }\href {https://doi.org/10.1016/S0079-6417(08)60077-3} {\emph
  {\bibinfo {title} {Progress in Low Temperature Physics, Application of
  Quantum Mechanics to Liquid Helium (Ch. 2)}}}\ (\bibinfo  {publisher}
  {Elsevier},\ \bibinfo {year} {1955})\BibitemShut {NoStop}%
\bibitem [{\citenamefont {Miller}\ \emph {et~al.}(1962)\citenamefont {Miller},
  \citenamefont {Pines},\ and\ \citenamefont {Nozieres}}]{Miller1962}%
  \BibitemOpen
  \bibfield  {author} {\bibinfo {author} {\bibfnamefont {A.}~\bibnamefont
  {Miller}}, \bibinfo {author} {\bibfnamefont {D.}~\bibnamefont {Pines}},\ and\
  \bibinfo {author} {\bibfnamefont {P.}~\bibnamefont {Nozieres}},\ }\bibfield
  {title} {\bibinfo {title} {Elementary excitations in liquid helium},\ }\href
  {https://doi.org/10.1103/PhysRev.127.1452} {\bibfield  {journal} {\bibinfo
  {journal} {Phys. Rev.}\ }\textbf {\bibinfo {volume} {127}},\ \bibinfo {pages}
  {1452} (\bibinfo {year} {1962})}\BibitemShut {NoStop}%
\bibitem [{\citenamefont {Nozieres}(2004)}]{Nozieres2004}%
  \BibitemOpen
  \bibfield  {author} {\bibinfo {author} {\bibfnamefont {P.}~\bibnamefont
  {Nozieres}},\ }\bibfield  {title} {\bibinfo {title} {Is the roton in
  superfluid 4he the ghost of a bragg spot},\ }\href
  {https://doi.org/10.1023/B:JOLT.0000044234.82957.2f} {\bibfield  {journal}
  {\bibinfo  {journal} {J. Low Temp. Phys}\ }\textbf {\bibinfo {volume}
  {137}},\ \bibinfo {pages} {45} (\bibinfo {year} {2004})}\BibitemShut
  {NoStop}%
\bibitem [{\citenamefont {Beauvois}\ \emph {et~al.}(2018)\citenamefont
  {Beauvois}, \citenamefont {Dawidowski}, \citenamefont {F\aa{}k},
  \citenamefont {Godfrin}, \citenamefont {Krotscheck}, \citenamefont
  {Ollivier},\ and\ \citenamefont {Sultan}}]{Beauvois2018}%
  \BibitemOpen
  \bibfield  {author} {\bibinfo {author} {\bibfnamefont {K.}~\bibnamefont
  {Beauvois}}, \bibinfo {author} {\bibfnamefont {J.}~\bibnamefont
  {Dawidowski}}, \bibinfo {author} {\bibfnamefont {B.}~\bibnamefont {F\aa{}k}},
  \bibinfo {author} {\bibfnamefont {H.}~\bibnamefont {Godfrin}}, \bibinfo
  {author} {\bibfnamefont {E.}~\bibnamefont {Krotscheck}}, \bibinfo {author}
  {\bibfnamefont {J.}~\bibnamefont {Ollivier}},\ and\ \bibinfo {author}
  {\bibfnamefont {A.}~\bibnamefont {Sultan}},\ }\bibfield  {title} {\bibinfo
  {title} {Microscopic dynamics of superfluid $^{4}\mathrm{He}$: A
  comprehensive study by inelastic neutron scattering},\ }\href
  {https://doi.org/10.1103/PhysRevB.97.184520} {\bibfield  {journal} {\bibinfo
  {journal} {Phys. Rev. B}\ }\textbf {\bibinfo {volume} {97}},\ \bibinfo
  {pages} {184520} (\bibinfo {year} {2018})}\BibitemShut {NoStop}%
\bibitem [{\citenamefont {Wilks}(1967)}]{Wilks1967}%
  \BibitemOpen
  \bibfield  {author} {\bibinfo {author} {\bibfnamefont {J.}~\bibnamefont
  {Wilks}},\ }\href@noop {} {\emph {\bibinfo {title} {The properties of liquid
  and solid helium}}}\ (\bibinfo  {publisher} {Clarendon Press Oxford},\
  \bibinfo {year} {1967})\BibitemShut {NoStop}%
\bibitem [{\citenamefont {Landau}\ and\ \citenamefont
  {Pomeranchuk}(1948)}]{Landau1948}%
  \BibitemOpen
  \bibfield  {author} {\bibinfo {author} {\bibfnamefont {L.}~\bibnamefont
  {Landau}}\ and\ \bibinfo {author} {\bibfnamefont {I.}~\bibnamefont
  {Pomeranchuk}},\ }\bibfield  {title} {\bibinfo {title} {On the motion of
  foreign particles in helium ii},\ }\href@noop {} {\bibfield  {journal}
  {\bibinfo  {journal} {Dokl. Akad. Nauk SSSR}\ }\textbf {\bibinfo {volume}
  {59}},\ \bibinfo {pages} {669} (\bibinfo {year} {1948})}\BibitemShut
  {NoStop}%
\bibitem [{\citenamefont {Haar}(1965)}]{TerHaar1965}%
  \BibitemOpen
  \bibfield  {author} {\bibinfo {author} {\bibfnamefont {D.~T.}\ \bibnamefont
  {Haar}},\ }\href@noop {} {\emph {\bibinfo {title} {Collected Papers of L.D
  Landau}}}\ (\bibinfo  {publisher} {Gordon and Breach, Science Publishers},\
  \bibinfo {year} {1965})\BibitemShut {NoStop}%
\bibitem [{\citenamefont {Baym}\ and\ \citenamefont
  {Pethick}(2004)}]{BaymBook2004}%
  \BibitemOpen
  \bibfield  {author} {\bibinfo {author} {\bibfnamefont {G.}~\bibnamefont
  {Baym}}\ and\ \bibinfo {author} {\bibfnamefont {C.}~\bibnamefont {Pethick}},\
  }\href@noop {} {\emph {\bibinfo {title} {Landau Fermi-Liquid Theory: Concept
  and Applications}}}\ (\bibinfo  {publisher} {Wiley-VCH},\ \bibinfo {year}
  {2004})\BibitemShut {NoStop}%
\bibitem [{\citenamefont {London}(1954)}]{LondonBook1954}%
  \BibitemOpen
  \bibfield  {author} {\bibinfo {author} {\bibfnamefont {F.}~\bibnamefont
  {London}},\ }\href@noop {} {\emph {\bibinfo {title} {Superfluid}}}\ (\bibinfo
   {publisher} {Dover Publications},\ \bibinfo {year} {1954})\BibitemShut
  {NoStop}%
\bibitem [{\citenamefont {Putterman}(1974)}]{PuttermanBook}%
  \BibitemOpen
  \bibfield  {author} {\bibinfo {author} {\bibfnamefont {S.}~\bibnamefont
  {Putterman}},\ }\href@noop {} {\emph {\bibinfo {title} {Superfluid
  hydrodynamics (North-Holland series in low temperature physics)}}}\ (\bibinfo
   {publisher} {North-Holland},\ \bibinfo {year} {1974})\BibitemShut {NoStop}%
\bibitem [{\citenamefont {Nozieres}\ and\ \citenamefont
  {Pines}(1994{\natexlab{b}})}]{NozieresBookCh7}%
  \BibitemOpen
  \bibfield  {author} {\bibinfo {author} {\bibfnamefont {P.}~\bibnamefont
  {Nozieres}}\ and\ \bibinfo {author} {\bibfnamefont {D.}~\bibnamefont
  {Pines}},\ }\href@noop {} {\emph {\bibinfo {title} {The Theory of Quantum
  Liquids (Vol. II), Superfluid Bose Liquid, Ch. 7}}}\ (\bibinfo  {publisher}
  {CRC Press},\ \bibinfo {year} {1994})\BibitemShut {NoStop}%
\bibitem [{\citenamefont {J{\"a}ckle}\ and\ \citenamefont
  {Kehr}(1971)}]{Jackle1971}%
  \BibitemOpen
  \bibfield  {author} {\bibinfo {author} {\bibfnamefont {J.}~\bibnamefont
  {J{\"a}ckle}}\ and\ \bibinfo {author} {\bibfnamefont {K.}~\bibnamefont
  {Kehr}},\ }\bibfield  {title} {\bibinfo {title} {Note on phonon lifetimes in
  he ii},\ }\href {https://doi.org/10.1016/0375-9601(71)90463-4} {\bibfield
  {journal} {\bibinfo  {journal} {Phys. Lett.}\ }\textbf {\bibinfo {volume}
  {37A}},\ \bibinfo {pages} {205} (\bibinfo {year} {1971})}\BibitemShut
  {NoStop}%
\bibitem [{\citenamefont {Dransfeld}(1962)}]{Dransfeld1962}%
  \BibitemOpen
  \bibfield  {author} {\bibinfo {author} {\bibfnamefont {K.}~\bibnamefont
  {Dransfeld}},\ }\bibfield  {title} {\bibinfo {title} {Ultrasonic absorption
  in liquid helium at temperatures below 0.6 k},\ }\bibfield  {journal}
  {\bibinfo  {journal} {Phys. Rev.}\ }\href
  {https://doi.org/10.1103/PhysRev.127.17} {10.1103/PhysRev.127.17} (\bibinfo
  {year} {1962})\BibitemShut {NoStop}%
\bibitem [{\citenamefont {Woodruff}(1962)}]{Woodruff1962}%
  \BibitemOpen
  \bibfield  {author} {\bibinfo {author} {\bibfnamefont {T.}~\bibnamefont
  {Woodruff}},\ }\bibfield  {title} {\bibinfo {title} {Sound absorption in
  liquid helium ii, $t<0.5$ k},\ }\href
  {https://doi.org/10.1103/PhysRev.127.682} {\bibfield  {journal} {\bibinfo
  {journal} {Phys. Rev.}\ }\textbf {\bibinfo {volume} {127}},\ \bibinfo {pages}
  {682} (\bibinfo {year} {1962})}\BibitemShut {NoStop}%
\bibitem [{\citenamefont {Pethick}\ and\ \citenamefont
  {Haar}(1966)}]{Pethick1966}%
  \BibitemOpen
  \bibfield  {author} {\bibinfo {author} {\bibfnamefont {C.}~\bibnamefont
  {Pethick}}\ and\ \bibinfo {author} {\bibfnamefont {D.~T.}\ \bibnamefont
  {Haar}},\ }\bibfield  {title} {\bibinfo {title} {On the attenuation of sound
  in liquid helium},\ }\href {https://doi.org/10.1016/0031-8914(66)90157-1}
  {\bibfield  {journal} {\bibinfo  {journal} {Physica}\ }\textbf {\bibinfo
  {volume} {32}},\ \bibinfo {pages} {1905} (\bibinfo {year}
  {1966})}\BibitemShut {NoStop}%
\bibitem [{\citenamefont {Disatnik}(1967)}]{Disatnik1967}%
  \BibitemOpen
  \bibfield  {author} {\bibinfo {author} {\bibfnamefont {Y.}~\bibnamefont
  {Disatnik}},\ }\bibfield  {title} {\bibinfo {title} {Single-collision-time
  theory of sound proyagation in liquid 4he below 0.6 k},\ }\href
  {https://doi.org/10.1103/PhysRev.158.162} {\bibfield  {journal} {\bibinfo
  {journal} {Phys. Rev.}\ }\textbf {\bibinfo {volume} {158}},\ \bibinfo {pages}
  {158} (\bibinfo {year} {1967})}\BibitemShut {NoStop}%
\bibitem [{\citenamefont {Eckstein}\ \emph {et~al.}(1970)\citenamefont
  {Eckstein}, \citenamefont {Eckstein}, \citenamefont {Ketterson},\ and\
  \citenamefont {Vignos}}]{PhysicalAcoustics6BookCh5}%
  \BibitemOpen
  \bibfield  {author} {\bibinfo {author} {\bibfnamefont {S.}~\bibnamefont
  {Eckstein}}, \bibinfo {author} {\bibfnamefont {Y.}~\bibnamefont {Eckstein}},
  \bibinfo {author} {\bibfnamefont {J.}~\bibnamefont {Ketterson}},\ and\
  \bibinfo {author} {\bibfnamefont {J.}~\bibnamefont {Vignos}},\ }\href@noop {}
  {\emph {\bibinfo {title} {Physical Acoustics \textit{Principles and Methods},
  Vol. VI, Ch. 5}}}\ (\bibinfo  {publisher} {Academic Press},\ \bibinfo {year}
  {1970})\BibitemShut {NoStop}%
\bibitem [{\citenamefont {Maris}(1973)}]{Maris1973b}%
  \BibitemOpen
  \bibfield  {author} {\bibinfo {author} {\bibfnamefont {H.}~\bibnamefont
  {Maris}},\ }\bibfield  {title} {\bibinfo {title} {Hydrodynamics of superfluid
  helium below 0.6 k. ii. velocity and attenuation of ultrasonic waves},\
  }\href {https://doi.org/10.1103/PhysRevA.8.2629} {\bibfield  {journal}
  {\bibinfo  {journal} {Phys. Rev. A}\ }\textbf {\bibinfo {volume} {8}},\
  \bibinfo {pages} {2629} (\bibinfo {year} {1973})}\BibitemShut {NoStop}%
\bibitem [{\citenamefont {Chase}\ and\ \citenamefont
  {Herlin}(1955)}]{Chase1955}%
  \BibitemOpen
  \bibfield  {author} {\bibinfo {author} {\bibfnamefont {C.}~\bibnamefont
  {Chase}}\ and\ \bibinfo {author} {\bibfnamefont {M.}~\bibnamefont {Herlin}},\
  }\bibfield  {title} {\bibinfo {title} {Ultrasonic propagation in magnetically
  cooled helium},\ }\href {https://doi.org/10.1103/PhysRev.97.1447} {\bibfield
  {journal} {\bibinfo  {journal} {Phys. Rev.}\ }\textbf {\bibinfo {volume}
  {97}},\ \bibinfo {pages} {1447} (\bibinfo {year} {1955})}\BibitemShut
  {NoStop}%
\bibitem [{\citenamefont {Abraham}\ \emph
  {et~al.}(1969{\natexlab{a}})\citenamefont {Abraham}, \citenamefont
  {Eckstein}, \citenamefont {Ketterson}, \citenamefont {Kuchnir},\ and\
  \citenamefont {Vignos}}]{Abraham1969a}%
  \BibitemOpen
  \bibfield  {author} {\bibinfo {author} {\bibfnamefont {B.~M.}\ \bibnamefont
  {Abraham}}, \bibinfo {author} {\bibfnamefont {Y.}~\bibnamefont {Eckstein}},
  \bibinfo {author} {\bibfnamefont {J.~B.}\ \bibnamefont {Ketterson}}, \bibinfo
  {author} {\bibfnamefont {M.}~\bibnamefont {Kuchnir}},\ and\ \bibinfo {author}
  {\bibfnamefont {J.}~\bibnamefont {Vignos}},\ }\bibfield  {title} {\bibinfo
  {title} {Sound propagation in liquid 4he},\ }\href
  {https://doi.org/10.1103/PhysRev.181.347} {\bibfield  {journal} {\bibinfo
  {journal} {Phys. Rev.}\ }\textbf {\bibinfo {volume} {181}},\ \bibinfo {pages}
  {347} (\bibinfo {year} {1969}{\natexlab{a}})}\BibitemShut {NoStop}%
\bibitem [{\citenamefont {Roach}\ \emph {et~al.}(1972)\citenamefont {Roach},
  \citenamefont {Ketterson},\ and\ \citenamefont {Kuchnir}}]{Roach1972}%
  \BibitemOpen
  \bibfield  {author} {\bibinfo {author} {\bibfnamefont {P.}~\bibnamefont
  {Roach}}, \bibinfo {author} {\bibfnamefont {J.}~\bibnamefont {Ketterson}},\
  and\ \bibinfo {author} {\bibfnamefont {M.}~\bibnamefont {Kuchnir}},\
  }\bibfield  {title} {\bibinfo {title} {Ultrasonic attenuation in liquid 4he
  under pressure},\ }\href {https://doi.org/10.1103/PhysRevA.5.2205} {\bibfield
   {journal} {\bibinfo  {journal} {Phys. Rev. A}\ }\textbf {\bibinfo {volume}
  {5}},\ \bibinfo {pages} {2205} (\bibinfo {year} {1972})}\BibitemShut
  {NoStop}%
\bibitem [{\citenamefont {Isihara}(1989)}]{Isihara1989}%
  \BibitemOpen
  \bibfield  {author} {\bibinfo {author} {\bibfnamefont {A.}~\bibnamefont
  {Isihara}},\ }\bibfield  {title} {\bibinfo {title} {Attenuation coefficient
  of first sound in liquid 4he},\ }\href
  {https://doi.org/doi.org/10.1103/PhysRevB.40.698} {\bibfield  {journal}
  {\bibinfo  {journal} {Phys. Rev. B}\ }\textbf {\bibinfo {volume} {40}},\
  \bibinfo {pages} {698} (\bibinfo {year} {1989})}\BibitemShut {NoStop}%
\bibitem [{\citenamefont {Kurkjian}\ \emph {et~al.}(2017)\citenamefont
  {Kurkjian}, \citenamefont {Castin},\ and\ \citenamefont
  {A.Sinatra}}]{Kurkjian2017}%
  \BibitemOpen
  \bibfield  {author} {\bibinfo {author} {\bibfnamefont {H.}~\bibnamefont
  {Kurkjian}}, \bibinfo {author} {\bibfnamefont {Y.}~\bibnamefont {Castin}},\
  and\ \bibinfo {author} {\bibnamefont {A.Sinatra}},\ }\bibfield  {title}
  {\bibinfo {title} {Three-phonon and four-phonon interaction processes in a
  pair-condensed fermi gas},\ }\href {https://doi.org/10.1002/andp.201600352}
  {\bibfield  {journal} {\bibinfo  {journal} {Ann. Phys. (Berlin)}\ }\textbf
  {\bibinfo {volume} {529}},\ \bibinfo {pages} {1600352} (\bibinfo {year}
  {2017})}\BibitemShut {NoStop}%
\bibitem [{\citenamefont {Beauvois}\ \emph {et~al.}(2019)\citenamefont
  {Beauvois}, \citenamefont {Godfrin}, \citenamefont {Krotscheck},\ and\
  \citenamefont {Zillich}}]{Beauvois2019}%
  \BibitemOpen
  \bibfield  {author} {\bibinfo {author} {\bibfnamefont {K.}~\bibnamefont
  {Beauvois}}, \bibinfo {author} {\bibfnamefont {H.}~\bibnamefont {Godfrin}},
  \bibinfo {author} {\bibfnamefont {E.}~\bibnamefont {Krotscheck}},\ and\
  \bibinfo {author} {\bibfnamefont {R.~E.}\ \bibnamefont {Zillich}},\
  }\bibfield  {title} {\bibinfo {title} {Transport and phonon damping in
  $^{4}\mathrm{He}$},\ }\href {https://doi.org/10.1007/s10909-019-02219-1}
  {\bibfield  {journal} {\bibinfo  {journal} {J. Low Temp. Phys}\ }\textbf
  {\bibinfo {volume} {197}},\ \bibinfo {pages} {113} (\bibinfo {year}
  {2019})}\BibitemShut {NoStop}%
\bibitem [{\citenamefont {Tucker}\ and\ \citenamefont
  {Wyatt}(1992)}]{Tucker1992}%
  \BibitemOpen
  \bibfield  {author} {\bibinfo {author} {\bibfnamefont {M.~A.~H.}\
  \bibnamefont {Tucker}}\ and\ \bibinfo {author} {\bibfnamefont {A.~F.~G.}\
  \bibnamefont {Wyatt}},\ }\bibfield  {title} {\bibinfo {title} {Four-phonon
  scattering in superfluid $^{4}\mathrm{He}$},\ }\href
  {https://doi.org/10.1088/0953-8984/4/38/008} {\bibfield  {journal} {\bibinfo
  {journal} {J. Phys.: Condens. Matter}\ }\textbf {\bibinfo {volume} {4}},\
  \bibinfo {pages} {7745} (\bibinfo {year} {1992})}\BibitemShut {NoStop}%
\bibitem [{\citenamefont {Souris}\ \emph {et~al.}(2014)\citenamefont {Souris},
  \citenamefont {Fefferman}, \citenamefont {Maris}, \citenamefont {Dauvois},
  \citenamefont {Jean-Baptiste}, \citenamefont {Beamish},\ and\ \citenamefont
  {Balibar}}]{Souris2014}%
  \BibitemOpen
  \bibfield  {author} {\bibinfo {author} {\bibfnamefont {F.}~\bibnamefont
  {Souris}}, \bibinfo {author} {\bibfnamefont {A.~D.}\ \bibnamefont
  {Fefferman}}, \bibinfo {author} {\bibfnamefont {H.~J.}\ \bibnamefont
  {Maris}}, \bibinfo {author} {\bibfnamefont {V.}~\bibnamefont {Dauvois}},
  \bibinfo {author} {\bibfnamefont {P.}~\bibnamefont {Jean-Baptiste}}, \bibinfo
  {author} {\bibfnamefont {J.~R.}\ \bibnamefont {Beamish}},\ and\ \bibinfo
  {author} {\bibfnamefont {S.}~\bibnamefont {Balibar}},\ }\bibfield  {title}
  {\bibinfo {title} {Movement of dislocations dressed with $^{3}\mathrm{He}$
  impurities in $^{4}\mathrm{He}$ crystals},\ }\href@noop {} {\bibfield
  {journal} {\bibinfo  {journal} {Phys. Rev. B}\ }\textbf {\bibinfo {volume}
  {90}},\ \bibinfo {pages} {180103(R)} (\bibinfo {year} {2014})}\BibitemShut
  {NoStop}%
\bibitem [{\citenamefont {Hendry}\ and\ \citenamefont
  {McClintock}(1987)}]{Hendry1987}%
  \BibitemOpen
  \bibfield  {author} {\bibinfo {author} {\bibfnamefont {P.~C.}\ \bibnamefont
  {Hendry}}\ and\ \bibinfo {author} {\bibfnamefont {P.~V.~E.}\ \bibnamefont
  {McClintock}},\ }\bibfield  {title} {\bibinfo {title} {Continuous flow
  apparatus for preparing isotopically pure $^{4}\mathrm{He}$},\ }\href@noop {}
  {\bibfield  {journal} {\bibinfo  {journal} {Cryogenics}\ }\textbf {\bibinfo
  {volume} {27}},\ \bibinfo {pages} {131} (\bibinfo {year} {1987})}\BibitemShut
  {NoStop}%
\bibitem [{\citenamefont {Bardeen}\ \emph {et~al.}(1966)\citenamefont
  {Bardeen}, \citenamefont {Baym},\ and\ \citenamefont {Pines}}]{Bardeen1966}%
  \BibitemOpen
  \bibfield  {author} {\bibinfo {author} {\bibfnamefont {J.}~\bibnamefont
  {Bardeen}}, \bibinfo {author} {\bibfnamefont {G.}~\bibnamefont {Baym}},\ and\
  \bibinfo {author} {\bibfnamefont {D.}~\bibnamefont {Pines}},\ }\bibfield
  {title} {\bibinfo {title} {Interactions between $^{3}\mathrm{He}$ atoms in
  dilute solutions of $^{3}\mathrm{He}$ in superfluid $^{4}\mathrm{He}$},\
  }\href {https://doi.org/10.1103/PhysRevLett.17.372} {\bibfield  {journal}
  {\bibinfo  {journal} {Phys. Rev. Lett.}\ }\textbf {\bibinfo {volume} {17}},\
  \bibinfo {pages} {372} (\bibinfo {year} {1966})}\BibitemShut {NoStop}%
\bibitem [{\citenamefont {Bardeen}\ \emph {et~al.}(1967)\citenamefont
  {Bardeen}, \citenamefont {Baym},\ and\ \citenamefont {Pines}}]{Bardeen1967}%
  \BibitemOpen
  \bibfield  {author} {\bibinfo {author} {\bibfnamefont {J.}~\bibnamefont
  {Bardeen}}, \bibinfo {author} {\bibfnamefont {G.}~\bibnamefont {Baym}},\ and\
  \bibinfo {author} {\bibfnamefont {D.}~\bibnamefont {Pines}},\ }\bibfield
  {title} {\bibinfo {title} {Effective interaction of $^{3}\mathrm{He}$ atoms
  in dilute solutions of $^{3}\mathrm{He}$ in $^{4}\mathrm{He}$ at low
  temperatures},\ }\href {https://doi.org/10.1103/PhysRev.156.207} {\bibfield
  {journal} {\bibinfo  {journal} {Phys. Rev.}\ }\textbf {\bibinfo {volume}
  {156}},\ \bibinfo {pages} {207} (\bibinfo {year} {1967})}\BibitemShut
  {NoStop}%
\bibitem [{\citenamefont {Baym}(1967)}]{Baym1967a}%
  \BibitemOpen
  \bibfield  {author} {\bibinfo {author} {\bibfnamefont {G.}~\bibnamefont
  {Baym}},\ }\bibfield  {title} {\bibinfo {title} {Theory of first sound in
  dilute solutions of $^{3}\mathrm{He}$ in $^{4}\mathrm{He}$ at very low
  temperatures},\ }\href {https://doi.org/10.1103/PhysRevLett.18.71} {\bibfield
   {journal} {\bibinfo  {journal} {Phys. Rev. Lett.}\ }\textbf {\bibinfo
  {volume} {18}},\ \bibinfo {pages} {71} (\bibinfo {year} {1967})}\BibitemShut
  {NoStop}%
\bibitem [{\citenamefont {Baym}\ and\ \citenamefont {Ebner}(1967)}]{Baym1967b}%
  \BibitemOpen
  \bibfield  {author} {\bibinfo {author} {\bibfnamefont {G.}~\bibnamefont
  {Baym}}\ and\ \bibinfo {author} {\bibfnamefont {C.}~\bibnamefont {Ebner}},\
  }\bibfield  {title} {\bibinfo {title} {Phonon-quasiparticle interactions in
  dilute solutions of $^{3}\mathrm{He}$ in superfluid $^{4}\mathrm{He}$: I.
  phonon thermal conductivity and ultrasonic attenuation},\ }\href
  {https://doi.org/10.1103/PhysRev.164.235} {\bibfield  {journal} {\bibinfo
  {journal} {Phys. Rev.}\ }\textbf {\bibinfo {volume} {164}},\ \bibinfo {pages}
  {235} (\bibinfo {year} {1967})}\BibitemShut {NoStop}%
\bibitem [{\citenamefont {Baym}\ \emph {et~al.}(1968)\citenamefont {Baym},
  \citenamefont {Saam},\ and\ \citenamefont {Ebner}}]{Baym1968}%
  \BibitemOpen
  \bibfield  {author} {\bibinfo {author} {\bibfnamefont {G.}~\bibnamefont
  {Baym}}, \bibinfo {author} {\bibfnamefont {W.~F.}\ \bibnamefont {Saam}},\
  and\ \bibinfo {author} {\bibfnamefont {C.}~\bibnamefont {Ebner}},\ }\bibfield
   {title} {\bibinfo {title} {Phonon-quasiparticle interactions in dilute
  solutions of $^{3}\mathrm{He}$ in superfluid $^{4}\mathrm{He}$: Iii.
  attenuation of first sound above 0.2 k},\ }\href@noop {} {\bibfield
  {journal} {\bibinfo  {journal} {Phys. Rev.}\ }\textbf {\bibinfo {volume}
  {173}},\ \bibinfo {pages} {306} (\bibinfo {year} {1968})}\BibitemShut
  {NoStop}%
\bibitem [{\citenamefont {Baym}\ \emph
  {et~al.}(2015{\natexlab{a}})\citenamefont {Baym}, \citenamefont {Beck},\ and\
  \citenamefont {Pethick}}]{Baym2015a}%
  \BibitemOpen
  \bibfield  {author} {\bibinfo {author} {\bibfnamefont {G.}~\bibnamefont
  {Baym}}, \bibinfo {author} {\bibfnamefont {D.~H.}\ \bibnamefont {Beck}},\
  and\ \bibinfo {author} {\bibfnamefont {C.~J.}\ \bibnamefont {Pethick}},\
  }\bibfield  {title} {\bibinfo {title} {Low-temperature transport properties
  of very dilut classical solutions of $^{3}\mathrm{He}$ in superfluid
  $^{4}\mathrm{He}$},\ }\href {https://doi.org/10.1007/s10909-014-1235-0}
  {\bibfield  {journal} {\bibinfo  {journal} {J. Low Temp. Phys}\ }\textbf
  {\bibinfo {volume} {178}},\ \bibinfo {pages} {200} (\bibinfo {year}
  {2015}{\natexlab{a}})}\BibitemShut {NoStop}%
\bibitem [{\citenamefont {Lorenzo}(2016)}]{DeLorenzoThesis}%
  \BibitemOpen
  \bibfield  {author} {\bibinfo {author} {\bibfnamefont {L.~A.~D.}\
  \bibnamefont {Lorenzo}},\ }\emph {\bibinfo {title} {Optomechanics with
  Superfluid Helium-4}},\ \href@noop {} {Ph.D. thesis},\ \bibinfo  {school}
  {California Institute of Technology} (\bibinfo {year} {2016})\BibitemShut
  {NoStop}%
\bibitem [{\citenamefont {Kerscher}\ \emph {et~al.}(2001)\citenamefont
  {Kerscher}, \citenamefont {Niemetz},\ and\ \citenamefont
  {Schoepe}}]{Kerscher2001}%
  \BibitemOpen
  \bibfield  {author} {\bibinfo {author} {\bibfnamefont {H.}~\bibnamefont
  {Kerscher}}, \bibinfo {author} {\bibfnamefont {M.}~\bibnamefont {Niemetz}},\
  and\ \bibinfo {author} {\bibfnamefont {W.}~\bibnamefont {Schoepe}},\
  }\bibfield  {title} {\bibinfo {title} {Viscosity and mean free path of very
  dilute solutions of $^{3}\mathrm{He}$ in $^{4}\mathrm{He}$},\ }\href
  {https://doi.org/10.1023/A:1017525901859} {\bibfield  {journal} {\bibinfo
  {journal} {J. Low Temp. Phys}\ }\textbf {\bibinfo {volume} {124}},\ \bibinfo
  {pages} {163} (\bibinfo {year} {2001})}\BibitemShut {NoStop}%
\bibitem [{\citenamefont {Baym}\ \emph {et~al.}(2013)\citenamefont {Baym},
  \citenamefont {Beck},\ and\ \citenamefont {Pethick}}]{Baym2013}%
  \BibitemOpen
  \bibfield  {author} {\bibinfo {author} {\bibfnamefont {G.}~\bibnamefont
  {Baym}}, \bibinfo {author} {\bibfnamefont {D.~H.}\ \bibnamefont {Beck}},\
  and\ \bibinfo {author} {\bibfnamefont {C.~J.}\ \bibnamefont {Pethick}},\
  }\bibfield  {title} {\bibinfo {title} {Transport in very dilute solutions of
  $^{3}\mathrm{He}$ in superfluid $^{4}\mathrm{He}$},\ }\href
  {https://doi.org/10.1103/PhysRevB.88.014512} {\bibfield  {journal} {\bibinfo
  {journal} {Phys. Rev. B}\ }\textbf {\bibinfo {volume} {88}},\ \bibinfo
  {pages} {0145512} (\bibinfo {year} {2013})}\BibitemShut {NoStop}%
\bibitem [{\citenamefont {Baym}\ \emph
  {et~al.}(2015{\natexlab{b}})\citenamefont {Baym}, \citenamefont {Beck},\ and\
  \citenamefont {Pethick}}]{Baym2015b}%
  \BibitemOpen
  \bibfield  {author} {\bibinfo {author} {\bibfnamefont {G.}~\bibnamefont
  {Baym}}, \bibinfo {author} {\bibfnamefont {D.~H.}\ \bibnamefont {Beck}},\
  and\ \bibinfo {author} {\bibfnamefont {C.~J.}\ \bibnamefont {Pethick}},\
  }\bibfield  {title} {\bibinfo {title} {Transport in ultradilute solutions of
  $^{3}\mathrm{He}$ in superfluid $^{4}\mathrm{He}$},\ }\href
  {https://doi.org/10.1103/PhysRevB.92.024504} {\bibfield  {journal} {\bibinfo
  {journal} {Phys. Rev. B}\ }\textbf {\bibinfo {volume} {92}},\ \bibinfo
  {pages} {024504} (\bibinfo {year} {2015}{\natexlab{b}})}\BibitemShut
  {NoStop}%
\bibitem [{\citenamefont {Boghosian}\ and\ \citenamefont
  {Meyer}(1967)}]{Boghosian1967}%
  \BibitemOpen
  \bibfield  {author} {\bibinfo {author} {\bibfnamefont {C.}~\bibnamefont
  {Boghosian}}\ and\ \bibinfo {author} {\bibfnamefont {H.}~\bibnamefont
  {Meyer}},\ }\bibfield  {title} {\bibinfo {title} {Density of a dilute
  $^{3}\mathrm{He}$ - $^{4}\mathrm{He}$ solution under pressure},\ }\href
  {https://doi.org/https://doi.org/10.1016/0375-9601(67)90693-7} {\bibfield
  {journal} {\bibinfo  {journal} {Phys. Lett.}\ }\textbf {\bibinfo {volume}
  {25A}},\ \bibinfo {pages} {352} (\bibinfo {year} {1967})}\BibitemShut
  {NoStop}%
\bibitem [{\citenamefont {Watson}\ \emph {et~al.}(1969)\citenamefont {Watson},
  \citenamefont {Reppy},\ and\ \citenamefont {Richardson}}]{Watson1969}%
  \BibitemOpen
  \bibfield  {author} {\bibinfo {author} {\bibfnamefont {G.}~\bibnamefont
  {Watson}}, \bibinfo {author} {\bibfnamefont {J.}~\bibnamefont {Reppy}},\ and\
  \bibinfo {author} {\bibfnamefont {R.}~\bibnamefont {Richardson}},\ }\bibfield
   {title} {\bibinfo {title} {Low-temperature density and solubility of
  $^{3}\mathrm{He}$ in liquid $^{4}\mathrm{He}$ under pressure},\ }\href
  {https://doi.org/10.1103/PhysRev.188.384} {\bibfield  {journal} {\bibinfo
  {journal} {Phys. Rev.}\ }\textbf {\bibinfo {volume} {188}},\ \bibinfo {pages}
  {384} (\bibinfo {year} {1969})}\BibitemShut {NoStop}%
\bibitem [{\citenamefont {Abraham}\ \emph
  {et~al.}(1969{\natexlab{b}})\citenamefont {Abraham}, \citenamefont {Brandt},
  \citenamefont {Eckstein}, \citenamefont {Munarin},\ and\ \citenamefont
  {G.Baym}}]{Abraham1969b}%
  \BibitemOpen
  \bibfield  {author} {\bibinfo {author} {\bibfnamefont {B.}~\bibnamefont
  {Abraham}}, \bibinfo {author} {\bibfnamefont {C.}~\bibnamefont {Brandt}},
  \bibinfo {author} {\bibfnamefont {Y.}~\bibnamefont {Eckstein}}, \bibinfo
  {author} {\bibfnamefont {J.}~\bibnamefont {Munarin}},\ and\ \bibinfo {author}
  {\bibnamefont {G.Baym}},\ }\bibfield  {title} {\bibinfo {title} {Relative
  molar volume and limiting solubility of $^{3}\mathrm{He}$ in superfluid
  $^{4}\mathrm{He}$},\ }\href {https://doi.org/10.1103/PhysRev.188.309}
  {\bibfield  {journal} {\bibinfo  {journal} {Phys. Rev.}\ }\textbf {\bibinfo
  {volume} {188}},\ \bibinfo {pages} {309} (\bibinfo {year}
  {1969}{\natexlab{b}})}\BibitemShut {NoStop}%
\bibitem [{\citenamefont {Casimir}(1938)}]{Casimir1938}%
  \BibitemOpen
  \bibfield  {author} {\bibinfo {author} {\bibfnamefont {H.}~\bibnamefont
  {Casimir}},\ }\bibfield  {title} {\bibinfo {title} {Note on the conduction of
  heat in crystals},\ }\href {https://doi.org/10.1016/S0031-8914(38)80162-2}
  {\bibfield  {journal} {\bibinfo  {journal} {Physica}\ }\textbf {\bibinfo
  {volume} {5}},\ \bibinfo {pages} {495} (\bibinfo {year} {1938})}\BibitemShut
  {NoStop}%
\bibitem [{\citenamefont {Ziman}(1960)}]{ZimanBook1960Ch11}%
  \BibitemOpen
  \bibfield  {author} {\bibinfo {author} {\bibfnamefont {J.}~\bibnamefont
  {Ziman}},\ }\href@noop {} {\emph {\bibinfo {title} {Electrons and Phonons
  \textit{The Theory of Transport Phenomena in Solids}, Ch. 11}}}\ (\bibinfo
  {publisher} {Clarendon Press Oxford},\ \bibinfo {year} {1960})\BibitemShut
  {NoStop}%
\bibitem [{\citenamefont {Tholen}\ and\ \citenamefont
  {Parpia}(1991)}]{Parpia1991}%
  \BibitemOpen
  \bibfield  {author} {\bibinfo {author} {\bibfnamefont {S.}~\bibnamefont
  {Tholen}}\ and\ \bibinfo {author} {\bibfnamefont {J.}~\bibnamefont
  {Parpia}},\ }\bibfield  {title} {\bibinfo {title} {Slip and the effect of 4he
  at the 3he-silicon interface},\ }\href
  {https://doi.org/10.1103/PhysRevLett.67.334} {\bibfield  {journal} {\bibinfo
  {journal} {Phys. Rev. Lett.}\ }\textbf {\bibinfo {volume} {67}},\ \bibinfo
  {pages} {334} (\bibinfo {year} {1991})}\BibitemShut {NoStop}%
\bibitem [{\citenamefont {Thomson}\ \emph {et~al.}(2014)\citenamefont
  {Thomson}, \citenamefont {Perron}, \citenamefont {Kimball}, \citenamefont
  {Mehta},\ and\ \citenamefont {Gasparini}}]{Thomson2014}%
  \BibitemOpen
  \bibfield  {author} {\bibinfo {author} {\bibfnamefont {S.}~\bibnamefont
  {Thomson}}, \bibinfo {author} {\bibfnamefont {J.}~\bibnamefont {Perron}},
  \bibinfo {author} {\bibfnamefont {M.~O.}\ \bibnamefont {Kimball}}, \bibinfo
  {author} {\bibfnamefont {S.}~\bibnamefont {Mehta}},\ and\ \bibinfo {author}
  {\bibfnamefont {F.}~\bibnamefont {Gasparini}},\ }\bibfield  {title} {\bibinfo
  {title} {Fabrication of uniform nanoscale cavities via silicon direct wafer
  bonding},\ }\href {https://doi.org/10.3791/51179} {\bibfield  {journal}
  {\bibinfo  {journal} {J. Vis. Exp.}\ }\textbf {\bibinfo {volume} {83}},\
  \bibinfo {pages} {e51179} (\bibinfo {year} {2014})}\BibitemShut {NoStop}%
\bibitem [{\citenamefont {Donnelly}\ and\ \citenamefont
  {Barenghi}(1998)}]{Donnelly1998}%
  \BibitemOpen
  \bibfield  {author} {\bibinfo {author} {\bibfnamefont {R.~J.}\ \bibnamefont
  {Donnelly}}\ and\ \bibinfo {author} {\bibfnamefont {C.~F.}\ \bibnamefont
  {Barenghi}},\ }\bibfield  {title} {\bibinfo {title} {The observed properties
  of liquid helium at the saturated vapor pressure},\ }\href
  {https://doi.org/10.1063/1.556028} {\bibfield  {journal} {\bibinfo  {journal}
  {J. Phys. Chem. Ref. Data}\ }\textbf {\bibinfo {volume} {27}},\ \bibinfo
  {pages} {1217} (\bibinfo {year} {1998})}\BibitemShut {NoStop}%
\bibitem [{Sem()}]{Semiwafer}%
  \BibitemOpen
  \href@noop {} {\bibinfo {title} {http://www.semiwafer.com/}}\BibitemShut
  {NoStop}%
\bibitem [{\citenamefont {Zhang}\ \emph {et~al.}(2006)\citenamefont {Zhang},
  \citenamefont {Pang}, \citenamefont {Yu},\ and\ \citenamefont
  {Kim}}]{Zhang2006}%
  \BibitemOpen
  \bibfield  {author} {\bibinfo {author} {\bibfnamefont {H.}~\bibnamefont
  {Zhang}}, \bibinfo {author} {\bibfnamefont {W.}~\bibnamefont {Pang}},
  \bibinfo {author} {\bibfnamefont {H.}~\bibnamefont {Yu}},\ and\ \bibinfo
  {author} {\bibfnamefont {E.}~\bibnamefont {Kim}},\ }\bibfield  {title}
  {\bibinfo {title} {High-tone bulk acoustic resonators on sapphire, crystal
  quartz, fused silica, and silicon substrates},\ }\href
  {https://doi.org/10.1063/1.2209029} {\bibfield  {journal} {\bibinfo
  {journal} {J. Appl. Phys.}\ }\textbf {\bibinfo {volume} {99}},\ \bibinfo
  {pages} {124911} (\bibinfo {year} {2006})}\BibitemShut {NoStop}%
\bibitem [{\citenamefont {Wang}\ \emph {et~al.}(2004)\citenamefont {Wang},
  \citenamefont {Hsu}, \citenamefont {Pu}, \citenamefont {Sung},\ and\
  \citenamefont {Hwa}}]{Wang2004}%
  \BibitemOpen
  \bibfield  {author} {\bibinfo {author} {\bibfnamefont {S.}~\bibnamefont
  {Wang}}, \bibinfo {author} {\bibfnamefont {Y.}~\bibnamefont {Hsu}}, \bibinfo
  {author} {\bibfnamefont {J.}~\bibnamefont {Pu}}, \bibinfo {author}
  {\bibfnamefont {J.~C.}\ \bibnamefont {Sung}},\ and\ \bibinfo {author}
  {\bibfnamefont {L.}~\bibnamefont {Hwa}},\ }\bibfield  {title} {\bibinfo
  {title} {Determination of acoustic wave velocities and elastic properties for
  diamond and other hard materials},\ }\href
  {https://doi.org/10.1016/j.matchemphys.2004.02.003} {\bibfield  {journal}
  {\bibinfo  {journal} {Mater. Chem. Phys.}\ }\textbf {\bibinfo {volume}
  {85}},\ \bibinfo {pages} {432} (\bibinfo {year} {2004})}\BibitemShut
  {NoStop}%
\bibitem [{\citenamefont {http://matweb.com/index.aspx}()}]{Matweb}%
  \BibitemOpen
  \bibfield  {author} {\bibinfo {author} {\bibnamefont
  {http://matweb.com/index.aspx}},\ }\href {http://matweb.com/index.aspx}
  {\bibinfo {title} {Material properties database}}\BibitemShut {NoStop}%
\bibitem [{\citenamefont {McGuigan}\ \emph {et~al.}(1978)\citenamefont
  {McGuigan}, \citenamefont {Lam}, \citenamefont {Gram}, \citenamefont
  {Hoffman},\ and\ \citenamefont {Douglass}}]{McGuigan1978}%
  \BibitemOpen
  \bibfield  {author} {\bibinfo {author} {\bibfnamefont {D.}~\bibnamefont
  {McGuigan}}, \bibinfo {author} {\bibfnamefont {C.}~\bibnamefont {Lam}},
  \bibinfo {author} {\bibfnamefont {R.}~\bibnamefont {Gram}}, \bibinfo {author}
  {\bibfnamefont {A.}~\bibnamefont {Hoffman}},\ and\ \bibinfo {author}
  {\bibfnamefont {D.}~\bibnamefont {Douglass}},\ }\bibfield  {title} {\bibinfo
  {title} {Measurements of the mechanical q of single-crystal silicon at low
  temperatures},\ }\href {https://doi.org/10.1007/BF00116202} {\bibfield
  {journal} {\bibinfo  {journal} {J. Low Temp. Phys}\ }\textbf {\bibinfo
  {volume} {30}},\ \bibinfo {pages} {621} (\bibinfo {year} {1978})}\BibitemShut
  {NoStop}%
\bibitem [{\citenamefont {Goryachev}\ \emph {et~al.}(2012)\citenamefont
  {Goryachev}, \citenamefont {Creedon}, \citenamefont {Ivanov}, \citenamefont
  {Galliou}, \citenamefont {Bourquin},\ and\ \citenamefont
  {Tobar}}]{Goryachev2012}%
  \BibitemOpen
  \bibfield  {author} {\bibinfo {author} {\bibfnamefont {M.}~\bibnamefont
  {Goryachev}}, \bibinfo {author} {\bibfnamefont {D.~L.}\ \bibnamefont
  {Creedon}}, \bibinfo {author} {\bibfnamefont {E.~N.}\ \bibnamefont {Ivanov}},
  \bibinfo {author} {\bibfnamefont {S.}~\bibnamefont {Galliou}}, \bibinfo
  {author} {\bibfnamefont {R.}~\bibnamefont {Bourquin}},\ and\ \bibinfo
  {author} {\bibfnamefont {M.~E.}\ \bibnamefont {Tobar}},\ }\bibfield  {title}
  {\bibinfo {title} {Extremely low-loss acoustic phonons in a quartz bulk
  acoustic waveresonator at millikelvin temperature},\ }\href
  {https://doi.org/10.1063/1.4729292} {\bibfield  {journal} {\bibinfo
  {journal} {Appl. Phys. Lett.}\ }\textbf {\bibinfo {volume} {100}},\ \bibinfo
  {pages} {243504} (\bibinfo {year} {2012})}\BibitemShut {NoStop}%
\bibitem [{\citenamefont {Penn}\ \emph {et~al.}(2001)\citenamefont {Penn},
  \citenamefont {Harry}, \citenamefont {Gretarsson}, \citenamefont
  {Kittelberger}, \citenamefont {Saulson}, \citenamefont {Schiller},
  \citenamefont {Smith},\ and\ \citenamefont {Swords}}]{Penn2001}%
  \BibitemOpen
  \bibfield  {author} {\bibinfo {author} {\bibfnamefont {S.~D.}\ \bibnamefont
  {Penn}}, \bibinfo {author} {\bibfnamefont {G.~M.}\ \bibnamefont {Harry}},
  \bibinfo {author} {\bibfnamefont {S.~M.}\ \bibnamefont {Gretarsson}},
  \bibinfo {author} {\bibfnamefont {S.~E.}\ \bibnamefont {Kittelberger}},
  \bibinfo {author} {\bibfnamefont {P.~R.}\ \bibnamefont {Saulson}}, \bibinfo
  {author} {\bibfnamefont {J.~J.}\ \bibnamefont {Schiller}}, \bibinfo {author}
  {\bibfnamefont {J.~R.}\ \bibnamefont {Smith}},\ and\ \bibinfo {author}
  {\bibfnamefont {S.~O.}\ \bibnamefont {Swords}},\ }\bibfield  {title}
  {\bibinfo {title} {High quality factor measured in fused silica},\ }\href
  {https://doi.org/10.1063/1.1394183} {\bibfield  {journal} {\bibinfo
  {journal} {Rev. Sci. Instrum.}\ }\textbf {\bibinfo {volume} {72}},\ \bibinfo
  {pages} {3670} (\bibinfo {year} {2001})}\BibitemShut {NoStop}%
\bibitem [{\citenamefont {Senkal}\ \emph {et~al.}(2015)\citenamefont {Senkal},
  \citenamefont {Ahamed}, \citenamefont {Ardakani}, \citenamefont {Askari},\
  and\ \citenamefont {Shkel}}]{Senkal2015}%
  \BibitemOpen
  \bibfield  {author} {\bibinfo {author} {\bibfnamefont {D.}~\bibnamefont
  {Senkal}}, \bibinfo {author} {\bibfnamefont {M.~J.}\ \bibnamefont {Ahamed}},
  \bibinfo {author} {\bibfnamefont {M.~H.~A.}\ \bibnamefont {Ardakani}},
  \bibinfo {author} {\bibfnamefont {S.}~\bibnamefont {Askari}},\ and\ \bibinfo
  {author} {\bibfnamefont {A.~M.}\ \bibnamefont {Shkel}},\ }\bibfield  {title}
  {\bibinfo {title} {Demonstration of 1 million q-factor onmicroglassblown
  wineglass resonators without-of-plane electrostatic transduction},\ }\href
  {https://doi.org/10.1109/JMEMS.2014.2365113} {\bibfield  {journal} {\bibinfo
  {journal} {J. Microelectromech S.}\ }\textbf {\bibinfo {volume} {24}},\
  \bibinfo {pages} {29} (\bibinfo {year} {2015})}\BibitemShut {NoStop}%
\bibitem [{\citenamefont {Yuan}\ \emph {et~al.}(2015)\citenamefont {Yuan},
  \citenamefont {Singh}, \citenamefont {Blanter},\ and\ \citenamefont
  {Steele}}]{Yuan2015}%
  \BibitemOpen
  \bibfield  {author} {\bibinfo {author} {\bibfnamefont {M.}~\bibnamefont
  {Yuan}}, \bibinfo {author} {\bibfnamefont {V.}~\bibnamefont {Singh}},
  \bibinfo {author} {\bibfnamefont {Y.~M.}\ \bibnamefont {Blanter}},\ and\
  \bibinfo {author} {\bibfnamefont {G.}~\bibnamefont {Steele}},\ }\bibfield
  {title} {\bibinfo {title} {Large cooperativity and microkelvin cooling with a
  three-dimensional optomechanical cavity},\ }\href
  {https://doi.org/10.1038/ncomms9491} {\bibfield  {journal} {\bibinfo
  {journal} {Nat. Commun.}\ }\textbf {\bibinfo {volume} {6}},\ \bibinfo {pages}
  {8491} (\bibinfo {year} {2015})}\BibitemShut {NoStop}%
\bibitem [{\citenamefont {Noguchi}\ \emph {et~al.}(2016)\citenamefont
  {Noguchi}, \citenamefont {Yamazaki}, \citenamefont {Ataka}, \citenamefont
  {Fujita}, \citenamefont {Tabuchi}, \citenamefont {Ishikawa}, \citenamefont
  {Usami},\ and\ \citenamefont {Nakamura}}]{Noguchi2016}%
  \BibitemOpen
  \bibfield  {author} {\bibinfo {author} {\bibfnamefont {A.}~\bibnamefont
  {Noguchi}}, \bibinfo {author} {\bibfnamefont {R.}~\bibnamefont {Yamazaki}},
  \bibinfo {author} {\bibfnamefont {M.}~\bibnamefont {Ataka}}, \bibinfo
  {author} {\bibfnamefont {H.}~\bibnamefont {Fujita}}, \bibinfo {author}
  {\bibfnamefont {Y.}~\bibnamefont {Tabuchi}}, \bibinfo {author} {\bibfnamefont
  {T.}~\bibnamefont {Ishikawa}}, \bibinfo {author} {\bibfnamefont
  {K.}~\bibnamefont {Usami}},\ and\ \bibinfo {author} {\bibfnamefont
  {Y.}~\bibnamefont {Nakamura}},\ }\bibfield  {title} {\bibinfo {title} {Ground
  state cooling of a quantum electromechanical system with a silicon nitride
  membrane in a 3d loop-gap cavity},\ }\href
  {https://doi.org/10.1088/1367-2630/18/10/103036} {\bibfield  {journal}
  {\bibinfo  {journal} {New J. Phys.}\ }\textbf {\bibinfo {volume} {18}},\
  \bibinfo {pages} {103036} (\bibinfo {year} {2016})}\BibitemShut {NoStop}%
\bibitem [{\citenamefont {Reagor}(2015)}]{ReagorThesis2015}%
  \BibitemOpen
  \bibfield  {author} {\bibinfo {author} {\bibfnamefont {M.}~\bibnamefont
  {Reagor}},\ }\emph {\bibinfo {title} {Superconducting Cavities for Circuit
  Quantum Electrodynamics}},\ \href@noop {} {Ph.D. thesis},\ \bibinfo  {school}
  {Yale University} (\bibinfo {year} {2015})\BibitemShut {NoStop}%
\bibitem [{\citenamefont {Harris-Lowe}\ and\ \citenamefont
  {Smee}(1970)}]{Harris-Lowe1970}%
  \BibitemOpen
  \bibfield  {author} {\bibinfo {author} {\bibfnamefont {R.}~\bibnamefont
  {Harris-Lowe}}\ and\ \bibinfo {author} {\bibfnamefont {K.}~\bibnamefont
  {Smee}},\ }\bibfield  {title} {\bibinfo {title} {Thermal expansion of liquid
  helium ii},\ }\href {https://doi.org/10.1103/PhysRevA.2.158} {\bibfield
  {journal} {\bibinfo  {journal} {Phys. Rev. A}\ }\textbf {\bibinfo {volume}
  {2}},\ \bibinfo {pages} {158} (\bibinfo {year} {1970})}\BibitemShut {NoStop}%
\bibitem [{\citenamefont {Pozar}(2012)}]{PozarBook}%
  \BibitemOpen
  \bibfield  {author} {\bibinfo {author} {\bibfnamefont {D.}~\bibnamefont
  {Pozar}},\ }\href@noop {} {\emph {\bibinfo {title} {Microwave Engineering}}}\
  (\bibinfo  {publisher} {Wiley},\ \bibinfo {year} {2012})\BibitemShut
  {NoStop}%
\bibitem [{\citenamefont {Bruch}\ and\ \citenamefont
  {Weinhold}(2000)}]{Bruch2000}%
  \BibitemOpen
  \bibfield  {author} {\bibinfo {author} {\bibfnamefont {L.}~\bibnamefont
  {Bruch}}\ and\ \bibinfo {author} {\bibfnamefont {F.}~\bibnamefont
  {Weinhold}},\ }\bibfield  {title} {\bibinfo {title} {Diamagnetism of
  helium},\ }\href {https://doi.org/10.1063/1.1318766} {\bibfield  {journal}
  {\bibinfo  {journal} {J. Chem. Phys.}\ }\textbf {\bibinfo {volume} {113}},\
  \bibinfo {pages} {8667} (\bibinfo {year} {2000})}\BibitemShut {NoStop}%
\bibitem [{\citenamefont {Niemela}\ and\ \citenamefont
  {Donnelly}(1995)}]{Niemela1995}%
  \BibitemOpen
  \bibfield  {author} {\bibinfo {author} {\bibfnamefont {J.}~\bibnamefont
  {Niemela}}\ and\ \bibinfo {author} {\bibfnamefont {R.}~\bibnamefont
  {Donnelly}},\ }\bibfield  {title} {\bibinfo {title} {Density and thermal
  expansion coefficient of liquid helium-4 from measurements of the dielectric
  constant},\ }\href {https://doi.org/10.1007/BF00754064} {\bibfield  {journal}
  {\bibinfo  {journal} {J. Low Temp. Phys}\ }\textbf {\bibinfo {volume} {98}},\
  \bibinfo {pages} {1} (\bibinfo {year} {1995})}\BibitemShut {NoStop}%
\end{thebibliography}%

\end{document}